%% file: Sequential-Detection-and-Estimation-of-Multipath-Channel-Parameters-Using-Belief-Propagation.tex
\documentclass[a4paper,conference]{IEEEtran}
\usepackage[table]{xcolor}
\pdfoutput=1

\usepackage{setspace}
\usepackage{bm}
\usepackage{mathtools}
\usepackage{graphicx}
\usepackage{relsize}
\usepackage{cite}
\usepackage{amsmath}
\usepackage{amsfonts}
\usepackage{color}
\usepackage{psfrag}
\usepackage{amssymb}
\usepackage{dblfloatfix}
\usepackage[font=small]{caption}
\usepackage{tabularx}
\usepackage{makecell}
\usepackage{lipsum}
\usepackage{float}
\usepackage{subfiles}
\usepackage{epstopdf}
\usepackage{verbatim}
\usepackage{subfig}
\usepackage{amsmath}
\usepackage{ifthen}
\usepackage{acronym}
\usepackage{notation}

\def\x{{\mathbf x}}

\acrodef{pnt}[PNT]{positioning, navigation and timing}
\acrodef{pa}[PA]{physical anchor}
\acrodef{bs}[BS]{base station}
\acrodef{awgn}[AWGN]{additive white Gaussian noise}
\acrodef{va}[VA]{virtual anchor}
\acrodef{mva}[MVA]{master virtual anchor}
\acrodef{mpc}[MPC]{multipath component}
\acrodef{nom}[NoM]{number-of-\acp{mpc}}
\acrodef{far}[MNFA]{mean number of false alarms}
\acrodef{pmpc}[PMPC]{potential \ac{mpc}}
\acrodef{smc}[SMC]{specular multipath component}
\acrodef{psmc}[PSMC]{potential \ac{smc}}
\acrodef{slam}[SLAM]{simultaneous localization and mapping}
\acrodef{pmf}[PMF]{probability mass function}
\acrodef{pdf}[PDF]{probability density function}
\acrodef{cdf}[CDF]{cummulative distribution function}
\acrodef{bp}[BP]{belief propagation}
\acrodef{spa}[SPA]{sum-product algorithm}
\acrodef{mmse}[MMSE]{minimum mean-square error}
\acrodef{simo}[SIMO]{single-input multiple-output}
\acrodef{miso}[MISO]{multiple-input single-output}
\acrodef{mimo}[MIMO]{multiple-input multiple-output}
\acrodef{ue}[UE]{mobile user}
\acrodef{tx}[Tx]{transmitter}
\acrodef{rx}[Rx]{receiver}
\acrodef{aoa}[AoA]{angle-of-arrival}
\acrodef{aod}[AoD]{angle-of-departure}
\acrodef{roi}[RoI]{region-of-interest}
\acrodef{snr}[SNR]{signal-to-noise ratio}
\acrodef{ospa}[OSPA]{optimal subpattern assignment}
\acrodef{mospa}[MOSPA]{mean \ac{ospa}}
\acrodef{dmc}[DMC]{dense multipath components}
\acrodef{sr}[SR]{super-resolution}
\acrodef{sbl}[SBL]{sparse Bayesian learning}
\acrodef{sde}[SDE]{sequential detection and estimation}
\acrodef{kest}[KEST]{Kalman enhanced super resolution tracking}
\acrodef{da}[DA]{data association}
\acrodef{fim}[FIM]{Fisher information matrix}
\acrodef{crlb}[CRLB]{Cram\'{e}r–Rao lower bound}
\acrodef{pcrlb}[PCRLB]{posterior \ac{crlb}}
\acrodef{va}[VA]{virtual anchor}
\acrodef{mnom}[MNoM]{mean estimated \ac{nom}}
\acrodef{uwb}[UWB]{ultra-wideband}

\acrodef{iid}[iid]{independent and identically distributed}

\newcommand{\ist}{\hspace*{.3mm}}
\newcommand{\rmv}{\hspace*{-.3mm}}
\newcommand{\iist}{\hspace*{1mm}}

\newcommand{\nn}{\nonumber}

\providecommand{\norm}[1]{\lVert#1\rVert}

\IEEEoverridecommandlockouts

\begin{document}
\allowdisplaybreaks
\frenchspacing

\title{Sequential Detection and Estimation of Multipath\\ Channel Parameters Using Belief Propagation}
\author{\normalsize \IEEEauthorblockN{Xuhong Li$^{\dagger}$,~\IEEEmembership{\normalsize Student Member,~IEEE}, Erik Leitinger$^{\dagger}$,~\IEEEmembership{\normalsize Member,~IEEE},\\ [1mm] Alexander Venus,~\IEEEmembership{\normalsize Student Member,~IEEE}, and Fredrik Tufvesson,~\IEEEmembership{\normalsize Fellow,~IEEE}\\ [1mm] {\small $^{\dagger}$ Have equally contributed as first authors.} }\\[-8mm]  
%
%
%
%

\thanks{
	Xuhong Li and and Fredrik Tufvesson are with the Department of Electrical and Information Technology, Lund University, 221 00 Lund, Sweden (e-mail: \{xuhong.li, fredrik.tufvesson\}@eit.lth.se). Erik Leitinger and Alexander Venus are with the Laboratory of Signal Processing and Speech Communication and the Christian Doppler Laboratory for Location-aware Electronic Systems, Graz University of Technology, 8010 Graz, Austria (e-mail: \{erik.leitinger, a.venus\}@tugraz.at). 
	
	This work was supported in part by the Swedish Research Council (VR), in part by the strategic research area Excellence Center at Link{\"{o}}ping---Lund in Information Technology (ELLIIT), in part by the Christian Doppler Research Association and in part by the TU Graz. 
	
	This article was presented in part at the 54th IEEE Asilomar Conference on Signals, Systems, and Computers, Pacific Grove, CA, USA, October 2020 (DOI: 10.1109/IEEECONF51394.2020.9443465). This article is accepted for publication in IEEE Transactions on Wireless Communications, 2022 (DOI:10.1109/TWC.2022.3165856). Compared to the IEEE version, this arXiv version provides additional simulation results in Appendix~\ref{app:AddSimResults}.

}

}

\maketitle

\begin{abstract}
	\subfile{./InputFiles/abstract}
\end{abstract}

\vspace*{3mm}

\IEEEpeerreviewmaketitle

\section{Introduction}
\subfile{./InputFiles/Introduction}

\section{Radio Signal Model}
\label{sec:SignalModel}
\subfile{./InputFiles/SignalModel}

\section{System Model}
\label{sec:SystemModel}
\subfile{./InputFiles/SystemModel}

\section{Joint Posterior \ac{pdf} and Problem Formulation}
\label{sec:JointPosteriorAndFG}
\subfile{./InputFiles/JointPosteriorAndFG}

\section{The Proposed Sum-Product Algorithm}
\label{sec:BPalgorithm}
\subfile{./InputFiles/BPalgorithm}

\section{Particle-based Implementation}
\label{sec:ParticleImplementation}
\subfile{./InputFiles/ParticleImplementation}

\section{Experimental Results}
\label{sec:ExperimentalResults}
\subfile{./InputFiles/ExperimentalResults}

\section{Conclusions}
\label{sec:Conclusions}
\subfile{./InputFiles/Conclusions}

\section{Acknowledgment}
\label{sec:Acknowledgment}
\subfile{./InputFiles/Acknowledgment}

\appendix
\section{Derivation of the Joint Posterior Distribution}
\label{sec:AppendixJointPosteriorDist}
\subfile{./InputFiles/AppendixJointPosteriorDist}


\bibliographystyle{IEEEtran}
\bibliography{IEEEabrv,./references}

\subfile{./InputFiles/bios}

\end{document}

%% file: InputFiles/abstract.tex
This paper proposes a \ac{bp}-based algorithm for sequential detection and estimation of \ac{mpc} parameters based on radio signals. Under dynamic channel conditions with moving transmitter/receiver, the number of \acp{mpc}, the \ac{mpc} dispersion parameters, and the number of false alarm contributions are unknown and time-varying. We develop a Bayesian model for sequential detection and estimation of \ac{mpc} dispersion parameters, and represent it by a factor graph enabling the use of \ac{bp} for efficient computation of the marginal posterior distributions. At each time step, a snapshot-based parametric channel estimator provides parameter estimates of a set of \acp{mpc} which are used as noisy measurements by the proposed \ac{bp}-based algorithm. It performs joint probabilistic data association, and estimation of the time-varying \ac{mpc} parameters and the mean number of false alarm measurements, by means of the sum-product algorithm rules. The algorithm also exploits amplitude information enabling the reliable detection of ``weak'' \acp{mpc} with very low component \acp{snr}. The performance of the proposed algorithm compares well to state-of-the-art algorithms for high \ac{snr} \acp{mpc}, but it significantly outperforms them for medium or low \ac{snr} \acp{mpc}. Results using real radio measurements demonstrate the excellent performance of the proposed algorithm in realistic and challenging scenarios.

%% file: InputFiles/Introduction.tex
The information of dispersive wireless radio channels and its temporal behavior in dynamic scenarios are of great importance for the development of radio channel models \cite{Morten_DynBirthDeath_2014, Pepe_TWC2020}, 5G wireless communication technologies \cite{Fredrik_SPMagazine2013,DiTaranto2014SPM,HuSha_TSP2018}, and multipath-based localization and mapping \cite{GentnerTWC2016, LeitingerICC2017, Erik_SLAM_TWC2019, MendrzikJSTSP2019,LeiMey:Asilomar2020_DataFusion}. The response of a non-static wireless radio channel is typically represented by superimposed weighted Dirac delta distributions with distinct and time-varying locations (or supports) in the respective dispersion domains (delay, \ac{aoa}, angle-of-departure, Doppler frequency, and combinations thereof). Each component is meant to represent a \acf{mpc}. In general, the channel response can be observed only within a finite aperture leading to some limitations on the ability to resolve \acp{mpc} closely spaced in the dispersion domains. The related time-varying \ac{mpc} parameters are usually estimated from multi-dimensional measurements using antenna arrays and multiple frequencies (wideband or \ac{uwb}) \cite{LeitingerGrebienTSP2021} using \ac{sr} algorithms that perform sequential estimation of \ac{mpc} parameters. 

\subsection{State-of-the-Art Methods}

If the \ac{nom} is known, subspace methods \cite{MUSIC_1986, ESPRIT_1989, HaardtTSP2008_TensorESPRIT} or maximum likelihood (ML) methods as for example \cite{Ottersten1993} are standard \ac{sr} methods to estimate time-invariant \ac{mpc} parameters. Expectation maximization-based methods \cite{FesslerTSP1994_SAGE}, have been proven a viable approximation of the computationally prohibitive ML methods \cite{SAGE_B_Fleury}. In recent years, a channel model has been introduced that also considers \ac{dmc} \cite{RichterPhD2005}. The \ac{dmc} incorporates \acp{mpc} that cannot be resolved due to finite observation aperture. Including the estimation of \ac{dmc} can improve the accuracy of the parameter estimation of distinct \acp{mpc}.

All afore mentioned methods have in common that they do not incorporate the estimation of the \ac{nom} into the estimation problem. One classical solution is to extend the methods with an outer stage for \ac{nom} detection using for example eigenvalue-based methods, or the generic information theoretic criteria, e.g., the Akaike/Bayesian information criterion, the minimum description length (MDL) principle \cite{ModelOrderSelection_SPM2004}. The outer stage schemes mostly tend to overestimate the \ac{nom}. Inspired by the ideas of sparse estimation and compressed sensing, some \ac{sr} sparse Bayesian parametric channel estimation algorithms \cite{ShutinTSP2011, BadiuTSP2017, HansenTSP2018, LeitingerGrebienTSP2021} have recently appeared which aim to reconstruct sparse signals from a reduced set of measurements specified by a sparse weight vector. By introducing a sparsity-promoting prior model for the weights, the estimation of the \ac{nom} and \ac{mpc} parameters can be jointly formulated in a Bayesian framework.

To capture the temporal behaviors of \ac{mpc} parameters in time-varying scenarios, many sequential estimation methods have been proposed, which can be grouped into two broad categories. Methods of the first category sequentially estimate the \ac{mpc} parameters directly based on the radio signals using for example an extended Kalman filter \cite{Jussi_TSP2009, Xuhong_TWC2019, ShutinCAMSAP2017, MeyerFusion2020}. Methods of the second category adopt a two-stage structure where the estimates of a snapshot-based channel estimator are used as measurements in a tracking filter \cite{KEST_TAP2012}. In this work, we focus on the two-stage methods. Due to the finite aperture of measurement systems and the resolution capability of snapshot-based parametric channel estimators, some measurements might not be well resolved hence incorporate contributions from more than one \acp{mpc} and false alarms may exist. In this case, to decide which measurement should be used for the update of which \ac{mpc} (i.e., \ac{da} problem) can be complicated. In general, existing sequential channel estimation methods adopt ``hard'' association which assumes that measurements are fine resolved and each of them originates from single \ac{mpc} that is specified by metrics such as the global nearest neighbor \cite{BarShalom_AlgorithmHandbook}. Probabilistic \ac{da}\cite{BarShalom_AlgorithmHandbook, Florian_Proceeding2018}, on the other hand, solves the origin uncertainty problem in a ``soft'' manner, in which the association probabilities for all current measurements are computed and used to form a mixture \ac{pdf} for the update of each \ac{mpc} state.

\subsection{Contributions} 

Here, we propose a \acf{bp}-based algorithm for sequential detection of \acp{mpc} and estimation of their dispersion parameters that uses the \ac{mpc} estimates from a snapshot-based parametric \ac{sr}-\ac{sbl} channel estimator (abbreviated as \ac{sr}-\ac{sbl}) as measurements \cite{ShutinCSTA2013, HansenTSP2018, BadiuTSP2017, LeitingerGrebienTSP2021}.{\footnote{This work is inspired by the \ac{bp}-based SLAM algorithm presented in \cite{Erik_SLAM_TWC2019, Erik_SSP2018, LeitingerICC2019}. The main novelties of the proposed algorithm over the previous work are threefold: (i) we present more detailed derivations for the joint prior \acp{pdf} and joint likelihoods; (ii) we extend the prior work with adaptive estimation of the \ac{far}; (iii) we apply the prior work successfully to the field of parametric radio channel estimation.}} To reduce the computational demand and to improve the convergence behavior of the snapshot-based estimator, statistical information of the dispersion parameters of the \ac{bp}-based algorithm is fed back to the \ac{sr}-\ac{sbl} channel estimator. The proposed algorithm jointly performs probabilistic \ac{da} and sequential estimation of \acf{pmpc} parameters by running the \ac{bp}-based algorithm, also known as the sum-product algorithm, on a factor graph \cite{Florian_Proceeding2018, Erik_SLAM_TWC2019, MeyWilJ21, MeyGemJ21}. Note that independence between \acp{mpc} is assumed throughout the work. We use a probabilistic model for \ac{mpc} existence where each \ac{pmpc} state is augmented by a binary existence variable and associated with an existence probability, which is also estimated and used for the detection of reliable \acp{mpc} modeling the birth and death of these components \cite{Florian_TSP2017, Erik_SLAM_TWC2019}. The complex amplitudes of \acp{mpc} are an integral part of the multipath channel model and must be estimated alongside with the dispersion parameters. Therefore, they are incorporated into the proposed algorithm. More specifically, the algorithm uses the statistics of \ac{mpc} amplitudes to determine the unknown and time-varying detection probabilities \cite{LerroACC1990,Erik_SSP2018, LeitingerICC2019, SoldiTSP2019}, which improves the detectability and maintenance of low \acf{snr} \acp{mpc}, and enables a better discrimination against false alarms. Knowing the correct \acf{far} is crucial for optimal performance of Bayesian detection and estimation algorithms. However, determining the \ac{far} in advance is not straightforward, especially, if the \ac{far} is time-varying. Using a fixed predefined value of \ac{far} that deviates largely from the true one, can potentially lead to decreased detection performance of \acp{mpc} and an increased number of detected false alarms. The proposed algorithm estimates the possibly time-varying \ac{far} to cope with false alarm measurements originating from the preprocessing step and clutter measurements originating from strongly fluctuating scattered \acp{mpc} (as for example \ac{dmc}). The main contributions of this paper are summarized as follows.
\begin{itemize}
	\item We introduce a Bayesian model for sequential detection and estimation of \ac{mpc} parameters, which uses the estimates from a snapshot-based channel estimator as measurements. Within this model, the death and birth of \acp{mpc} and \acp{da} are formulated probabilistically, and adaptive detection probabilities are incorporated by exploiting amplitude information. 
	\item We further present a \ac{bp}-based algorithm based on the factor graph representation of the estimation problem, where the \ac{pmpc} states and the \ac{far} are estimated jointly and sequentially. 
	\item The performance of the proposed algorithm is demonstrated with both synthetic and real measurements. Moreover, the results are compared with the \ac{kest} \cite{KEST_TAP2012} algorithm (a state-of-the-art sequential parametric channel estimation method), and \ac{pcrlb}.
\end{itemize}

This paper advances over our preliminary conference publication \cite{XuhongAsilomar2020} by (i) presenting a detailed derivations of the factor graph, (ii) including the detailed derivations of the factors related to the \ac{far}, (iii) establishing a particle-based implementation of the proposed algorithm, (iv) validating the performance with additional simulated and real scenarios, and (v) demonstrating a comparison with a state-of-the-art method and the \ac{pcrlb}.

\textit{Notations}: Column vectors and matrices are denoted by boldface lowercase and uppercase letters. Random variables are displayed in san serif, upright fonts as for example $\rv{x}$ and $\RV{x}$ and their realizations in serif, italic font as for example $x$ and $\V{x}$. $f(\V{x})$ denotes the \ac{pdf} or \ac{pmf} of continuous or discrete random vector. $(\cdot)^{\mathrm{T}}$, $(\cdot)^\ast$, and $(\cdot)^{\text{H}}$ denote matrix transpose, complex conjugation and Hermitian transpose, respectively. $ \norm{\cdot} $ is the Euclidean norm. $ \vert\cdot\vert $ represents the cardinality of a set. $ \mathrm{diag}\{\V{x}\} $ denotes a diagonal matrix with entries in $ \V{x} $. $ \mathrm{tr}\{\cdot\} $ denotes the trace of a matrix. $\M{I}_{[\cdot]}$ is an identity matrix of dimension given in the subscript. $[\V{X}]_{n,n}$ denotes the $n$th diagonal entry of $ \V{X} $. $ [\V{X}]_{1:n} $ denotes a sub-matrix containing $1\rmv\rmv:\rmv\rmv n$ columns and rows of $ \V{X} $. Furthermore, $ \bar{1}(e) $ denotes the function of the event $ e = 0 $ (i.e., $ \bar{1}(e) = 1 $ if $ e = 0 $ and 0 otherwise). ${1}_{\mathbb{A}}(\V{x})$ denotes the indicator function that is ${1}_{\mathbb{A}}(\V{x}) = 1$ if $\x \in \mathbb{A}$ and $ 0 $ otherwise. For any function $g(\V{x})$ we define the integral $\langle g(\V{x})\rangle_{f(\V{x})} = \int g(\V{x}) f(\V{x}) \mathrm{d}\V{x}$. As for example $\langle \V{x}\rangle_{f(\V{x})} = \int \V{x} f(\V{x}) \mathrm{d}\V{x}$ and $\langle f(\V{x})\rangle_{{1}_{\mathbb{A}}(\V{x})} = \int_{\mathbb{A}} f(\V{x})\mathrm{d}\V{x}$ denote the expected value and the integral over $f(\V{x})$ of the random vector $\V{x}$, respectively.

%% file: InputFiles/SignalModel.tex
We consider a \ac{simo} \ac{uwb} system, where the \ac{tx} uses a single antenna, and the \ac{rx} is equipped with an antenna array made of $ H $ elements located at $\V{p}_{n}^{(h)} \in \mathbb{R}^{2\times1}$ with $h \in \{1,\dots,H\}$.\footnote{Note that the extension of the algorithm to a \ac{miso} or a \ac{mimo} system considering an antenna array at the \ac{tx} is straightforward.} We define $d_{n}^{(h)} = \|\V{p}_{n}^{(h)} - \V{p}_{n}\|$ and $\varphi_{n}^{(h)} = \angle{(\V{p}_{n}^{(h)}-\V{p}_{n})}-\psi$, as the distance of the $h$th element to the reference location $\V{p}_{n} \in \mathbb{R}^{2\times1}$, i.e., the array's center of gravity, and its angle relative to the array orientation $\psi$, respectively. For the sake of brevity, we assume a two-dimensional scenario with horizontal-only propagation.\footnote{An extension to three-dimensional scenarios with horizontal and vertical propagation is straightforward, but it would lead to a cumbersome notation and one would not gain any further insights.} 

\subsection{Received Signal}

Signals are represented by their complex envelopes with respect to a center frequency $f_\mathrm{c}$. The single antenna at the transmitter emits a periodic signal $\tilde{s}(t)$. The received signal at each antenna element $h$ at the discrete observation time $n$ is given as \cite{LeitingerGrebienTSP2021,LeitingerAsilomar2020,RichterPhD2005}
\begin{align}
s_{\mathrm{RX},n}^{(h)}(t) &= \sum_{l=1}^{L_n} \tilde{\alpha}_{l,n} s\big(t;\tilde{d}_{l,n},\tilde{\varphi}_{l,n},\V{p}_{n}^{(h)}\big) + \omega_n^{(h)}(t)
\label{eq:SignalModel_continuous} 
\end{align}
where the first term comprises $ L_n $ \acp{mpc}, with each being characterized by its complex amplitude $ \tilde{\alpha}_{l,n} \in \mathbb{C}$, distance $ \tilde{d}_{l,n} = c\tilde{\tau}_{l,n} \in \mathbb{R}$ to the array's center of gravity directly related to the time delay via the speed of light $c$, and \ac{aoa} $ \tilde{\varphi}_{l,n} \in [-\pi,\pi)$ w.r.t. the array orientation. Under the far-field plane-wave assumption, the signal $s\big(t;\tilde{d},\tilde{\varphi},\V{p}^{(h)}\big)$ is given as $s\big(t; \tilde{d}, \tilde{\varphi},\V{p}^{(h)}\big) = \text{e}^{j2\pi f_\mathrm{c} g(\tilde{\varphi},\V{p}^{(h)})} \tilde{s}\big(t - \tilde{d}/c + g\big(\tilde{\varphi},\V{p}^{(h)}\big)\big)$. The function $g\big(\tilde{\varphi},\V{p}^{(h)}\big) = \frac{d^{(h)} \cos{( \tilde{\varphi} - \psi - \varphi^{(h)} )}}{c}$ gives the delay shift of a plane-wave incident with \ac{aoa} $\tilde{\varphi}$ being measured relatively to the array orientation $\psi$, at the $h$th antenna position w.r.t. the array's center of gravity. We assume that the \ac{tx} and \ac{rx} are time synchronized and the array orientation is known. The measurement noise process $\omega_n^{(h)}(t)$ in (\ref{eq:SignalModel_continuous}) is independent \ac{awgn} with double-sided power spectral density $ N_{0}/2$. 

The received signal in \eqref{eq:SignalModel_continuous} observed over a duration $ T $ is sampled with frequency $ f_s = 1/T_s $ at each time $n$, yielding a length $ N_{\mathrm{s}} = T/T_{\mathrm{s}} $ sample vector $ \RV{s}_{\mathrm{RX},n}^{(h)} \in \mathbb{C}^{ N_{\mathrm{s}} \times 1} $ from each array element. By stacking $ \RV{s}_{\mathrm{RX},n}^{(h)} $ from $ H $ array elements, the discrete time signal vector $ \RV{s}_{\mathrm{RX},n} \triangleq [\RV{s}_{\mathrm{RX},n}^{(1)\ist \mathrm{T}} \iist \cdots \iist \RV{s}_{\mathrm{RX},n}^{(H)\ist \mathrm{T}}]^{\mathrm{T}} \in \mathbb{C}^{ N_{\mathrm{s}}H \times 1} $ is given by 
\begin{align}
\V{s}_{\mathrm{RX},n} = \V{S}(\tilde{\V{\theta}}_{n}) \tilde{\V{\alpha}}_{n} + \V{\omega}_n.
\label{eq:SignalModel_sampled} 
\end{align}
In the first summand $ \tilde{\V{\alpha}}_{n} \triangleq [\tilde{\alpha}_{1,n} \ist \cdots \ist \tilde{\alpha}_{L_{n},n}]^{\mathrm{T}}\in \mathbb{C}^{ L_{n} \times 1} $, $ \tilde{\V{\theta}}_{n} \triangleq [\tilde{\V{\theta}}^{\ist\mathrm{T}}_{1,n} \ist \cdots \ist \tilde{\V{\theta}}^{\ist\mathrm{T}}_{L_{n},n} ]^{\mathrm{T}} \in \mathbb{R}^{2L_{n}\times1}$ with $ \tilde{\V{\theta}}_{l,n} \triangleq [\tilde{d}_{l,n} \iist \tilde{\varphi}_{l,n}]^{\mathrm{T}} \in \mathbb{R}^{2\times1} $ denoting the vector comprising the \ac{mpc} parameters, and $ \V{S}(\tilde{\V{\theta}}_{n}) \triangleq [\V{s}(\tilde{\V{\theta}}_{1,n}) \ist \cdots \ist $ $ \V{s}(\tilde{\V{\theta}}_{L_{n},n})] \in \mathbb{C}^{ N_{\mathrm{s}}H\times L_{n}} $ with columns given by $ \V{s}(\tilde{\V{\theta}}_{l,n}) = [\V{s}_{1}(\tilde{\V{\theta}}_{l,n})^\mathrm{T}\iist \cdots \iist \V{s}_H(\tilde{\V{\theta}}_{l,n})^\mathrm{T}]^\mathrm{T} \in \mathbb{C}^{ N_{\mathrm{s}}H\times 1}$. The $h$th entry of $\V{s}(\tilde{\V{\theta}}_{l,n})$ reads $\V{s}_h(\tilde{\V{\theta}}_{l,n}) \triangleq [s\big(-[(N-1)/2]T_\mathrm{s};\\ \tilde{d}_{l,n}, \tilde{\varphi}_{l,n},\V{p}^{(h)}\big) \iist \cdots \iist s\big([(N-1)/2] T_\mathrm{s}; \tilde{d}_{l,n}, \tilde{\varphi}_{l,n},\V{p}^{(h)}\big)]^\mathrm{T} \in \mathbb{C}^{N_{\mathrm{s}}\times 1}$. The measurement noise vector $\RV{\omega}_n \in \mathbb{C}^{ N_{\mathrm{s}}H \times 1} $ is a complex circularly symmetric Gaussian random vector with covariance matrix $\V{C} = \sigma^2 \M{I}_{N_{\mathrm{s}}H}$, where $\sigma^2 = N_0/T_{\mathrm{s}}$ is the noise variance. The component \ac{snr} of each \ac{mpc} is $ \widetilde{\mathrm{SNR}}_{l,n} = \frac{|\tilde{\alpha}_{l,n}|^2 \norm{\V{s}(\tilde{\V{\theta}}_{l,n})}^2} {\sigma^2}$ and the according normalized amplitude equals $\tilde{u}_{l,n} = \sqrt{\widetilde{\mathrm{SNR}}_{l,n}}$. An \ac{mpc} exists only when its associated geometric feature is visible at the \ac{rx} position. The true number $ L_n $ of \acp{mpc} as well as their individual parameters $\tilde{u}_{l,n}$, $\tilde{d}_{l,n}$, and $\tilde{\varphi}_{l,n}$ are unknown and time-varying in dynamic scenarios. We propose an algorithm to sequentially detect and estimate these parameters. 

\subsection{Parametric Channel Estimation}
\label{sec:paramCE}


Based on the discrete time signal vector in \eqref{eq:SignalModel_sampled}, a snapshot-based \ac{sr}-\ac{sbl} estimator \cite{ShutinTSP2011, ShutinCSTA2013, BadiuTSP2017, HansenTSP2018, LeitingerGrebienTSP2021} provides estimated dispersion parameters of $ M_{n}$ \acfp{mpc} stacked into the vector $ \V{z}_{n} \rmv\triangleq\rmv [\V{z}_{1,n}^{\mathrm{T}} \ist \cdots \ist \V{z}_{M_n,n}^{\mathrm{T}}]^{\mathrm{T}} \rmv\in\rmv \mathbb{R}^{3M_{n}\times1}$ at each time $n$. An entry $ \V{z}_{m,n} \rmv\triangleq\rmv [{z_\mathrm{d}}_{m,n} \iist {z_\mathrm{\varphi}}_{m,n} \iist {z_\mathrm{u}}_{m,n}]^{\mathrm{T}} \rmv\in\rmv \mathbb{R}^{3\times1}$ comprises the distance estimate ${z_\mathrm{d}}_{m,n}$, the \ac{aoa} estimate ${z_\mathrm{\varphi}}_{m,n}$, and the normalized amplitude estimate ${z_\mathrm{u}}_{m,n}$ of an \ac{mpc}. The normalized amplitude estimate is given by $ {z_\mathrm{u}}_{m,n} \triangleq |{\mu_{\alpha}}_{m,n}|/{\sigma_{\alpha}}_{m,n}$ where ${\mu_{\alpha}}_{m,n}$ and ${\sigma_{\alpha}}_{m,n}$ denote, respectively, the estimated mean and standard deviation of the complex amplitudes. Note that all estimated component \acp{snr} given by ${z^2_\mathrm{u}}_{m,n}$ are above a predefined detection threshold $u_{\mathrm{de}}$. The estimated amplitudes ${z_\mathrm{u}}_{m,n}$ are directly related to the detection probabilities of the corresponding \acp{mpc} \cite{Erik_SSP2018, LeitingerICC2019} (see Section~\ref{subsec:LikelihoodFunctions}). At time $ n = 1 $, the initial prior \ac{pdf} of the dispersion parameters considered in the snapshot-based estimator is uniform over the validation region. For each time $ n \geq 2 $, the snapshot-based estimator is initialized with $ \hat{L}_{n-1} $ detected \acp{mpc} returned from the \ac{bp}-based algorithm at time $ n-1 $ (see Section~\ref{subsec:detection_estimation}). The vector $ \V{z}_{n} $ is used as noisy measurement by the proposed algorithm. Note that the snapshot-based estimator decomposes $\V{s}_{\mathrm{RX},n}$ into individual, decorrelated components with parameters stacked into ${\V{z}_{n}}$, reducing the number of dimensions (as $ M_{n}$ is usually much smaller than $N_{\mathrm{s}}H$).

%% file: InputFiles/SystemModel.tex
\subsection{\ac{pmpc} States}
\label{subsec:PSMC_States}

Following \cite{Florian_Proceeding2018, Erik_SLAM_TWC2019}, we account for the time-varying and unknown \ac{nom} by introducing \acp{pmpc} indexed by $k \in \{1,\dots, K_{n}\}$. The number of \acp{pmpc} $K_n$ is the maximum number of actual \acp{mpc} that have produced measurements so far \cite{Florian_Proceeding2018}. The existence/non-existence of \ac{pmpc} $k$ as an actual \ac{mpc} is modeled by a binary random variable $ \rv{r}_{k,n} \in \mathbb{B} = \{0,1\} $ in the sense that a \ac{pmpc} exists if and only if $r_{k,n} = 1$. Augmented states of \acp{pmpc} are denoted as $ \RV{y}_{k,n} \triangleq [\RV{x}_{k,n}^{\mathrm{T}} \iist \rv{r}_{k,n}]^{\mathrm{T}} \in \mathbb{R}^{5\times1} \times \mathbb{B}$, where $ \RV{x}_{k,n} = [\RV{\theta}_{k,n}^{\mathrm{T}} \iist \rv{u}_{k,n} \iist {\rv{v}_{\mathrm{d}}}_{k,n}\iist {\rv{v}_\mathrm{\varphi}}_{k,n}]^{\mathrm{T}} \rmv\rmv\in\rmv\rmv \mathbb{R}^{5\times1}$, $ \RV{\theta}_{k,n} \rmv\rmv= [\rv{d}_{k,n} \iist \rv{\varphi}_{k,n}]^{\mathrm{T}} \rmv\rmv\in \rmv\rmv \mathbb{R}^{2\times1}$, and $ {\rv{v}_{\mathrm{d}}}_{k,n} $ and $ {\rv{v}_\mathrm{\varphi}}_{k,n} $ are the distance velocity and angular velocity, respectively.

Formally, \ac{pmpc} $k$ is also considered even if it is non-existent, i.e., $r_{k,n} = 0$. The states $\RV{\V{x}}_{k,n}$ of non-existent \acp{pmpc} are obviously irrelevant and have no influence on the \ac{pmpc} detection and state estimation. Therefore, all \acp{pdf} defined for \ac{pmpc} states, $ f(\V{y}_{k,n}) = f(\V{x}_{k,n},r_{k,n}) $, are of the form $ f(\V{x}_{k,n},r_{k,n}=0) = f_{k,n} f_{\mathrm{D}}(\V{x}_{k,n}) $, where $ f_{\mathrm{D}}(\V{x}_{k,n}) $ is an arbitrary ``dummy \ac{pdf}'' and $ f_{k,n} \in [0,1]$ is a constant representing the probability of non-existence \cite{Florian_Proceeding2018, Florian_TSP2017, Erik_SLAM_TWC2019}. 

\subsection{State-Transition Model}

For each \ac{pmpc} with state $\RV{y}_{k,n-1} $ with $k \rmv\in\rmv \{1,\dots, K_{n-1}\}$ at time $n-1$, there is one ``legacy'' \ac{pmpc} with state $ \underline{\RV{y}}_{k,n} \rmv\rmv\triangleq\rmv\rmv [\underline{\RV{x}}_{k,n}^{\mathrm{T}} \iist \underline{\rv{r}}_{k,n}]^{\mathrm{T}} $ with $k \rmv\rmv\in\rmv\rmv \{1,\dots, K_{n-1}\}$ at time $n$. Assuming that \ac{pmpc} states evolve independently across $ k $ and $ n $, the corresponding state-transition \ac{pdf} of the joint states $ \RV{y}_{n-1} \rmv\rmv\triangleq\rmv\rmv [\RV{y}_{1,n-1}^{\mathrm{T}} \ist \cdots \ist \RV{y}_{K_{n-1},n-1}^{\mathrm{T}} ]^{\mathrm{T}} $ and $ \underline{\RV{y}}_{n} \rmv\rmv\triangleq\rmv\rmv [\underline{\RV{y}}_{1,n}^{\mathrm{T}} \ist \cdots \ist  \underline{\RV{y}}_{K_{n-1},n}^{\mathrm{T}} ]^{\mathrm{T}} $ factorizes as \cite{Florian_Proceeding2018}
\begin{align}
	f(\underline{\V{y}}_{n}|\V{y}_{n-1}) = \prod_{k=1}^{K_{n-1}} f(\underline{\V{y}}_{k,n}|\V{y}_{k,n-1})
	\label{eq:StateTransPDF_PSMC}
\end{align}
where $ f(\underline{\V{y}}_{k,n}|\V{y}_{k,n-1})\rmv\rmv=\rmv\rmv f(\underline{\V{x}}_{k,n}, \underline{r}_{k,n}|\V{x}_{k,n-1}, r_{k,n-1}) $ is the single \ac{pmpc} state-transition \ac{pdf}. If a \ac{pmpc} did not exist at time $ n\rmv\rmv-\rmv\rmv1 $, i.e., $ r_{k,n-1}\rmv\rmv=\rmv\rmv 0 $, it cannot exist at time $ n $ as a legacy \ac{pmpc}. This means that 
\begin{align}
	f(\underline{\V{x}}_{k,n}, \underline{r}_{k,n}|\V{x}_{k,n-1}, 0) =
	\begin{cases}
		f_{\mathrm{D}}(\underline{\V{x}}_{k,n}), 	&\underline{r}_{k,n}= 0\\
		0, 											&\underline{r}_{k,n}= 1 \, .
	\end{cases}
	\label{eq:ST_pdf1} 
\end{align}
If a \ac{pmpc} existed at time $ n-1 $, i.e., $ r_{k,n-1} = 1 $, at time $n$ it either dies i.e., $ \underline{r}_{k,n} = 0 $ or it still exists i.e., $ \underline{r}_{k,n} = 1 $, with the survival probability denoted as $ p_{\mathrm{s}} $. If it does survive, the state $ \underline{\RV{x}}_{k,n} $ is distributed according to the state-transition \ac{pdf} $ f(\underline{\V{x}}_{k,n}|\V{x}_{k,n-1}) $. Thus, 
\begin{align}
	f(\underline{\V{x}}_{k,n}, \underline{r}_{k,n}|\V{x}_{k,n-1}, 1) = 
	\begin{cases}
		(1-p_{\mathrm{s}})f_{\mathrm{D}}(\underline{\V{x}}_{k,n}),       			&\underline{r}_{k,n}= 0\\
		p_{\mathrm{s}}f(\underline{\V{x}}_{k,n}|\V{x}_{k,n-1}),	&\underline{r}_{k,n} = 1. \, 
	\end{cases}
	\label{eq:ST_pdf2} 
\end{align}
We also define the state vector for all times up to $n$ of legacy \acp{pmpc} as $ \underline{\RV{y}}_{1:n} \triangleq [\underline{\RV{y}}_{1}^{\mathrm{T}} \ist \cdots \ist  \underline{\RV{y}}_{n}^{\mathrm{T}} ]^{\mathrm{T}} $.

\subsection{Measurement Model}
\label{subsec:LikelihoodFunctions} 

Before the current measurements $\RV{z}_{n}$ are observed, the number of measurements $\rv{M}_{n}$ is a random variable. The vector collecting the number of measurements is defined as $\RV{m}_{1:n} \triangleq [\rv{M}_{1}\ist\cdots\ist \rv{M}_{n}]^{\mathrm{T}}$. The conditional \ac{pdf} $ f(\V{z}_{m,n}|\V{x}_{k,n})$ of $\RV{z}_{n}$ assumes that the individual measurements $ \RV{z}_{m,n} $ are conditionally independent given the state $ \RV{x}_{k,n}$. At each time $n$, a snapshot-based channel estimator provides the current observed measurement vector $ \V{z}_{n} = [\V{z}_{1,n}^{\mathrm{T}} \ist \cdots \ist  \V{z}_{M_n,n}^{\mathrm{T}}]^{\mathrm{T}}$ (see Section~\ref{sec:SignalModel}), which is not random anymore and with fixed $M_{n} $. If $ \V{z}_{m,n} $ is a \ac{pmpc}-oriented measurement, we assume that the conditional \ac{pdf} $ f(\V{z}_{m,n}|\V{x}_{k,n}) $ is conditionally independent across ${z_\mathrm{d}}_{m,n}$, ${z_\mathrm{\varphi}}_{m,n}$, and ${z_\mathrm{u}}_{m,n}$ given the states $\rv{d}_{k,n}$, $\rv{\varphi}_{k,n}$, and $\rv{u}_{k,n}$, thus it factorizes as
\begin{align}
	f(\V{z}_{m,n}|\V{x}_{k,n}) & = f({z_\mathrm{d}}_{m,n} | d_{k,n}, u_{k,n}) f({z_\mathrm{\varphi}}_{m,n} | \varphi_{k,n}, u_{k,n}) \nonumber \\
	& \quad \times f({z_\mathrm{u}}_{m,n} | u_{k,n})
	\label{eq:LHF5}
\end{align}
where the individual likelihood functions of the distance measurement $ f({z_\mathrm{d}}_{m,n} | d_{k,n}, u_{k,n}) $ and the \ac{aoa} measurement $ f({z_\mathrm{\varphi}}_{m,n} | \varphi_{k,n}, u_{k,n}) $ are modeled by Gaussian \acp{pdf}, i.e.,
\begin{align}
	f({z_\mathrm{d}}_{m,n} | d_{k,n}, u_{k,n}) = \big(2\pi{\sigma^2_\mathrm{d}}_{k,n} \big)^{-\frac{1}{2}} \mathrm{e}^{ \frac{-({z_\mathrm{d}}_{m,n} - d_{k,n})^2}{2{\sigma^2_\mathrm{d}}_{k,n}}}
	\label{eq:pdf_distance}
\end{align}
and 
\begin{align}
	f({z_\mathrm{\varphi}}_{m,n} | \varphi_{k,n}, u_{k,n}) = \big(2\pi{\sigma^2_\mathrm{\varphi}}_{k,n}\big)^{-\frac{1}{2}} \mathrm{e}^{\frac{-({z_\mathrm{\varphi}}_{m,n} - \varphi_{k,n})^2}{2{\sigma^2_\mathrm{\varphi}}_{k,n}}}.
	\label{eq:pdf_AoA}
\end{align}
The variances depend on $u_{k,n}$ and are determined based on the Fisher information, respectively, given by $ {\sigma^2_\mathrm{d}}_{k,n}\rmv\rmv=\rmv\rmv c^2/(8\pi^2 \beta_{\mathrm{bw}}^2 u_{k,n}^2) $ and $ {\sigma^2_\mathrm{\varphi}}_{k,n}\rmv\rmv=\rmv\rmv c^2/(8\pi^2f_{\mathrm{c}}^2 u_{k,n}^2 D^2(\varphi_{k,n})) $, where the latter is a function of the \ac{pmpc} state $\rv{\varphi}_{k,n}$ \cite{Thomas_Asilomar2018, LeitingerICC2019}. Here, $ \beta_{\mathrm{bw}}^2$ is the mean square bandwidth of the signal $\tilde{s}(t)$ and $ D^2(\varphi_{k,n}) $ is the squared array aperture. The likelihood function $ f({z_\mathrm{u}}_{m,n} | u_{k,n}) $ of the normalized amplitude measurement $ {z_\mathrm{u}}_{m,n} $ is modeled by a truncated Rician \ac{pdf} \cite[Ch.\,1.6.7]{BarShalom_AlgorithmHandbook}, i.e.,
\begin{align}
	& f({z_\mathrm{u}}_{m,n} | u_{k,n}) = \frac{  \frac{{z_\mathrm{u}}_{m,n}}{{\sigma^2_\mathrm{u}}_{k,n}} \mathrm{e}^{\big(\frac{ -({z^2_\mathrm{u}}_{m,n} + u_{k,n}^2)}{2{\sigma^2_\mathrm{u}}_{k,n}} \big)} \mathrm{I}_{0}(\frac{{z_\mathrm{u}}_{m,n} u_{k,n}}{{\sigma^2_\mathrm{u}}_{k,n}}) } { p_{\mathrm{d}}(u_{k,n}) }
	\label{eq:pdf_normAmplitude}
\end{align} 
for $ {z_\mathrm{u}}_{m,n} > \sqrt{u_{\mathrm{de}}} $, where the squared scale parameter $ {\sigma^2_\mathrm{u}}_{k,n} $ depends on $u_{k,n}$ and is determined based on the Fisher information given by $ {\sigma^2_\mathrm{u}}_{k,n} = \frac{1}{2} + \frac{1}{4N_{\mathrm{s}}H}u_{k,n}^2 $, $ \mathrm{I}_{0}(\cdot) $ represents the $ 0 $th-order modified first-kind Bessel function, and $u_{\mathrm{de}}$ is the detection threshold of the snapshot-based estimator. The derivation of the squared scale parameter is given in Appendix~\ref{app:normAmplitude}. The detection probability (i.e., the probability that a \ac{pmpc} $ \RV{y}_{k,n} $ generates a measurement $ \V{z}_{m,n} $) is modeled by a Rician \ac{cdf}, i.e., $ p_{\mathrm{d}}(u_{k,n}) = Q_{1}(u_{k,n}/{\sigma_\mathrm{u}}_{k,n}, \sqrt{u_{\mathrm{de}}}/{\sigma_\mathrm{u}}_{k,n}) $ \cite{LerroACC1990,Erik_SSP2018, LeitingerICC2019}, where the \ac{cdf} $ Q_{1}(\cdot,\cdot) $ denotes the Marcum Q-function \cite[Ch.\, 1.6.7]{BarShalom_AlgorithmHandbook}. Note that $p_{\mathrm{d}}(u_{k,n})$ is directly related to the \ac{mpc}'s visibility and the component \ac{snr} as well as to the snapshot-based estimator.

False alarm measurements originating from the snapshot-based parametric channel estimator are assumed statistically independent of \ac{pmpc} states. They are modeled by a Poisson point process with mean $ {\rv{\mu}_{\mathrm{fa}}}_{n} $ and \ac{pdf} $ f_{\mathrm{fa}}(\V{z}_{m,n}) $, which is assumed to factorize as $ f_{\mathrm{fa}}(\V{z}_{m,n}) = {f_{\mathrm{fa}}}_{\mathrm{d}}({z_\mathrm{d}}_{m,n}) {f_{\mathrm{fa}}}_{\mathrm{\varphi}}({z_\varphi}_{m,n}){f_{\mathrm{fa}}}_{\mathrm{u}}({z_\mathrm{u}}_{m,n}) $, where $ {f_{\mathrm{fa}}}_{\mathrm{d}}({z_\mathrm{d}}_{m,n})\rmv\rmv=\rmv\rmv 1/d_{\mathrm{max}} $ and $ {f_{\mathrm{fa}}}_{\mathrm{\varphi}}({z_\mathrm{\varphi}}_{m,n})\rmv\rmv =\rmv\rmv 1/2\pi $ are assumed to be uniform on $[0,\ist d_{\mathrm{max}}]$ and on $[0,\ist 2\pi)$, respectively. With noise only, the Rician \ac{pdf} in \eqref{eq:pdf_normAmplitude} degenerates to a Rayleigh \ac{pdf}, thus the false alarm \ac{pdf} $ {f_{\mathrm{fa}}}_{\mathrm{u}}({z_\mathrm{u}}_{m,n}) $ of the normalized amplitude is given as $ {f_{\mathrm{fa}}}_{\mathrm{u}} ({z_\mathrm{u}}_{m,n})\rmv\rmv=\rmv\rmv 2{z_\mathrm{u}}_{m,n} \exp (-{z^2_\mathrm{u}}_{m,n} )/p_{\mathrm{fa}}$ for ${z_\mathrm{u}}_{m,n}\rmv\rmv >  \sqrt{u_{\mathrm{de}}}$ where $ p_{\mathrm{fa}} = \exp(-u_{\mathrm{de}}) $ denotes the false alarm probability \cite[Ch.\, 1.6.7]{BarShalom_AlgorithmHandbook}. The amplitude information can significantly improve the detectability of \acp{mpc} with low component \acp{snr} as shown in Section~\ref{sec:resultSynRadioMeas} using very challenging simulation setups (see Fig.~\ref{fig:aveMOSPA_eachSNR}).

The \ac{far} $ {\rv{\mu}_{\mathrm{fa}}}_{n} $ is assumed unknown, time-varying, and automatically adapted online in the proposed algorithm. It evolves across time according to the state-transition \ac{pdf} $f({\mu_{\mathrm{fa}}}_{n}|{\mu_{\mathrm{fa}}}_{n-1})$. The state vector for all times up to $n$ is given as ${\RV{\mu}_{\mathrm{fa}}}_{1:n} \triangleq [{\rv{\mu}_{\mathrm{fa}}}_{1} \ist\cdots\ist {\rv{\mu}_{\mathrm{fa}}}_{n}]^{\mathrm{T}}$.

\subsection{New \acp{pmpc}}
\label{sec:NEWPSMC}

Newly detected \acp{pmpc} at time $n$, i.e., \acp{pmpc} that generate measurements for the first time at time $n$, are modeled by a Poisson point process with mean $\rv{\mu}_{\mathrm{n}}$ and \ac{pdf} $f_{\mathrm{n}}(\overline{\V{x}}_{m,n})$. The mean $\rv{\mu}_{\mathrm{n}}$ is assumed to be a known constant. Following \cite{Florian_Proceeding2018, Erik_SLAM_TWC2019}, newly detected \acp{pmpc} are represented by new \ac{pmpc} states $ \overline{\RV{y}}_{m,n} \triangleq [\overline{\RV{x}}_{m,n}^{\mathrm{T}}\iist \overline{\rv{r}}_{m,n}]^{\mathrm{T}} $, $ m \in \{1,\dots, M_n\} $. Each new \ac{pmpc} $ \overline{\RV{y}}_{m,n} $ corresponds to a measurement $\V{z}_{m,n}$, thus the number of new \acp{pmpc} at time $n$ equals to the number of measurements $M_{n}$. Here, $\overline{r}_{m,n} = 1$ means that the measurement $\V{z}_{m,n}$ was generated by a newly detected \ac{pmpc}. The state vector of all new \acp{pmpc} at time $n$ is given by $ \overline{\RV{y}}_{n} \triangleq [\overline{\RV{y}}_{1,n}^{\mathrm{T}} \ist\cdots\ist \overline{\RV{y}}_{\rv{M}_{n},n}^{\mathrm{T}} ]^{\mathrm{T}} $ and state vector for all times up to $n$ by $ \overline{\RV{y}}_{1:n} \triangleq [\overline{\RV{y}}_{1}^{\mathrm{T}} \ist\cdots\ist \overline{\RV{y}}_{n}^{\mathrm{T}} ]^{\mathrm{T}} $. The new \acp{pmpc} become legacy \acp{pmpc} at time $n+1$, accordingly the number of legacy \acp{pmpc} is updated as $K_{n} = K_{n-1} + M_{n}$. (The number of \acp{pmpc} is bounded by a pruning operation as it is detailed in Section~\ref{subsec:detection_estimation}.) The vector containing all \ac{pmpc} states at time $n$ is given by $ \RV{y}_{n} \triangleq [\underline{\RV{y}}_{n}^{\mathrm{T}} \iist \overline{\RV{y}}_{n}^{\mathrm{T}} ]^{\mathrm{T}} $, where $\RV{y}_{k,n}$ with $ k \in \{1,\dots, K_{n}\} $, and state vector for all times up to $n$ by $ \RV{y}_{1:n} \triangleq [\RV{y}_{1}^{\mathrm{T}} \ist\cdots\ist \RV{y}_{n}^{\mathrm{T}} ]^{\mathrm{T}} $.

\subsection{Data Association Uncertainty}
\label{sec:DA}

Estimation of multiple \ac{pmpc} states is complicated by the \ac{da} uncertainty, i.e., it is unknown which measurement $\RV{z}_{m,n} $ originated from which \ac{pmpc}. Furthermore, it is not known if a measurement did not originate from a \ac{pmpc} (false alarm), or if a \ac{pmpc} did not generate any measurement (missed detection). The associations between measurements and legacy \acp{pmpc} are described by the \ac{pmpc}-oriented association vector $ \RV{a}_{n} \triangleq [\rv{a}_{1,n} \ist \cdots \ist  \rv{a}_{\rv{K}_{n-1},n}]^{\mathrm{T}} $ with entries $ a_{k,n} \triangleq m \rmv\in\rmv \{1,\dots, M_n\}$, if legacy \ac{pmpc} $ k $ generates measurement $ m $, or $ a_{k,n} \triangleq 0 $, if legacy \ac{pmpc} $ k $ does not generate any measurement. In line with \cite{WilliamsLauTAE2014, Florian_Proceeding2018, Erik_SLAM_TWC2019}, the associations can be equivalently described by a measurement-oriented association vector $ \RV{b}_{n} \triangleq [\rv{b}_{1,n} \ist \cdots \ist \rv{b}_{\rv{M}_{n},n}]^{\mathrm{T}} $ with entries $ b_{m,n} \triangleq k \rmv\in\rmv \{1,\dots, K_{n-1}\} $, if measurement $ m $ is generated by legacy \ac{pmpc} $ k $, or $ b_{m,n} \triangleq 0 $, if measurement $ m $ is not generated by any legacy \ac{pmpc}. Furthermore, we assume that at any time $ n $, each \ac{pmpc} can generate at most one measurement, and each measurement can be generated by at most one \ac{pmpc} \cite{WilliamsLauTAE2014, Florian_Proceeding2018, Erik_SLAM_TWC2019}. This is enforced by the exclusion functions $ \Psi(\V{a}_n,\V{b}_n) $ and $ \Gamma_{\V{a}_{n}}(\overline{r}_{m,n}) $. The function $ \Psi(\V{a}_n,\V{b}_n) = \prod_{k = 1}^{K_{n-1}}\prod_{m = 1}^{M_{n}}\psi(a_{k,n},b_{m,n})$ is defined as $ \psi(a_{k,n},b_{m,n}) = 0 $, if $ a_{k,n} = m $ and $ b_{m,n} \neq k $ or $ b_{m,n} = k $ and $ a_{k,n} \neq m $, otherwise it equals $ 1 $. The function $ \Gamma_{\V{a}_{n}}(\overline{r}_{m,n}) = 0 $, if $ \overline{r}_{m,n} = 1 $ and $ a_{k,n} = m $, otherwise it equals $ 1 $. The ``redundant formulation'' of using $ \RV{a}_{n} $ together with $ \RV{b}_{n} $ is the key to make the algorithm scalable for large numbers of \acp{pmpc} and measurements. The association vectors for all times up to $n$ are given by $\RV{a}_{1:n} \triangleq [\RV{a}_{1}^{\mathrm{T}} \ist\cdots\ist \RV{a}_{n}^{\mathrm{T}} ]^{\mathrm{T}}$ and $ \RV{b}_{1:n} \triangleq [\RV{b}_{1}^{\mathrm{T}} \ist\cdots\ist \RV{b}_{n}^{\mathrm{T}} ]^{\mathrm{T}} $. A figure that conceptually illustrates the \acp{da} between measurements and \acp{pmpc} is given in \cite{XuhongAsilomar2020}.

%% file: InputFiles/JointPosteriorAndFG.tex
\subsection{Joint Posterior \ac{pdf} and Factor Graph}
\label{subsec:JointPosteriorAndFG}
Here, we assume that the measurements $\V{z}_{1:n} \triangleq [\V{z}_{1}^{\mathrm{T}}, \dots, \V{z}_{n}^{\mathrm{T}}]^{\mathrm{T}}$ for all times up to $n$ are observed and thus fixed. By using common assumptions \cite{BarShalom_AlgorithmHandbook, Florian_Proceeding2018, Erik_SLAM_TWC2019}, the joint posterior \ac{pdf} of $\RV{y}_{1:n}$, $\RV{a}_{1:n}$, $\RV{b}_{1:n}$, ${\RV{\mu}_{\mathrm{fa}}}_{1:n}$, and $\RV{m}_{1:n}$ conditioned on the observed measurement vector $\V{z}_{1:n}$ is given by
\begin{align}
& f(\V{y}_{1:n}, \V{a}_{1:n}, \V{b}_{1:n}, {\V{\mu}_{\mathrm{fa}}}_{1:n}, \V{m}_{1:n} | \V{z}_{1:n}) \nn \\
& \qquad = f(\underline{\V{y}}_{1:n}, \overline{\V{y}}_{1:n}, \V{a}_{1:n}, \V{b}_{1:n}, {\V{\mu}_{\mathrm{fa}}}_{1:n}, \V{m}_{1:n} | \V{z}_{1:n}) \nn \\
& \qquad \propto f(\V{z}_{1:n}|\underline{\V{y}}_{1:n}, \overline{\V{y}}_{1:n}, \V{a}_{1:n}, \V{b}_{1:n}, \V{m}_{1:n}) \nn \\
& \qquad \quad \times f(\underline{\V{y}}_{1:n}, \overline{\V{y}}_{1:n}, \V{a}_{1:n}, \V{b}_{1:n}, {\V{\mu}_{\mathrm{fa}}}_{1:n}, \V{m}_{1:n}).
\label{eq:TheJointPosteriorPDF_1} 
\end{align}
After inserting the expressions \eqref{eq:jointPriorPDF_global} and \eqref{eq:LHF_global} and performing some simple manipulations, the joint posterior \ac{pdf} in \eqref{eq:TheJointPosteriorPDF_1} can be rewritten as 
\begin{align}
& f(\underline{\V{y}}_{1:n}, \overline{\V{y}}_{1:n}, \V{a}_{1:n}, \V{b}_{1:n}, {\V{\mu}_{\mathrm{fa}}}_{1:n}, \V{m}_{1:n} | \V{z}_{1:n}) \nn \\
& \quad \propto f({\mu_{\mathrm{fa}}}_{1}) \prod_{l = 1}^{M_{1}} h(\overline{\V{y}}_{l,1}, b_{l,1}, {\mu_{\mathrm{fa}}}_{1}; \V{z}_{1}) \nn \\
& \qquad \times \prod_{n' = 2}^{n} f({\mu_{\mathrm{fa}}}_{n'} | {\mu_{\mathrm{fa}}}_{n'-1}) \left( \prod_{k' = 1}^{K_{n'-1}} f(\underline{\V{y}}_{k',n'}|\V{y}_{k',n'-1}) \right) \nn \\
& \qquad \times \left(\prod_{k = 1}^{K_{n'-1}} g(\underline{\V{y}}_{k,n'}, a_{k,n'}, {\mu_{\mathrm{fa}}}_{n'}; \V{z}_{n'}) \prod_{m = 1}^{M_{n'}} \psi(a_{k,n'},b_{m,n'})\right) \nn \\
& \qquad \times \left(\prod_{m' = 1}^{M_{n'}} h(\overline{\V{y}}_{m',n'}, b_{m',n'}, {\mu_{\mathrm{fa}}}_{n'}; \V{z}_{n'})\right)
\label{eq:TheJointPosteriorPDF_2}
\end{align}
where the functions $ g(\underline{\V{y}}_{k,n}, a_{k,n}, {\mu_{\mathrm{fa}}}_{n}; \V{z}_{n}) $ and $ h(\overline{\V{y}}_{m,n}, b_{m,n},\\ {\mu_{\mathrm{fa}}}_{n}; \V{z}_{n}) $ will be discussed next.

\begin{figure*}[t]
	\centering
	\includegraphics[width=0.8\textwidth, height=0.4\textwidth]{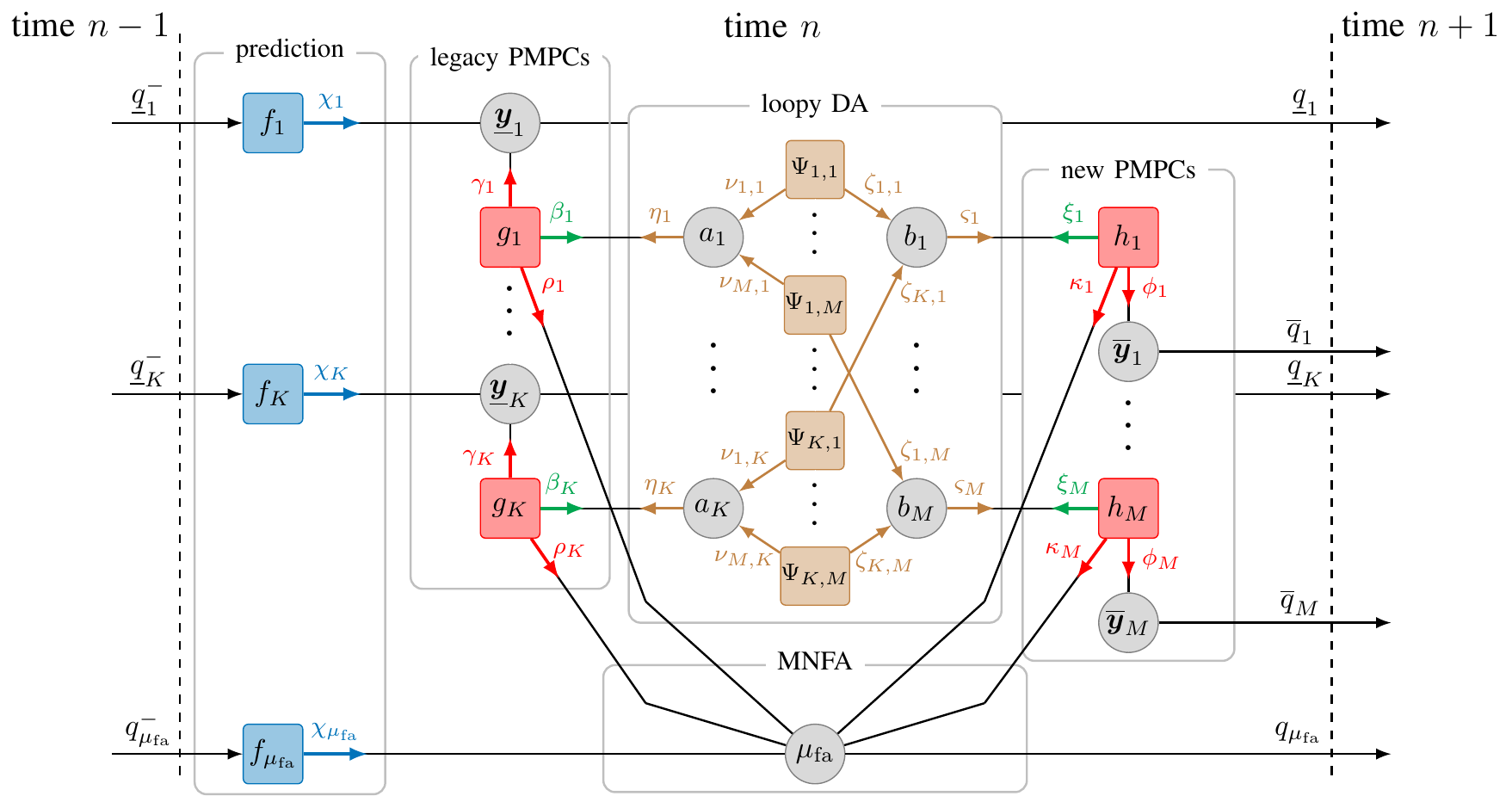}\\[1mm]
	\caption{Factor graph representation of the factorized joint posterior \ac{pdf} (\ref{eq:TheJointPosteriorPDF_2}). For simplicity, the following short notations are used: $ K \triangleq K_{n-1} $, $ M \triangleq M_{n} $; \emph{variable nodes}: $ a_{k} \triangleq a_{k,n} $, $ b_{m} \triangleq b_{m,n} $, $ \mu_{\mathrm{fa}} \triangleq {\mu_{\mathrm{fa}}}_{n} $, $ \underline{\V{y}}_{k} \triangleq \underline{\V{y}}_{k,n} $, $ \overline{\V{y}}_{m} \triangleq \overline{\V{y}}_{m,n} $; \emph{factor nodes}: $ f_{k} \triangleq f(\underline{\V{y}}_{k,n}|\V{y}_{k,n-1}) $, $ g_{k} \triangleq g(\underline{\V{x}}_{k,n}, \underline{r}_{k,n}, a_{k,n}, {\mu_{\mathrm{fa}}}_{n}; \V{z}_{n}) $, $ h_{m} \triangleq h(\overline{\V{x}}_{m,n}, \overline{r}_{m,n}, b_{m,n}, {\mu_{\mathrm{fa}}}_{n}; \V{z}_{n}) $, $ f_{\mu_{\mathrm{fa}}} \triangleq f( {\mu_{\mathrm{fa}}}_{n} | {\mu_{\mathrm{fa}}}_{n-1} ) $, $ \Psi_{k,m} \triangleq \Psi({a_{k,n},b_{m,n}}) $; \emph{prediction}: $ \chi_{k} \triangleq \chi(\underline{\V{x}}_{k,n}, \underline{r}_{k,n}) $, $ \chi_{\mu_{\mathrm{fa}}} \triangleq \chi({\mu_{\mathrm{fa}}}_{n}) $; \emph{measurement evaluation}: $ \beta_{k} \triangleq \beta(a_{k}) $, $ \xi_{m} \triangleq \xi(b_{m}) $; \emph{loopy DA}: $ \nu_{m,k} \triangleq \nu_{m \rightarrow k}(a_{k,n}) $, $ \zeta_{k,m} \triangleq \zeta_{k \rightarrow m}(b_{m,n}) $, $ \eta_{k} \triangleq \eta(a_{k}) $, $ \varsigma_{m} \triangleq \varsigma(b_{m}) $; \emph{measurement update}: $ \gamma_{k} \triangleq \gamma(\underline{\V{x}}_{k,n}, \underline{r}_{k,n}) $, $ \rho_{k} \triangleq \rho_{k}({\mu_{\mathrm{fa}}}_{n}) $, $ \phi_{m} \triangleq \phi(\overline{\V{x}}_{m,n}, \overline{r}_{m,n}) $, $ \kappa_{m} \triangleq \kappa_{m}({\mu_{\mathrm{fa}}}_{n}) $; belief calculation: $ \underline{q}^{-}_{k} \triangleq \underline{q}(\V{x}_{k,n-1}, r_{k,n-1}) $, $ \underline{q}_{k} \triangleq \underline{q}(\underline{\V{x}}_{k,n}, \underline{r}_{k,n}) $, $ \overline{q}_{m} \triangleq \overline{q}(\overline{\V{x}}_{m,n}, \overline{r}_{m,n}) $, $ q^{-}_{\mu_{\mathrm{fa}}} \triangleq q({\mu_{\mathrm{fa}}}_{n-1}) $, $ q_{\mu_{\mathrm{fa}}} \triangleq q({\mu_{\mathrm{fa}}}_{n}) $.}	 
	\label{fig:factorGraph}
\end{figure*}

The pseudo likelihood functions $ g(\underline{\V{y}}_{k,n}, a_{k,n}, {\mu_{\mathrm{fa}}}_{n}; \V{z}_{n})$ $ = g(\underline{\V{x}}_{k,n}, \underline{r}_{k,n}, a_{k,n}, {\mu_{\mathrm{fa}}}_{n}; \V{z}_{n}) $ and $ h(\overline{\V{y}}_{m,n}, b_{m,n}, {\mu_{\mathrm{fa}}}_{n}; \V{z}_{n})$ $ = h(\overline{\V{x}}_{m,n}, \overline{r}_{m,n}, b_{m,n}, {\mu_{\mathrm{fa}}}_{n}; \V{z}_{n}) $ are given by 
\begin{align}
& g(\underline{\V{x}}_{k,n}, \underline{r}_{k,n} = 1, a_{k,n}, {\mu_{\mathrm{fa}}}_{n}; \V{z}_{n}) \nn \\
& \quad =
\begin{cases}
\dfrac{ n({\mu_{\mathrm{fa}}}_{n}) f(\V{z}_{m,n}|\underline{\V{x}}_{k,n}) p_{\mathrm{d}}(\underline{u}_{k,n}) } {{\mu_{\mathrm{fa}}}_{n}f_{\mathrm{fa}}(\V{z}_{m,n})}, 										& a_{k,n} = m \\
n({\mu_{\mathrm{fa}}}_{n}) \big(1 - p_{\mathrm{d}}(\underline{u}_{k,n})\big),													 & a_{k,n} = 0
\end{cases}
\label{eq:g} 
\end{align}
and $ g(\underline{\V{x}}_{k,n}, \underline{r}_{k,n}=0, a_{k,n}, {\mu_{\mathrm{fa}}}_{n}; \V{z}_{n}) = \bar{1}(a_{k,n})n({\mu_{\mathrm{fa}}}_{n}) $ with a factor related to \ac{far} $ n({\mu_{\mathrm{fa}}}_{n}) \triangleq (e^{-{\mu_{\mathrm{fa}}}_{n}}{\mu_{\mathrm{fa}}}_{n}^{M_{n}})^{1/(K_{n-1} + M_{n})} $ and by
\begin{align}
& \hspace*{-4mm} h(\overline{\V{x}}_{m,n}, \overline{r}_{m,n} = 1, b_{m,n}, {\mu_{\mathrm{fa}}}_{n}; \V{z}_{n}) \nn \\
& \hspace*{-4mm} \quad =
\begin{cases} 
0, 																				& b_{m,n} = k \\
\dfrac{n({\mu_{\mathrm{fa}}}_{n}) \mu_{\mathrm{n}} f_{\mathrm{n}}(\overline{\V{x}}_{m,n}) f(\V{z}_{m,n}|\overline{\V{x}}_{m,n})} { {\mu_{\mathrm{fa}}}_{n} f_{\mathrm{fa}}(\V{z}_{m,n}) }, 								& b_{m,n} = 0
\end{cases}
\label{eq:h} 
\end{align} 
and $ h(\overline{\V{x}}_{m,n}, \overline{r}_{m,n} = 0, b_{m,n}, {\mu_{\mathrm{fa}}}_{n}; \V{z}_{n}) = n({\mu_{\mathrm{fa}}}_{n}) $, respectively. The factor graph \cite{FG_SPA_TIT2001,Loeliger2004SPM} representing the factorization in \eqref{eq:TheJointPosteriorPDF_2} is shown in Fig.~\ref{fig:factorGraph}. The detailed derivations of the joint posterior \ac{pdf} in \eqref{eq:TheJointPosteriorPDF_2} and the pseudo likelihood functions in \eqref{eq:g} and \eqref{eq:h} are provided in the Appendix in Section~\ref{app:JointPrior}, Section~\ref{app:JointLHF}, and Section~\ref{app:JointPosterior}.

\subsection{\ac{pmpc} Detection and State Estimation}
\label{subsec:detection_estimation}
The problem considered is the sequential detection of \acp{pmpc} and estimation of their states $\RV{y}_{k,n}$, $k \in \{1,\dots, K_{n-1} + M_n\}$ along with the estimation of the \ac{far} $ {\rv{\mu}_{\mathrm{fa}}}_{n} $ based on the measurement vector $\V{z}_{1:n}$. This relies on the marginal posterior existence probabilities $ p(r_{k,n} = 1 | \V{z}_{1:n}) $, the marginal posterior \acp{pdf} $ f(\V{x}_{k,n} | r_{k,n} = 1,\V{z}_{1:n}) $ and $ f({\mu_{\mathrm{fa}}}_{n}|\V{z}_{1:n}) $. More specifically, a \ac{pmpc} is detected if $ p(r_{k,n} = 1|\V{z}_{1:n}) > p_\mathrm{de} $ \cite{Kay1998}, where $ p_\mathrm{de} $ is the existence probability threshold not to be confused with $ u_{\mathrm{de}} $ the detection threshold of the snapshot-based estimator. The probabilities $ p(r_{k,n} = 1|\V{z}_{1:n}) $ are obtained from the marginal posterior \acp{pdf} of the \ac{pmpc} states, $ f(\V{y}_{k,n}| \V{z}_{1:n}) = f(\V{x}_{k,n}, r_{k,n} | \V{z}_{1:n}) $, according to 
\begin{align}
\hspace*{-2mm} p(r_{k,n} = 1 | \V{z}_{1:n}) = \big\langle f(\V{x}_{k,n}, r_{k,n} = 1 | \V{z}_{1:n}) \big\rangle_{1_{\mathbb{R}^{5\times1}}(\V{x}_{k,n})}
\label{eq:existProb} 
\end{align}
and the marginal posterior \acp{pdf} $ f(\V{x}_{k,n} | r_{k,n} = 1,\V{z}_{1:n}) $ are obtained from $ f(\V{x}_{k,n}, r_{k,n} | \V{z}_{1:n}) $ as 
\begin{align}
f(\V{x}_{k,n} | r_{k,n} = 1,\V{z}_{1:n}) = \dfrac{f(\V{x}_{k,n}, r_{k,n} = 1 | \V{z}_{1:n})} {p(r_{k,n} = 1 | \V{z}_{1:n})}\ist.
\label{eq:PSMC_existProb} 
\end{align}
The number of detected \acp{pmpc} represents the estimated \ac{nom} given by $ \hat{L}_{n} $. The states $ \rv{u}_{k,n} $ and $\RV{\theta}_{k,n}$ of the detected \acp{pmpc} are estimated by means of the \ac{mmse} estimator \cite{Kay_EstimationTheory}, i.e., 
\begin{align}	
u_{k,n}^{\,\mathrm{MMSE}} \triangleq  \big\langle u_{k,n}  \big\rangle_{f(\V{x}_{k,n} | r_{k,n} = 1,\V{z}_{1:n})}
\label{eq:MMSE_x1}  
\end{align} 
\begin{align}
\V{\theta}_{k,n}^{\,\mathrm{MMSE}} \triangleq  \big\langle\V{\theta}_{k,n} \big\rangle_{f(\V{x}_{k,n} | r_{k,n} = 1,\V{z}_{1:n})} 
\label{eq:MMSE_x} 
\end{align}
with $\V{\theta}_{k,n}^{\,\mathrm{MMSE}} = [d_{k,n}^{\,\mathrm{MMSE}}\iist \varphi_{k,n}^{\,\mathrm{MMSE}}]^{\mathrm{T}} \in \mathbb{R}^{2\times1}$, respectively. Note that the estimated component \acp{snr} is given by $ \mathrm{SNR}^{\,\mathrm{MMSE}}_{k,n} = \big(u_{k,n}^{\,\mathrm{MMSE}}\big)^2 $. Finally, the estimate of the \ac{far} $ {\rv{\mu}_{\mathrm{fa}}}_{n} $ is given by 
\begin{align} 
	{\mu_{\mathrm{fa}}}_{n}^{\,\mathrm{MMSE}} \triangleq  \big\langle {\mu_{\mathrm{fa}}}_{n} \big\rangle_{f({\mu_{\mathrm{fa}}}_{n}|\V{z}_{1:n})} \label{eq:MMSE_meanClutterRate}\ist. 
\end{align}

To initialize the dispersion parameters of the snapshot-based channel estimator (see Section~\ref{sec:paramCE}), the prior \ac{pdf} is assumed to be Gaussian with vector-valued mean given by the \ac{mmse} estimate $ \V{\theta}_{k,n-1}^{\,\mathrm{MMSE}} \in \mathbb{R}^{2 \times 1} $ and covariance matrix $ \V{\Sigma}_{k,n-1} = \mathrm{diag}\{ {\sigma_\mathrm{d}}^{\,\mathrm{MMSE}}_{k,n-1}, \iist {\sigma_\mathrm{\varphi}}^{\,\mathrm{MMSE}}_{k,n-1} \} \in \mathbb{R}^{2 \times 2} $ of the parameters with $k \in \{1,\dots,\hat{L}_{n-1}\}$, where ${\sigma_\mathrm{d}}^{\,\mathrm{MMSE}}_{k,n} = \big( \big\langle (d_{k,n}-d_{k,n}^{\,\mathrm{MMSE}})^2 \big\rangle_{f(\V{x}_{k,n} | r_{k,n} = 1,\V{z}_{1:n})}\big)^{\frac{1}{2}}$ and ${\sigma_\mathrm{\varphi}}^{\,\mathrm{MMSE}}_{k,n} = \big(\big\langle (\varphi_{k,n} - \varphi_{k,n}^{\,\mathrm{MMSE}})^2 \big\rangle_{f(\V{x}_{k,n} | r_{k,n} = 1,\V{z}_{1:n})} \big)^{\frac{1}{2}} $.

\vspace*{2mm}

As the number of \acp{pmpc} grows with time $n$ (at each time by $ K_{n} = K_{n-1} + M_{n}$), \acp{pmpc} with $ p(r_{k,n} = 1 | \V{z}_{1:n})$ below a threshold $p_\mathrm{pr}$ are removed from the state space (``pruned'').

%% file: InputFiles/BPalgorithm.tex
The posterior \acp{pdf} $ f(\underline{\bm{x}}_{k,n}, \underline{r}_{k,n} | \bm{z}_{1:n}) $, $ f(\overline{\bm{x}}_{m,n}, \overline{r}_{m,n} | \bm{z}_{1:n}) $, and $ f({\mu_{\mathrm{fa}}}_{n}|\bm{z}_{1:n}) $ involved in \eqref{eq:existProb}, \eqref{eq:PSMC_existProb} and \eqref{eq:MMSE_meanClutterRate} are marginal \acp{pdf} of the joint posterior \ac{pdf} $f(\V{y}_{1:n}, \V{a}_{1:n}, \V{b}_{1:n}, {\V{\mu}_{\mathrm{fa}}}_{1:n}, \V{m}_{1:n} | \V{z}_{1:n})$. Since direct marginali- zation of the joint posterior \ac{pdf} is infeasible, we use loopy (iterative) \ac{bp} \cite{FG_SPA_TIT2001} by means of the sum-product algorithm rules \cite{FG_SPA_TIT2001,Loeliger2004SPM} on the factor graph shown in Fig.~\ref{fig:factorGraph}. Due to the loops inside the factor graph, the resulting beliefs $ \underline{q}(\underline{\V{y}}_{k,n}) = \underline{q}(\underline{\V{x}}_{k,n}, \underline{r}_{k,n}) $, $ \overline{q}(\overline{\V{y}}_{m,n}) = \overline{q}(\overline{\V{x}}_{m,n}, \overline{r}_{m,n}) $, and $ q({\mu_{\mathrm{fa}}}_{n}) $ are only approximations of the respective posterior marginal \acp{pdf}, and there is no canonical order in which the messages should be computed \cite{FG_SPA_TIT2001}. For the proposed algorithm, we specify the following order: (i) messages are not sent backward in time; (ii) iterative message passing is only performed for probabilistic \ac{da} at each time $ n $. Combining the specified order with the generic \ac{bp} rules for calculating messages and beliefs yields the following calculations at each time $ n $ (which are in parts in line with \cite[Ch.~III]{Florian_Proceeding2018, Erik_SLAM_TWC2019}):
\begin{enumerate}
	\item \emph{Prediction}: First, a prediction step is performed. The prediction for the \ac{far} is given by 
	\begin{align}
	\chi({\mu_{\mathrm{fa}}}_{n}) = \big\langle f({\mu_{\mathrm{fa}}}_{n}|{\mu_{\mathrm{fa}}}_{n-1})\big\rangle_{q({\mu_{\mathrm{fa}}}_{n-1})}
	\label{eq:prediction_meanClutterRate}
	\end{align}
	and for all the legacy \acp{pmpc} is given by
	\begin{align}
	& \chi(\underline{\V{x}}_{k,n}, \underline{r}_{k,n}) \nonumber \\ 
	& = \hspace*{-1mm} \sum\limits_{\substack{r_{k,n-1} \\ \in \{0,1\}}} \big\langle f(\underline{\V{x}}_{k,n}, \underline{r}_{k,n}|\V{x}_{k,n-1}, r_{k,n-1})\big\rangle_{\underline{q}(\V{x}_{k,n-1}, r_{k,n-1})}\nn \\[-3mm]
	\label{eq:prediction_PSMCs}\\[-7mm]\nn
	\end{align} 
	where $ q({\mu_{\mathrm{fa}}}_{n-1}) $ and $ \underline{q}(\V{x}_{k,n-1}, r_{k,n-1}) $ were calculated at the previous time $ n-1 $. After substituting the \ac{pmpc} state-transition \acp{pdf} in \eqref{eq:prediction_PSMCs} with \eqref{eq:ST_pdf1} and \eqref{eq:ST_pdf2}, respectively, we obtain the prediction messages for legacy \acp{pmpc} as 
	\begin{align}
	\chi(\underline{\V{x}}_{k,n}, 1) & = p_{\mathrm{s}} \big\langle f(\underline{\V{x}}_{k,n} | \V{x}_{k,n-1} ) \big\rangle_{\underline{q}(\V{x}_{k,n-1}, 1)}
	\label{eq:Predict_LegacyPSMCs_existing} 
	\end{align}
	and $ \chi(\underline{\V{x}}_{k,n}, 0) = \chi_{k,n} f_{\mathrm{D}}(\underline{\V{x}}_{k,n}) $ with 
	\begin{align}
	\chi_{k,n} = (1-p_{\mathrm{s}}) \big\langle \underline{q}(\V{x}_{k,n-1}, 1) \big\rangle_{1_{\mathbb{R}^{5\times1}}(\V{x}_{k,n-1})} + \underline{q}_{k,n-1} \rule[-1.1em]{0pt}{0pt}
	\label{eq:Predict_LegacyPSCs_Nonexisting} 
	\end{align}
	where $ \chi_{k,n} \triangleq \big\langle \chi(\underline{\V{x}}_{k,n}, 0) \big\rangle_{1_{\mathbb{R}^{5\times1}}(\underline{\V{x}}_{k,n})} $ and $ \underline{q}_{k,n-1} \triangleq \big\langle \underline{q}(\V{x}_{k,n-1}, 0) \big\rangle_{1_{\mathbb{R}^{5\times1}}(\V{x}_{k,n-1})} $. After the prediction step, the following steps are performed for all legacy and new \acp{pmpc} in parallel:
	
	\item \emph{Measurement Evaluation}: 
	 For legacy \acp{pmpc}, the messages $ \beta(a_{k,n}) $ passed from the factor nodes $ g(\underline{\V{x}}_{k,n},$ $ \underline{r}_{k,n},a_{k,n}, {\mu_{\mathrm{fa}}}_{n}; \V{z}_{n}) $ to the \ac{pmpc}-oriented \ac{da} variable nodes $ a_{k,n} $ are calculated by 	 
	\begin{align}
	\beta(a_{k,n}) & = \big\langle \big\langle g(\underline{\V{x}}_{k,n}, 1, a_{k,n}, {\mu_{\mathrm{fa}}}_{n}; \V{z}_{n}) \big\rangle_{\chi(\underline{\V{x}}_{k,n}, 1)}\big\rangle_{\chi({\mu_{\mathrm{fa}}}_{n})} \nn\\
	& \hspace*{3mm} + \bar{1}(a_{k,n})\big\langle \big\langle \chi(\underline{\V{x}}_{k,n}, 0) \nn\\
	& \hspace*{3mm} \times n({\mu_{\mathrm{fa}}}_{n})\big\rangle_{1_{\mathbb{R}^{5\times1}}(\underline{\V{x}}_{k,n})}\big\rangle_{\chi({\mu_{\mathrm{fa}}}_{n})}\ist.
	\label{eq:meaEval_legacyPSMCs}
	\end{align}
	For new \acp{pmpc}, the messages $ \xi(b_{m,n}) $ passed from the factor nodes $ h(\overline{\V{x}}_{m,n}, \overline{r}_{m,n},$ $b_{m,n}, {\mu_{\mathrm{fa}}}_{n}; \V{z}_{n}) $ to the measurement-oriented DA variable nodes $ b_{m,n} $ are calculated according to 
	\begin{align}
	\xi(b_{m,n}) & = \sum_{\overline{r}_{m,n} \in \{0,1\}} \hspace*{0mm} \big\langle \big\langle h(\overline{\V{x}}_{m,n}, \overline{r}_{m,n}, b_{m,n}, \nn \\
	& \hspace*{15mm} {\mu_{\mathrm{fa}}}_{n}; \V{z}_{n}) \big\rangle_{1_{\mathbb{R}^{5\times1}}(\overline{\V{x}}_{m,n})}\big\rangle_{\chi({\mu_{\mathrm{fa}}}_{n})}\ist.
	\label{eq:meaEval_newPSCs} 
	\end{align}
	More specifically, for $ b_{m,n} = k $ it becomes $ \xi(b_{m,n}) = \big\langle n({\mu_{\mathrm{fa}}}_{n}) \chi({\mu_{\mathrm{fa}}}_{n}) \big\rangle_{1_{\mathbb{R}}({\mu_{\mathrm{fa}}}_{n})} $ and for $ b_{m,n} = 0 $ it becomes
	\begin{align}
	\xi(b_{m,n}) & = \big\langle n({\mu_{\mathrm{fa}}}_{n}) \chi({\mu_{\mathrm{fa}}}_{n}) \big\rangle_{1_{\mathbb{R}}({\mu_{\mathrm{fa}}}_{n})} \nn \\
	&\hspace*{2mm} + \Big\langle \Big\langle \dfrac{\mu_{\mathrm{n}}f_{\mathrm{n}}(\overline{\V{x}}_{m,n})n({\mu_{\mathrm{fa}}}_{n})}{{\mu_{\mathrm{fa}}}_{n}} \nn \\
	& \hspace*{2mm} \times \dfrac{f(\V{z}_{m,n}|\overline{\V{x}}_{m,n})}{f_{\mathrm{fa}}(\V{z}_{m,n})} \Big\rangle_{1_{\mathbb{R}^{5\times1}}(\overline{\V{x}}_{m,n})}\Big\rangle_{\chi({\mu_{\mathrm{fa}}}_{n})}\ist.
	\label{eq:meaEval_newPSMCs_k}
	\end{align}
	\item \emph{Iterative Probabilistic DA}: With the messages $ \beta(a_{k,n}) $ and $ \xi(b_{m,n}) $, the probabilistic DA messages $ \eta(a_{k,n}) $ and $ \varsigma(b_{m,n}) $ are obtained with an efficient loopy \ac{bp}-based algorithm as shown in \cite{WilliamsLauTAE2014, Florian_TSP2017, Erik_SLAM_TWC2019} 
	\begin{align}
	\eta(a_{k,n}) & = \prod_{m=1}^{M_{n}}\nu_{\psi_{m\rightarrow k}}^{(p)}(a_{k,n})
	\label{eq:IterativeDA1} 
	\end{align} 
	\begin{align}
	\varsigma(b_{m,n}) & = \prod_{k=1}^{K_{n-1}}\zeta_{\psi_{k\rightarrow m}}^{(p)}(b_{m,n}) \ist
	\label{eq:IterativeDA2}
	\end{align}	
	where $ \nu_{\psi_{m\rightarrow k}}^{(p)}(a_{k,n}) $ and $ \zeta_{\psi_{k\rightarrow m}}^{(p)}(b_{m,n}) $ denote the messages passed from $ \psi(a_{k,n}, b_{m,n}) $ to the variable nodes $ a_{k,n} $ and $ b_{m,n} $ at each iteration $ p \in \{1,\dots, P\} $, respectively. 	
		
	\item \emph{Measurement Update}: For legacy \acp{pmpc}, the messages $ \gamma(\underline{\V{x}}_{k,n}, \underline{r}_{k,n}) $ passed from the factor nodes $ g(\underline{\V{x}}_{k,n}, \underline{r}_{k,n}, a_{k,n}, {\mu_{\mathrm{fa}}}_{n}; \V{z}_{n}) $ to the variable nodes $ \underline{\V{y}}_{k,n} $ are calculated by 
	\begin{align}
	\gamma(\underline{\V{x}}_{k,n}, 1) & = \sum_{a_{k,n}=0}^{M_n} \eta(a_{k,n}) \big\langle g(\underline{\V{x}}_{k,n}, 1, a_{k,n}, \nn \\
	& \hspace*{27mm}
	{\mu_{\mathrm{fa}}}_{n}; \V{z}_{n}) \big\rangle_{\chi({\mu_{\mathrm{fa}}}_{n})}
	\label{eq:meaUpdate_legacyPSMCs_exist} 
	\end{align}
	and $ \gamma(\underline{\V{x}}_{k,n}, 0) = \gamma_{k,n}f_{\mathrm{D}}(\underline{\V{x}}_{k,n}) $ with 
	\begin{align}
	\gamma_{k,n} & = \big\langle \gamma(\underline{\V{x}}_{k,n}, 0) \big\rangle_{1_{\mathbb{R}^{5\times1}}(\underline{\V{x}}_{k,n})} \nonumber\\
	& = \eta(0) \big\langle n({\mu_{\mathrm{fa}}}_{n}) \big\rangle_{\chi({\mu_{\mathrm{fa}}}_{n})}\ist.
	\label{eq:meaUpdate_legacyPSMCs_nonexist} 
	\end{align}
	For new \acp{pmpc}, the messages $ \phi(\overline{\V{x}}_{m,n}, \overline{r}_{m,n}) $ passed from the factor nodes $ h(\overline{\V{x}}_{m,n}, \overline{r}_{m,n}, b_{m,n}, {\mu_{\mathrm{fa}}}_{n}; \V{z}_{n}) $ to the variable nodes $ \overline{\V{y}}_{m,n} $ are calculated by
	\begin{align}
	\phi(\overline{\V{x}}_{m,n}, 1) = \varsigma(0)\big\langle h(\overline{\V{x}}_{m,n}, 1, b_{m,n}, {\mu_{\mathrm{fa}}}_{n}; \V{z}_{n}) \big\rangle_{\chi({\mu_{\mathrm{fa}}}_{n})} \rule[-1.1em]{0pt}{0pt}
	\label{eq:meaUpdate_newPSMCs_exist} 
	\end{align}
	and $ \phi(\overline{\V{x}}_{m,n}, 0) = \phi_{m,n}f_{\mathrm{D}}(\overline{\V{x}}_{m,n}) $ with
	\begin{align}
	\phi_{m,n} & \triangleq \big\langle \phi(\overline{\V{x}}_{m,n}, 0) \big\rangle_{1_{\mathbb{R}^{5\times1}}(\overline{\V{x}}_{m,n})} \nonumber\\
	& = \sum_{b_{m,n}=0}^{K_{n-1}} \varsigma(b_{m,n}) \big\langle n({\mu_{\mathrm{fa}}}_{n}) \big\rangle_{\chi({\mu_{\mathrm{fa}}}_{n})}\ist.
	\label{eq:meaUpdate_newPSMCs_nonexist} 
	\end{align}
	For the \ac{far} ${\rv{\mu}_{\mathrm{fa}}}_{n}$, the messages $ \rho_{k}({\mu_{\mathrm{fa}}}_{n}) $ and $ \kappa_{m}({\mu_{\mathrm{fa}}}_{n}) $ passed from the factor nodes $ g(\underline{\V{x}}_{k,n}, \underline{r}_{k,n}, a_{k,n}, {\mu_{\mathrm{fa}}}_{n}; \V{z}_{n}) $\,\,\, and\,\,\, $ h(\overline{\V{x}}_{m,n}, \overline{r}_{m,n},b_{m,n},\\ {\mu_{\mathrm{fa}}}_{n}; \V{z}_{n}) $, respectively, to the variable node ${\mu_{\mathrm{fa}}}_{n}$ are calculated by 
	\begin{align}
	\rho_{k}({\mu_{\mathrm{fa}}}_{n}) & = \sum_{a_{k,n}=0}^{M_n} \eta(a_{k,n})\hspace*{-1.5mm}
	\sum_{\underline{r}_{k,n} \in \{0,1\}} \hspace*{-1.5mm} \big\langle g(\underline{\V{x}}_{k,n}, \underline{r}_{k,n},  \nonumber \\ 
	& \hspace*{18mm} a_{k,n}, {\mu_{\mathrm{fa}}}_{n}; \V{z}_{n}) \big\rangle_{\chi(\underline{\V{x}}_{k,n}, \underline{r}_{k,n})}
	\label{eq:meaUpdate_meanClutterRate_legacy} 
	\end{align}
	and 
	\begin{align}
	\kappa_{m}({\mu_{\mathrm{fa}}}_{n}) & = n({\mu_{\mathrm{fa}}}_{n}) \varsigma(0) \Big\langle\dfrac{\mu_{\mathrm{n}} f(\V{z}_{m,n}|\overline{\V{x}}_{m,n}) } {{\mu_{\mathrm{fa}}}_{n} f_{\mathrm{fa}}(\V{z}_{m,n})}\Big\rangle_{f_{\mathrm{n}}(\overline{\V{x}}_{m,n})} \nonumber\\
	& \hspace*{17mm} + \sum_{b_{m,n}=0}^{K_{n-1}} \varsigma(b_{m,n}) n({\mu_{\mathrm{fa}}}_{n}) \ist.
	\label{eq:meaUpdate_meanClutterRate_new} 
	\end{align}
	
	\item \emph{Belief Calculation}:
	With all the messages above, the approximations of the marginal posterior \acp{pdf} needed for the MMSE estimations in Section~\ref{subsec:detection_estimation} are calculated as follows. The beliefs $ \underline{q}(\underline{\V{y}}_{k,n}) = \underline{q}(\underline{\V{x}}_{k,n}, \underline{r}_{k,n}) $ approximating the marginal posterior \acp{pdf} $ f(\underline{\V{y}}_{k,n} | \V{z}_{1:n}) = f(\underline{\V{x}}_{k,n}, \underline{r}_{k,n}| \V{z}_{1:n}) $ for legacy \acp{pmpc} are obtained as 
	\begin{align}
	\underline{q}(\underline{\V{x}}_{k,n}, 1) & = \dfrac{1}{\underline{C}_{k,n}} \chi(\underline{\V{x}}_{k,n}, 1) \gamma(\underline{\V{x}}_{k,n}, 1) \label{eq:belief_legacyPSMCs_exist}
	\end{align}
	and $ \underline{q}(\underline{\V{x}}_{k,n}, 0) = \underline{q}_{k,n} f_{\mathrm{D}}(\underline{\V{x}}_{k,n}) $ with $ \underline{q}_{k,n} = \dfrac{1}{\underline{C}_{k,n}} \chi_{k,n}\gamma_{k,n} $. The normalization constant is given as $ \underline{C}_{k,n}= \big\langle\gamma(\underline{\V{x}}_{k,n}, 1)\rangle_{\chi(\underline{\V{x}}_{k,n}, 1)}  + \chi_{k,n}\gamma_{k,n} $. The beliefs $ \overline{q}(\overline{\V{y}}_{m,n}) = \overline{q}(\overline{\V{x}}_{m,n}, \overline{r}_{m,n}) $ approximating the marginal posterior \acp{pdf} $ f(\overline{\V{y}}_{m,n} | \V{z}_{1:n}) = f(\overline{\V{x}}_{m,n}, \overline{r}_{m,n}| \V{z}_{1:n}) $ for new \acp{pmpc} are obtained as 
	\begin{align}
	\overline{q}(\overline{\V{x}}_{m,n}, 1) = \dfrac{1}{\overline{C}_{m,n}} \phi(\overline{\V{x}}_{m,n}, 1) \label{eq:belief_newPSMCs_exist}
	\end{align}
	and $ \overline{q}(\overline{\V{x}}_{m,n}, 0) = \overline{q}_{m,n} f_{\mathrm{D}}(\overline{\V{x}}_{m,n}) $ with $ \overline{q}_{m,n} = \dfrac{1}{\overline{C}_{m,n}} \phi_{m,n} $. The normalization constant is given as $ \overline{C}_{m,n} = \big\langle \phi(\overline{\V{x}}_{m,n}, 1) \big\rangle_{1_{\mathbb{R}^{5\times1}}(\overline{\V{x}}_{m,n})} +  \phi_{m,n} $. Finally, the belief $ q({\mu_{\mathrm{fa}}}_{n}) $ approximating the marginal posterior \ac{pdf} $ f({\mu_{\mathrm{fa}}}_{n}|\V{z}_{1:n}) $ for the \ac{far} is obtained as 
	\begin{align}
	q({\mu_{\mathrm{fa}}}_{n}) & = \chi({\mu_{\mathrm{fa}}}_{n}) \prod_{k = 1}^{K_{n-1}} \rho_{k}({\mu_{\mathrm{fa}}}_{n}) \prod_{m = 1}^{M_{n}} \kappa_{m}({\mu_{\mathrm{fa}}}_{n}).
	\label{eq:belief_meanClutterRate}
	\end{align}
\end{enumerate} 


%% file: InputFiles/ParticleImplementation.tex
Since integrations involved in the calculations of the messages and beliefs cannot be obtained analytically, we use a computationally efficient sequential particle-based message passing implementation which provides approximate computation. In what follows, we present particle-based implementations for \ac{far} related steps \eqref{eq:prediction_meanClutterRate}, \eqref{eq:meaUpdate_meanClutterRate_legacy}, \eqref{eq:meaUpdate_meanClutterRate_new} and \eqref{eq:belief_meanClutterRate}. The implementation of all other steps in Section~\ref{sec:BPalgorithm}, and the calculation of posterior existence probabilities of \acp{pmpc} are performed in line with \cite[Section~VI]{Florian_TSP2017}. Similarly to \cite{Erik_SLAM_TWC2019}, our implementation uses a ``stacked state'' \cite{MeyerTSPIN2016} comprising the \ac{pmpc} states and the \ac{far} state.

\begin{enumerate}
	\item \emph{Prediction}: The belief $ q({\mu_{\mathrm{fa}}}_{n-1}) $ calculated at the previous time $ n-1 $ is represented by $ J $ particles and weights, i.e., $ \{ {{}\hat{\mu}_{\mathrm{fa}}}_{n-1}^{(j)}, {{}\hat{w}_{\mathrm{fa}}}_{n-1}^{(j)} \}_{j=1}^{J} $. At time $ n $, for each particle $ {{}\hat{\mu}_{\mathrm{fa}}}_{n-1}^{(j)} $, $ j \in \{1,\dots, J\} $ one particle $ {\mu^{'}_{\mathrm{fa}}}_{n}^{(j)} $ with corresponding weight $ {w^{'}_{\mathrm{fa}}}_{n}^{(j)} = {{}\hat{w}_{\mathrm{fa}}}_{n-1}^{(j)} $ is drawn from $ f({\mu_{\mathrm{fa}}}_{n} | {\mu_{\mathrm{fa}}}_{n-1}) $, and $ \{ {\mu^{'}_{\mathrm{fa}}}_{n}^{(j)}, {w^{'}_{\mathrm{fa}}}_{n}^{(j)} \}_{j=1}^{J} $ represent the prediction message $ \chi({\mu_{\mathrm{fa}}}_{n}) $ in \eqref{eq:prediction_meanClutterRate}. Note that the proposal distribution underlying the weight calculation is $ f({\mu_{\mathrm{fa}}}_{n} | {\mu_{\mathrm{fa}}}_{n-1}) $.
	\item \emph{Measurement Update}: The non-normalized weights corresponding to the messages for legacy \acp{pmpc} $ \rho_{k}({\mu_{\mathrm{fa}}}_{n}) $ in \eqref{eq:meaUpdate_meanClutterRate_legacy} and new \acp{pmpc} $ \kappa_{m}({\mu_{\mathrm{fa}}}_{n}) $ in \eqref{eq:meaUpdate_meanClutterRate_new} are calculated by 
	\begin{align}
	{\underline{w}_{\mathrm{fa}}}_{k,n}^{(j)} & = \sum_{a_{k,n}=0}^{M_n} \eta(a_{k,n}) g(\underline{\V{x}}_{k,n}^{(j)}, 1, a_{k,n}, {\mu_{\mathrm{fa}}}_{n}^{(j)}; \V{z}_{n}) \nonumber \\
	& \quad \times w_{k,n}^{(j)} +  n({\mu_{\mathrm{fa}}}_{n}^{(j)}) \dfrac{(1-\sum_{j=0}^{J} w_{k,n}^{(j)})}{J} 
	\label{eq:Imp_meaUpdate_FAR_legacy} 
	\end{align}
	and 
	\begin{align}
	{\overline{w}_{\mathrm{fa}}}_{m,n}^{(j)} &=  n({\mu_{\mathrm{fa}}}_{n}^{(j)}) \dfrac{\mu_{\mathrm{n}} f(\V{z}_{m,n}|\overline{\V{x}}_{m,n}^{(j)}) } {{\mu_{\mathrm{fa}}}_{n}^{(j)} f_{\mathrm{fa}}(\V{z}_{m,n})}f_{\mathrm{n}}(\overline{\V{x}}_{m,n}^{(j)}) \nonumber \\
	& \hspace*{17mm} + \sum_{b_{m,n}=0}^{K_{n-1}} \varsigma(b_{m,n}) n({\mu_{\mathrm{fa}}}_{n}^{(j)})
	\label{eq:Imp_meaUpdate_FAR_new} 
	\end{align}
	respectively. The weighted particles $ \{\underline{\V{x}}_{k,n}^{(j)}, \underline{w}_{k,n}^{(j)}\}_{j=1}^{J} $ represent the prediction messages $ \chi(\underline{\V{x}}_{k,n}, \underline{r}_{k,n}) $ of legacy \acp{pmpc} in \eqref{eq:prediction_PSMCs}, thus their predicted existence probabilities can be approximated as $ \sum_{j=0}^{J} \underline{w}_{k,n}^{(j)} $, and $ (1-\sum_{j=0}^{J} \underline{w}_{k,n}^{(j)})/J $ is the weight of particles representing $ \chi(\underline{\V{x}}_{k,n}, 0) $. Moreover, the weighted particles $ \{\overline{\V{x}}_{m,n}^{(j)},\overline{w}_{k,n}^{(j)} = 1/J\}_{j=1}^{J} $ with equal weights represent the states of new \acp{pmpc}. 
	
	\item \emph{Belief Calculation and State Estimation}: The above approximate messages are further used for calculating the non-normalized weights corresponding to the belief $ q({\mu_{\mathrm{fa}}}_{n}) $ in \eqref{eq:belief_meanClutterRate}, given by
	\begin{align}
	{w_{\mathrm{fa}}}_{n}^{(j)} &= {w^{'}_{\mathrm{fa}}}_{n}^{(j)} \prod_{k = 1}^{K_{n-1}} {\underline{w}_{\mathrm{fa}}}_{k,n}^{(j)} \prod_{m = 1}^{M_{n}} {\overline{w}_{\mathrm{fa}}}_{m,n}^{(j)}.
	\label{eq:Imp_belief_FAR}
	\end{align}
	After normalization $ {w_{\mathrm{fa}}}_{n}^{(j)} = {w_{\mathrm{fa}}}_{n}^{(j)}/\sum_{j=0}^{J}{w_{\mathrm{fa}}}_{n}^{(j)} $, an approximation of the \ac{mmse} state estimate $ {\mu_{\mathrm{fa}}}_{n}^{\,\mathrm{MMSE}} $ in \eqref{eq:MMSE_meanClutterRate} is given by 
	\begin{align}
	{\mu_{\mathrm{fa}}}_{n}^{\,\mathrm{MMSE}} \approx \sum_{j=0}^{J} {w_{\mathrm{fa}}}_{n}^{(j)} {\mu^{'}_{\mathrm{fa}}}_{n}^{(j)}\ist .
	\label{eq:Imp_MMSE_FAR}
\end{align}
	
\end{enumerate} 
To avoid the particle degeneracy effect, a resampling step \cite{Florian_TSP2017} is performed as a final preparation for the next time $ n+1 $ leading to $ \{ {{}\hat{\mu}_{\mathrm{fa}}}_{n}^{(j)}, {{}\hat{w}_{\mathrm{fa}}}_{n}^{(j)} = 1/J \}_{j=1}^{J} $ representing the belief $ q({\mu_{\mathrm{fa}}}_{n}) $. Assuming a fixed number $P$ of message passing iterations for \ac{da}, the computational complexity of calculating the (approximate) marginal posterior \acp{pdf} scales only linearly in the number of particles. Moreover, the complexity of the iterative \ac{da} given by the operations in \eqref{eq:IterativeDA1} and \eqref{eq:IterativeDA2} scales as $O(K_{n-1}M_{n})$, i.e., quadratically in the number of \acp{pmpc} \cite{WilliamsLauTAE2014, Florian_TSP2017, Florian_Proceeding2018}.

%% file: InputFiles/ExperimentalResults.tex
The performance of the proposed algorithm is validated using both synthetic and real radio measurements. For synthetic measurements, the performance is further compared with the \ac{pcrlb} \cite{Tichavsky_PosteriorCRLB_1998} and that of the \ac{kest} algorithm \cite{KEST_TAP2012}. 

\subsection{Analysis Setup}
\label{subsec:AnalysisSettings}

\subsubsection{Common Simulation Setup}
\label{subsubsec:SimulationParameters}
For synthetic and real measurements, the following setups and parameters are commonly used. We assume that \ac{mpc} dispersion parameters with according velocities and normalized amplitudes evolve independently across time and to each other. More specifically, the state-transition \ac{pdf} of $\underline{\RV{\theta}}_{k,n}$ (with according velocities ${\underline{\rv{v}}_{\mathrm{d}}}_{k,n}$ and ${\underline{\rv{v}}_\mathrm{\varphi}}_{k,n}$) is chosen to be a nearly-constant velocity model. The state-transition \ac{pdf} of the normalized amplitude $\underline{\rv{u}}_{k,n}$ is chosen to be $ \underline{u}_{k,n} = u_{k,n-1} + {\epsilon_{\mathrm{u}}}_{k,n} $, where the noise ${\rv{\epsilon}_{\mathrm{u}}}_{k,n}$ is \ac{iid} across $ k $ and $ n $, zero-mean, and Gaussian with variance ${\underline{\sigma}^2_{\mathrm{u}}}_{k,n}$. Based on the models above, the state-transition \ac{pdf} of legacy \ac{pmpc} state $ \RV{\underline{\bm{x}}}_{k,n} $ is collectively given as $ \underline{\bm{x}}_{k,n} = \V{F}\bm{x}_{k,n-1} + \V{\Gamma}\V{\epsilon}_{n} $, where the transition matrices $ \V{F} \in \mathbb{R}^{5\times5} $ and $ \V{\Gamma} \in \mathbb{R}^{5\times3} $ are formulated as in \cite[Section 6.3.2]{BarShalom2002EstimationTracking, Jussi_TSP2009} with sampling period $\Delta T = 1$\,s. The driving process $ \RV{\epsilon}_{n} \in \mathbb{R}^{3\times1} $ is \ac{iid} across $ k $ and $ n $, zero-mean and Gaussian with covariance matrix $ \mathrm{diag}\{{\sigma^2_\mathrm{d}}, {\sigma^2_\mathrm{\varphi}}, {\underline{\sigma}^2_{\mathrm{u}}}_{k,n}\} $. In addition, the state-transition \ac{pdf} of the \ac{far} $ {\rv{\mu}_{\mathrm{fa}}}_{n} $ is given as $ {\mu_{\mathrm{fa}}}_{n} = {\mu_{\mathrm{fa}}}_{n-1} + {\epsilon_{\mathrm{fa}}}_{n} $, where $ {\rv{\epsilon}_{\mathrm{fa}}}_{n} $ is \ac{iid} across $ n $, zero-mean, and Gaussian with variance $\sigma^2_{\mathrm{fa}}$.\footnote{Incorporating the interacting multiple model (IMM) \cite{SoldiTSP2019} into the algorithm can help to resolve the motion uncertainties of the unknown variable states, therefore the demand on presetting and tunning of noise variances for the state-transition \acp{pdf} can be relaxed. However, it is out of the scope of this paper and can be considered as a future extension.} For computational efficiency of the particle-based implementation, the likelihood function \eqref{eq:pdf_normAmplitude} of the normalized amplitude is approximated by a truncated Gaussian \ac{pdf}, i.e., 
\begin{align} 
f({z_\mathrm{u}}_{m,n} | u_{k,n}) = \frac{ \frac{1}{{\sigma_\mathrm{u}}_{k,n}\sqrt{2\pi}} \exp\big({\frac{ -({z_\mathrm{u}}_{m,n} - u_{k,n})^2}{2{\sigma^2_\mathrm{u}}_{k,n}}}\big) } { p_{\mathrm{d}}(u_{k,n}) }
\label{eq:pdf_normAmplitude_gaussianApprox} 
\end{align}
for $ {z_\mathrm{u}}_{m,n} > \sqrt{u_{\mathrm{de}}} $, where $ p_{\mathrm{d}}(u_{k,n}) = Q(( \sqrt{u_{\mathrm{de}}} - u_{k,n})/{\sigma_\mathrm{u}}_{k,n})$ is the \ac{cdf} of Gaussian distribution. The birth \ac{pdf} of a new \acp{pmpc} $ f_{\mathrm{n}}(\overline{\V{x}}_{m,n}) = 1/(2\pi d_{\mathrm{max}})$ is assumed to be uniform on $[0,\ist 2\pi d_{\mathrm{max}})$. The particles for the initial state $ \RV{\overline{\bm{x}}}_{m,n} $ of a new \ac{pmpc} are drawn from a 5-D Gaussian \ac{pdf} with means $ [{z_\mathrm{d}}_{m,n}\ist {z_\mathrm{\varphi}}_{m,n}\iist {z_\mathrm{u}}_{m,n} \iist 0\iist 0]^{\mathrm{T}} $ and variances $ [{\sigma^2_\mathrm{d}}_{m,n} \ist {\sigma^2_\mathrm{\varphi}}_{m,n} \ist {\sigma^2_\mathrm{u}}_{m,n} \ist \sigma^2_{\mathrm{v_{d}}} \ist \sigma^2_{\mathrm{v_{\varphi}}} ]^{\mathrm{T}} $, where ${\sigma^2_\mathrm{d}}_{m,n}$, $ {\sigma^2_\mathrm{\varphi}}_{m,n} $ and $ {\sigma^2_\mathrm{u}}_{m,n} $ are calculated using the amplitude measurements ${z_\mathrm{u}}_{m,n}$ (see Section~\ref{subsec:LikelihoodFunctions}). The particles for the initial \ac{far} $ {\rv{\mu}_{\mathrm{fa}}}_{n} $ state are drawn from a Gaussian PDF with mean $ M_{1}/2 $ and variance $\sigma^2_{\mathrm{fa}_\mathrm{ini}}$. The other simulation parameters are as follows: the survival probability $ p_{\mathrm{s}} = 0.999 $, the existence probability threshold $ p_{\mathrm{de}} = 0.5 $, the pruning threshold $ p_{\mathrm{pr}} = 10^{-4} $, the mean number of newly detected \acp{mpc} $ \mu_{\mathrm{n}} = 0.008 $, the maximum number of message passing iterations for DA $ P = 5000 $ and the \acp{pdf} of the states are represented by $ J = 10000 $ particles each. 

For each simulation run, \ac{awgn} is generated with noise variance $ \sigma^2 $ specified by the \ac{snr} output defined as $ \mathrm{SNR}_{\mathrm{1m}} = 10\log_{10}\big(\frac{|\alpha_{\mathrm{LoS}}|^2 \norm{\V{s}_{\mathrm{LoS}}}^2}{\sigma^2} \big) $, where the amplitude $ \alpha_{\mathrm{LoS}} $ and the signal vector $ \V{s}_{\mathrm{LoS}} $ of the line-of-sight (LoS) path are computed at $ 1 $\,m distance. For comparability with other papers presenting parametric channel estimation methods as for example \cite{ShutinCAMSAP2017, KEST_TAP2012, BadiuTSP2017}, we also define the input \ac{snr}, i.e., $\mathrm{SNR}_{\mathrm{1m}}^{\mathrm{in}}\rmv\rmv =\rmv\rmv  \mathrm{SNR}_{\mathrm{1m}}\rmv\rmv - \rmv\rmv 10\log_{10}(N_{\mathrm{s}}H)$, the input component \acp{snr} $\mathrm{SNR}_{l,n}^{\mathrm{in}}\rmv\rmv =\rmv\rmv  10\log_{10}\big(\widetilde{\mathrm{SNR}}_{l,n}/(N_{\mathrm{s}}H)\big)$ and the input detection threshold $u_{\mathrm{de}}^{\mathrm{in}} \rmv\rmv =\rmv\rmv 10\log_{10}\big(u_{\mathrm{de}}/(N_{\mathrm{s}}H)\big)$ excluding the array and frequency sample gain. 

\begin{figure*}[t]
	\centering
	\hspace*{8mm}\subfloat[\label{subfig:fullSynMea_BP_distance_singleMC}]
	{\hspace*{-10.5mm}\includegraphics[width=0.31\textwidth,height=0.21\textwidth]{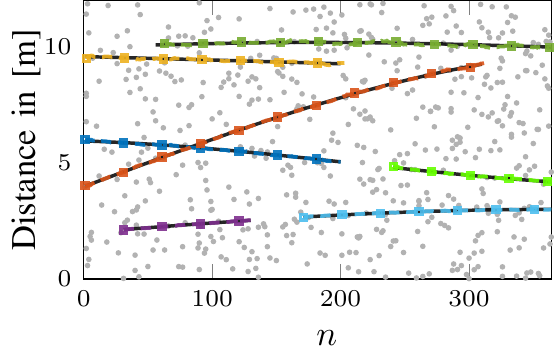}}
	\hspace*{13mm}\subfloat[\label{subfig:fullSynMea_BP_azimuth_angle_singleMC}]
	{\hspace*{-10mm}\includegraphics[width=0.31\textwidth,height=0.215\textwidth]{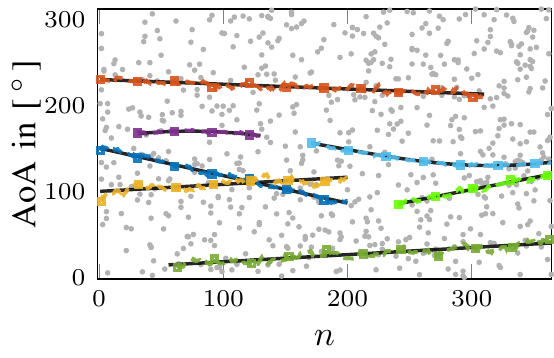}}
	\hspace*{14mm}\subfloat[\label{subfig:fullSynMea_BP_SNR_singleMC}]
	{\hspace*{-10.5mm}\includegraphics[width=0.31\textwidth,height=0.222\textwidth]{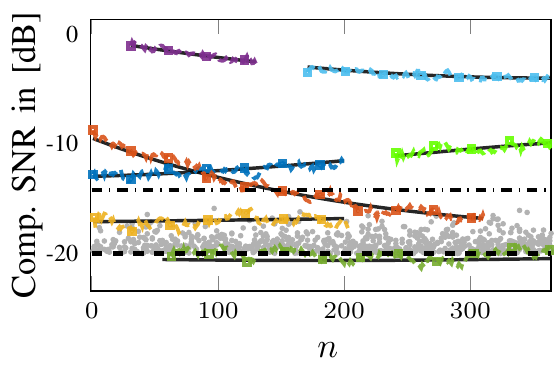}}
	\caption{Results for fully synthetic measurements with the proposed algorithm given $ \mathrm{SNR}_{\mathrm{1m}}^{\mathrm{in}} = 5.4$\,dB. The estimates of (a) distances, (b) \acp{aoa}, and (c) input component \acp{snr}. The black solid lines denote the true \ac{mpc} parameters. The gray dots denote the false alarm measurements. The estimates of different \acp{mpc} are denoted with densely-dashed lines with square markers in different colors. The horizontal dashed and dash-dotted lines in (c) indicate the detection thresholds of $ -20 $\,dB and $ -14.4 $\,dB, respectively.}	 
	\label{fig:fullSynMea_BP_singleMC}
\end{figure*}

\subsubsection{Metrics}
The performance of different methods is measured using the \ac{ospa} metric \cite{OSPA_TSP2008}, which can efficiently capture the estimation errors of the \ac{nom} and \ac{mpc} states when comparing with the true \ac{mpc} states at each time. We use \ac{ospa} metric order $ 2 $ and cutoff parameters $ 0.1 $\,m and $ 10^{\circ} $ for distance and \ac{aoa} respectively, where cutoff parameters denote the weightings of how the metric penalizes the \ac{nom} estimation errors as opposed to \ac{mpc} state estimation errors. In addition, we compute the \ac{pcrlb} \cite{Tichavsky_PosteriorCRLB_1998} as a performance benchmark. The error bounds for distance and \ac{aoa} at time $ n $ are given as $ {\varepsilon_{\mathrm{d}}}_{n} = (\frac{\mathrm{tr} [{\V{J}_{\mathrm{p}}}_{n}^{-1}]_{1:L_{n}}}{L_{n}})^{\frac{1}{2}} $ and $ {\varepsilon_{\mathrm{\varphi}}}_{n} = (\frac{\mathrm{tr}[{\V{J}_{\mathrm{p}}}_{n}^{-1}]_{L_{n}+1:2L_{n}}} {L_{n}})^{\frac{1}{2}} $, where $ {\V{J}_{\mathrm{p}}}_{n} $ denotes the posterior \ac{fim}. The \ac{mospa} errors, mean error bounds (MEBs), and the mean estimate of each unknown variable are obtained by averaging over all simulation runs.

\subsection{Performance for Synthetic Measurements}

We assume that the static single-antenna \ac{tx} transmits $\tilde{s}(t)$ to the \ac{rx} equipped with a $ 3 \times 3 $ uniform rectangular array with inter-element spacing of $ 2 $\,cm. Over $ 364 $ time steps, $ 7 $ \acp{mpc} with different lifetimes and time-varying distances and \acp{aoa} were synthesized. The amplitude of each \ac{mpc} is assumed to follow free-space pathloss and is attenuated by $ 3 $\,dB after each reflection. Furthermore, we have designed two intersections between \acp{mpc} in their distance and \ac{aoa} parameters at time $ n = 83 $ and $n = 125$, respectively. For the intersection at time $ n = 83 $, the distances intersect simultaneously with the amplitudes. For the transmit signal $ \tilde{s}(t) $, we use a root-raised-cosine pulse with a roll-off factor of $ 0.6 $ and pulse duration of $ T_{\mathrm{p}} = 2$\,ns (bandwidth of $ 500 $\,MHz) at center frequency of $ f_{\mathrm{c}} = 6$\,GHz. With sampling period $ T_{\mathrm{s}} = 1.25 $\,ns, the number of samples per array element is $ N_{\mathrm{s}} = 46 $. In \cite{LeitingerAsilomar2020, LeitingerGrebienTSP2021}, a detection threshold $u_{\mathrm{de}}^{\mathrm{in}}$ is determined for \ac{sr}-\ac{sbl} channel estimation algorithms for single-snapshot wideband \ac{mimo} measurements. Given the signal parameters, number of antennas, and array-layout as defined above, this detection threshold $u_{\mathrm{de}}^{\mathrm{in}}$ for a chosen false alarm probability of $10^{-2}$ is given as $ -14.4 $\,dB. \acp{mpc} below $u_{\mathrm{de}}^{\mathrm{in}}$ are mostly miss detected and therefore barely utilized as measurements in the proposed algorithm. To support the detection and estimation of low \ac{snr} \acp{mpc}, the detection threshold can be relaxed. This will inevitably bring more false alarm measurements, but the proposed algorithm can efficiently filter them out even under low $ \mathrm{SNR}_{\mathrm{1m}}^{\mathrm{in}} $ conditions as shown in the following experimental results. We performed $ 100 $ simulation runs for each $ \mathrm{SNR}_{\mathrm{1m}}^{\mathrm{in}} \in \{-0.6, 2.4, 3.9, 5.4, 8.4, 13.4, 15.4, 18.4\} $\,dB with parameters: $ d_{\mathrm{max}} = 17 $\,m, $ \sigma_{\mathrm{fa}_\mathrm{ini}} = 0.5 $, $ \sigma_{\mathrm{fa}} = 0.15 $, $ \sigma_{\mathrm{d}} = 0.002\,\mathrm{m}/{\mathrm{s}^2} $, $ \sigma_{\mathrm{\varphi}} = 0.17^{\circ}/\mathrm{s}^2 $, $ {\underline{\sigma}_{\mathrm{u}}}_{k,n} = 0.02\, u_{k,n-1}^{\,\mathrm{MMSE}} $, $ \sigma_{\mathrm{v_{d}}} = 0.01\,\mathrm{m}/{\mathrm{s}} $, $ \sigma_{\mathrm{v_{\varphi}}} = 0.6^{\circ}/{\mathrm{s}} $.\footnote{The heuristic approach to scale the standard deviation ${\underline{\sigma}_{\mathrm{u}}}_{k,n}$ by the \ac{mmse} estimates $u_{k,n-1}^{\,\mathrm{MMSE}}$ was chosen since the range of normalized amplitude values tends to be very large.} Note that the first four $ \mathrm{SNR}_{\mathrm{1m}}^{\mathrm{in}} $ values represent extreme testing setups, where the true component \acp{snr} (output) of some \acp{mpc} are even below or close to $ 0\,$dB \ac{snr} output.

\subsubsection{Fully Synthetic Measurements}
\label{subsec:fullSynMea}

\begin{figure*}[t]
	\centering
	\hspace*{12mm}\subfloat[\label{subfig:fullySynMea_BP_MOSPA_distance_allSNRs}]
	{\hspace*{-10mm}\includegraphics[width=0.31\textwidth,height=0.212\textwidth]{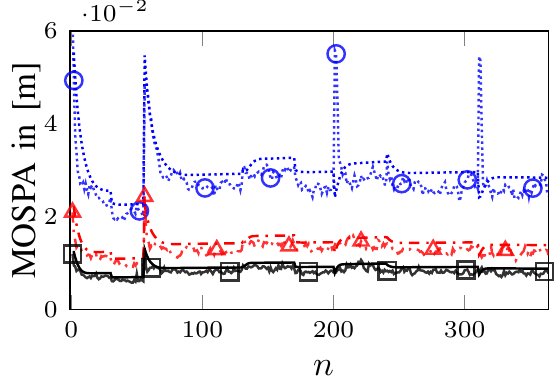}}
	\hspace*{12mm}\subfloat[\label{subfig:fullySynMea_BP_MOSPA_AoA_allSNRs}]
	{\hspace*{-10mm}\includegraphics[width=0.31\textwidth,height=0.195\textwidth]{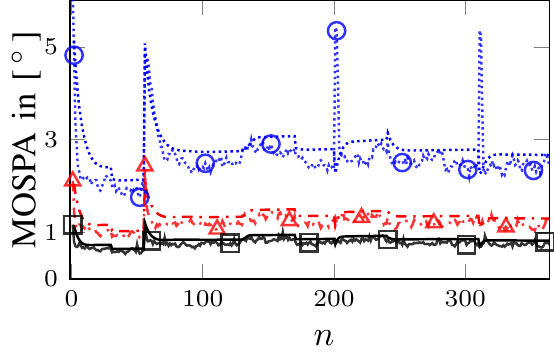}}
	\hspace*{12mm}\subfloat[\label{subfig:fullySynMea_BP_MOSPA_SNR_allSNRs}]
	{\hspace*{-10mm}\includegraphics[width=0.31\textwidth,height=0.195\textwidth]{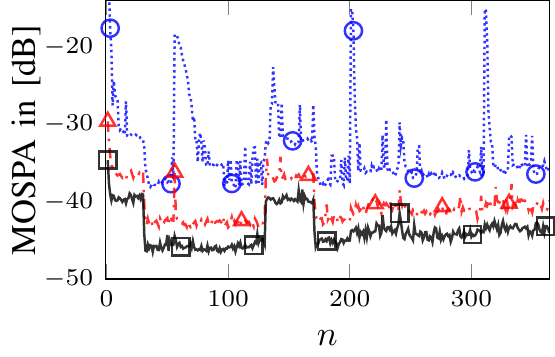}}\\[-2mm]
	\hspace*{12mm}\subfloat[\label{subfig:fullySynMea_BP_meanDetectedNumMPC_allSNRs}]
	{\hspace*{-11mm}\includegraphics[width=0.38\textwidth,height=0.19\textwidth]{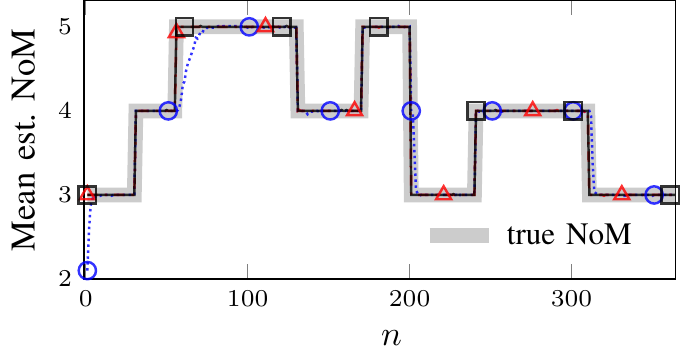}}
	\hspace*{12mm}\subfloat[\label{subfig:fullySynMea_BP_MeanFAR_allSNRs}]
	{\hspace*{-11mm}\includegraphics[width=0.38\textwidth,height=0.19\textwidth]{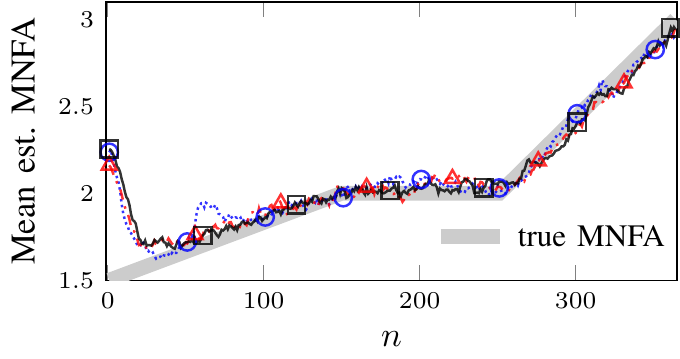}}\\[1mm]
	\centering
	\includegraphics[]{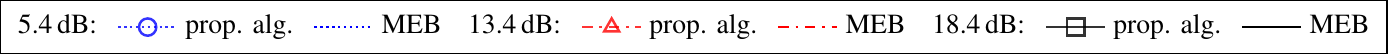}\\
	\caption{Results for fully synthetic measurements with the proposed algorithm (prop. alg.) given $ \mathrm{SNR}_{\mathrm{1m}}^{\mathrm{in}} = \{5.4, 13.4, 18.4\} $\,dB. MOSPA errors of the estimated (a) distances, (b) \acp{aoa}, and (c) input component \acp{snr}. Mean estimates of the \ac{nom} (d) and the \ac{far} (e). }	 
	\label{fig:fullSynMea_BP_MOSPA_allSNRs}
\end{figure*}

\begin{figure*}[t]
	\centering
	\hspace*{11mm}\subfloat[\label{subfig:MOSPA_d}]
	{\hspace*{-9mm}\includegraphics[width=0.235\textwidth,height=0.228\textwidth]{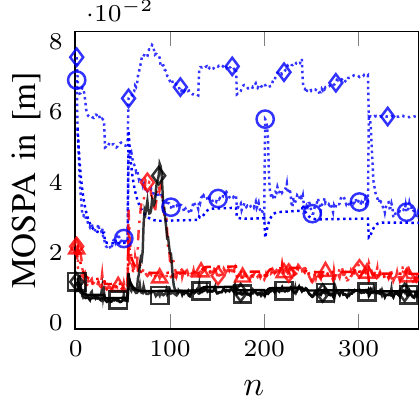}}
	\hspace*{11mm}\subfloat[\label{subfig:MOSPA_AoA}]
	{\hspace*{-9mm}\includegraphics[width=0.235\textwidth,height=0.21\textwidth]{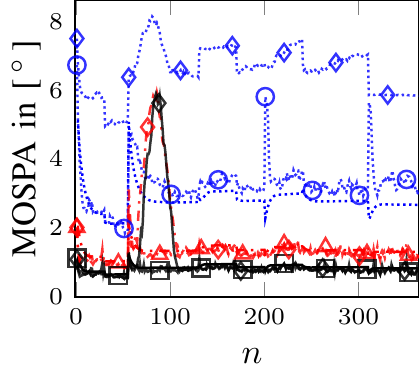}}
	\hspace*{11mm}\subfloat[\label{subfig:MOSPA_SNR}]
	{\hspace*{-9.5mm}\includegraphics[width=0.235\textwidth,height=0.21\textwidth]{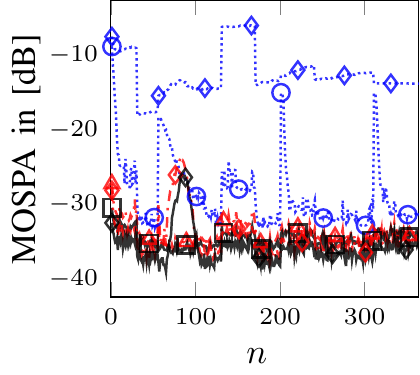}}
	\hspace*{11mm}\subfloat[\label{subfig:ModelOrder}]
	{\hspace*{-10mm}\includegraphics[width=0.23\textwidth,height=0.21\textwidth]{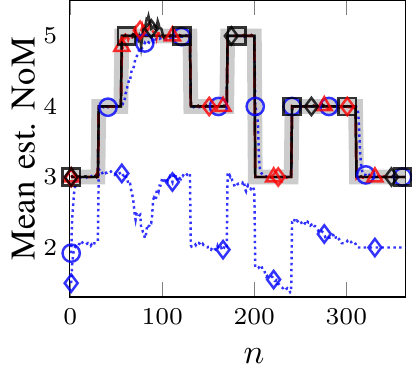}}\\[1mm]
	\centering
	\includegraphics[]{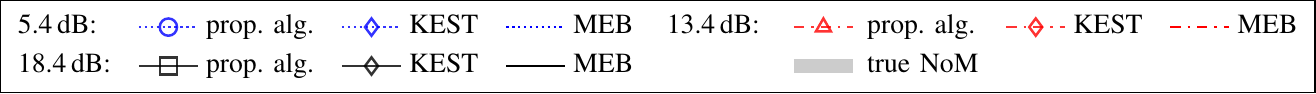}\\
	\caption{Results for synthetic radio measurements given $ \mathrm{SNR}_{\mathrm{1m}}^{\mathrm{in}} = \{5.4, 13.4, 18.4\}$\,dB. MOSPA errors of the estimated (a) distances, (b) \acp{aoa}, and (c) input component \acp{snr}. Mean estimates of the \ac{nom} (d). }	 
	\label{fig:MOSPA_eachSNR}
\end{figure*}

First, we present the simulation results using fully synthetic measurements without involving the snapshot-based channel estimator. In each simulation run, \ac{mpc}-oriented measurements were generated by adding Gaussian noises to the true \ac{mpc} parameters, where the noise variances were state-dependent and computed as in Section \ref{subsec:LikelihoodFunctions}. In addition, false alarm measurements were generated with increasing \ac{far} $ {\mu_{\mathrm{fa}}}_{n} $ from $ 1.5 $ to $ 3 $. More specifically, the distances and \acp{aoa} of false alarm measurements were drawn from uniform distributions in the validation region, and the norm amplitudes were generated with Rayleigh distribution with squared scale parameter $ 1/2 $ using a detection threshold of $ u_{\mathrm{de}}^{\mathrm{in}} = -20$\,dB.

Fig.~\ref{fig:fullSynMea_BP_singleMC} shows the results of an exemplary simulation run given $ \mathrm{SNR}_{\mathrm{1m}}^{\mathrm{in}} = 5.4\,$dB. It is seen that the proposed algorithm exhibits high detection and estimation accuracy for medium and high \ac{snr} \acp{mpc}. The ``weakest'' \ac{mpc}---with component SNR below the detection threshold of $ -20 $\,dB---is stably detected shortly after the beginning although the related measurements are mostly miss detected in the \ac{sr}-\ac{sbl}. In addition, the proposed algorithm excellently copes with intersecting \acp{mpc}. Fig.~\ref{fig:fullSynMea_BP_MOSPA_allSNRs} further presents the MOSPA errors and the mean estimates of the \ac{nom} and the \ac{far} given $ \mathrm{SNR}_{\mathrm{1m}}^{\mathrm{in}} = \{5.4, 13.4, 18.4\} $\,dB. Given medium and high $ \mathrm{SNR}_{\mathrm{1m}}^{\mathrm{in}} $ values, the \ac{nom} is accurately estimated and the MOSPA errors attain the MEBs. For distance, \ac{aoa} and component \ac{snr}, the MOSPA errors are mostly below $ 2 $\,cm, $ 2^{\circ} $ and $ -35 $\,dB, respectively. Given $ \mathrm{SNR}_{\mathrm{1m}}^{\mathrm{in}} = 5.4\,$dB, the MOSPA errors remain on the MEB-levels mostly, despite a few peaks due to the underestimated \ac{nom}. Furthermore, the slow varying \ac{far} is accurately estimated for all conditions. Additional simulation results for fully synthetic measurements with fast varying \ac{far} are provided in Appendix~\ref{app:AddSimResults_MNFA}.

\subsubsection{Synthetic Radio Measurements} \label{sec:resultSynRadioMeas}
Next, we show the overall performance of the proposed two-stage algorithm by involving the snapshot-based channel estimator. In each simulation run, radio measurements were synthesized by applying true \ac{mpc} parameters to the radio signal model \eqref{eq:SignalModel_sampled} with given $ \mathrm{SNR}_{\mathrm{1m}}^{\mathrm{in}} $. The measurements $ \bm{z}_{m,n} $ at each time $ n $ were provided by a snapshot-based \ac{sr}-\ac{sbl} channel estimator in line with the implementation in \cite{ShutinCSTA2013}. We relaxed the detection threshold to $u_\mathrm{de}^{\mathrm{in}} = -18\,$dB for $ \mathrm{SNR}_{\mathrm{1m}}^{\mathrm{in}} $ values above $ 5\, $dB and $u_\mathrm{de}^{\mathrm{in}} = -20\,$dB otherwise. For comparison, we implemented the \ac{kest} algorithm according to \cite{ ModelOrderSelection_SPM2004} and \cite{KEST_TAP2012} which performs detection of \acp{mpc} and sequential estimation of their distances, \acp{aoa} and amplitudes. For \ac{nom} estimation in the \ac{kest} algorithm, the penalty factor for a \ac{nom} change was set to $ p_{\mathrm{chg}} = 0.1 $, and the penalty $ p_{\mathrm{ord}} $ for a higher \ac{nom} was chosen according to the MDL principle \cite{KEST_TAP2012, ModelOrderSelection_SPM2004}.

\begin{figure*}[t]
	\centering
	\hspace*{10mm}\subfloat[\label{subfig:aveSNR_eachMPC}]
	{\hspace*{-12mm}\includegraphics[width=0.38\textwidth,height=0.22\textwidth]{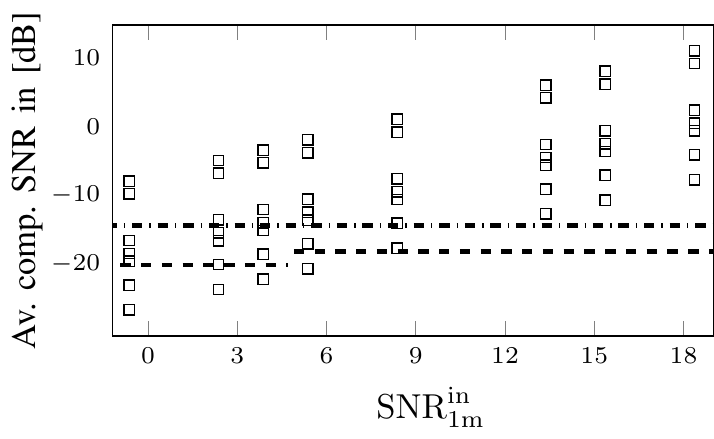}}
	\hspace*{18mm}\subfloat[\label{subfig:aveME_NoM_eachSNR}]
	{\hspace*{-11mm}\includegraphics[width=0.367\textwidth,height=0.217\textwidth]{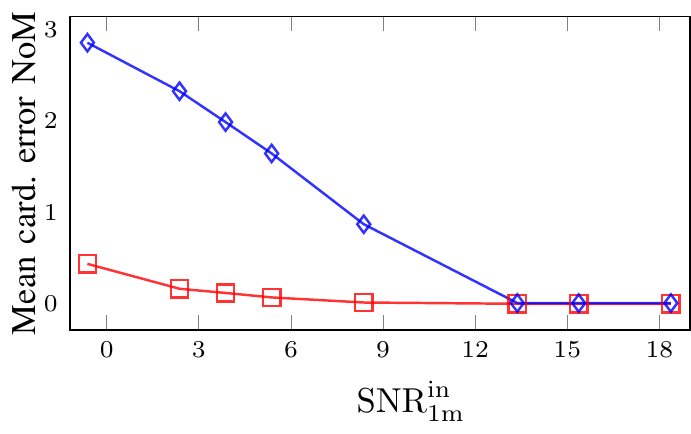}}\\[-5mm] 
	\hspace*{11mm}\subfloat[\label{subfig:aveMOSPA_d}]
	{\hspace*{-9mm}\includegraphics[width=0.37\textwidth,height=0.225\textwidth]{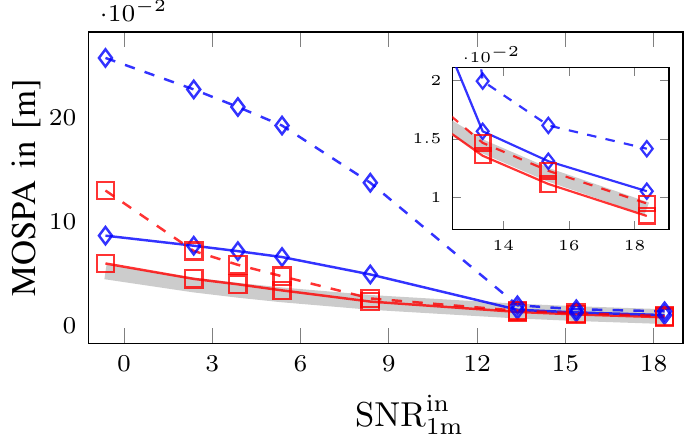}}
	\hspace*{18mm}\subfloat[\label{subfig:aveMOSPA_AoA}]
	{\hspace*{-14mm}\includegraphics[width=0.38\textwidth,height=0.208\textwidth]{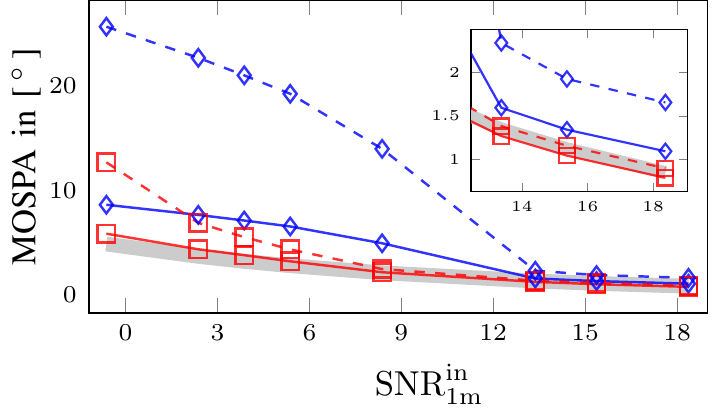}}\\[1mm]
	\centering
	\includegraphics[]{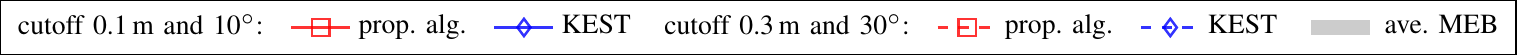}\\
	\caption{Results for synthetic radio measurements: The average (averaged over all time steps) (a) true input component \acp{snr} of $ 7 $ \acp{mpc}, (b) mean cardinality error of the \ac{nom}, and (c) and (d) MOSPA errors and MEBs. In (a), the dash-dotted line indicates the theoretical detection threshold of $ -14.4 $\,dB, and the dashed lines indicate the relaxed detection thresholds of $ -20 $\,dB and $ -18 $\,dB, respectively.}	 
	\label{fig:aveMOSPA_eachSNR}
\end{figure*}

For comparison between the proposed algorithm and the \ac{kest} algorithm of an exemplary simulation run, the reader is referred to Appendix~\ref{app:AddSimResults_synMea}. Fig.~\ref{fig:MOSPA_eachSNR} shows the MOSPA errors, MEBs and the mean estimate of the \ac{nom} for each $ \mathrm{SNR}_{\mathrm{1m}}^{\mathrm{in}} \in \{5.4, 13.4, 18.4\} $\,dB. The peaks of MOSPA errors indicate the \ac{nom} estimation errors which mostly happen when there is \ac{mpc} birth or death. For medium and high $ \mathrm{SNR}_{\mathrm{1m}}^{\mathrm{in}}  $, it shows that the MOSPA errors of the proposed algorithm are mostly below $ 2 $\,cm, $2^{\circ} $ and $-35$\,dB respectively for distance, \ac{aoa} and input component \ac{snr}. The \ac{kest} algorithm shows comparable results except for the high error peaks around the intersecting \acp{mpc} at time $ n = 83 $. At low \ac{snr}, the MOSPA errors versus time $n$ of the proposed algorithm are slightly above the MEBs since the ``weakest'' \acp{mpc}---with component \acp{snr} well below the relaxed detection thresholds---are occasionally not detected, however, the \ac{kest} algorithm has much larger MOSPA errors, since it almost never detects these \acp{mpc}.

Finally, Fig.~\ref{fig:aveMOSPA_eachSNR} shows component \acp{snr} of the \acp{mpc}, the mean cardinality error \cite{OSPA_TSP2008} in the \ac{nom}, and MOSPA errors averaged over all time steps versus \ac{snr} ($ \mathrm{SNR}_{\mathrm{1m}}^{\mathrm{in}} \in \{-0.6, 2.4, 3.9,$ $ 5.4, 8.4, 13.4, 15.4, 18.4\} $\,dB). Fig.~\ref{subfig:aveSNR_eachMPC} shows that quite a few of the \acp{mpc} have component \acp{snr} that are below the relaxed detection thresholds (some even far below the theoretical threshold) for the first five $ \mathrm{SNR}_{\mathrm{1m}}^{\mathrm{in}} $ values. These \acp{mpc} are mostly miss detected by the \ac{kest} algorithm, leading to a much higher mean cardinality error \cite{OSPA_TSP2008} as shown in Fig.~\ref{subfig:aveME_NoM_eachSNR} and therefore also to much higher MOSPA errors, as shown in Fig.~\ref{subfig:aveMOSPA_d} and Fig.~\ref{subfig:aveMOSPA_AoA}, than that of the proposed algorithm. By setting the cutoff parameters to larger values, i.e., $ 0.3 $\,m and $ 30^{\circ} $, the mean cardinality error and MOSPA errors are better visualized. At high $ \mathrm{SNR}_{\mathrm{1m}}^{\mathrm{in}} $ values both algorithms mostly detected all \acp{mpc}. However, the proposed algorithm still outperforms the \ac{kest} algorithm since its MOSPA errors are lower when \acp{mpc} are intersecting. In summary, the proposed algorithm outperforms the \ac{kest} algorithm in all testing conditions and attains the MEBs for $ \mathrm{SNR}_{\mathrm{1m}}^{\mathrm{in}} $ above $ 5\,$dB. A study of the runtimes for both algorithms can be found in Appendix~\ref{app:runtimeAnalysis}. For instance, the runtimes per time step (for MATLAB implementations) of both algorithms for $ \mathrm{SNR}_{\mathrm{1m}}^{\mathrm{in}} = 18.4\,$dB, obtained by averaging over all simulation runs and time steps, are both approximately 2\,s\,/\,step.

\subsection{Performance for Real \ac{uwb} Radio Measurements}

\begin{figure*}[t]
	\centering
	\includegraphics[width=0.65\textwidth,height=0.29\textwidth]{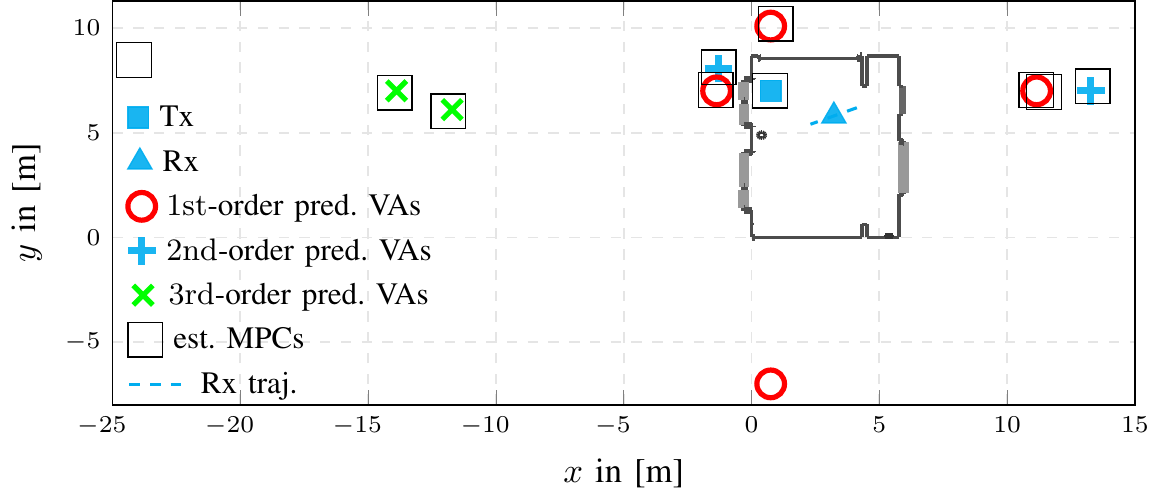}\\[0mm] 
	\caption{Floor plan of the measurement environment and the results of the proposed algorithm using real radio measurements of \emph{setup}-2 ($ 1\,$GHz bandwidth, $ 5\times5 $ array) at time $ n = 50 $. For each of the $ 10 $ detected \acp{mpc}, the estimated state is transformed into a two-dim. coordinate and associated with a geometrically calculated \ac{va}.}	 
	\label{fig:Floorplan_realData}
\end{figure*}

For further validation of the proposed algorithm, we use real radio measurements collected in a seminar room at TU Graz, Austria. The floor plan is depicted in Fig.~\ref{fig:Floorplan_realData}, including the positions of the static \ac{tx} and a few mirror images of the \ac{tx}, i.e., \acp{va}, which model the \acp{mpc} due to specular reflections. More details about the measurement environment and \ac{va} calculations can be found in \cite{MeasureMINT2013,Erik_SLAM_TWC2019,LeitingerJSAC2015}. On the \ac{tx} side, a dipole-like antenna with an approximately uniform radiation pattern in the azimuth plane and zeros in the floor and ceiling directions was used. At each \ac{rx} position, a same antenna was mounted on a plotter and moved yielding a virtual uniform rectangular array with an inter-element spacing of $ 2$\,cm. The \ac{uwb} signals are measured at $ 100 $ \ac{rx} positions along a $ 2 $\,m trajectory using an M-sequence correlative channel sounder with frequency range $3.1–10.6\,$GHz. We selected a subband with bandwidth $ B = 1/T_{\mathrm{p}}$ using filtering with a root-raised-cosine pulse with a roll-off factor of $ 0.6 $ and pulse duration $ T_{\mathrm{p}} $ at center frequency of $ f_{\mathrm{c}} = 6$\,GHz. The following two measurement setups are used: (i) \emph{setup}-1: $ T_{\mathrm{p}} = 2$\,ns, $ B = 0.5\,$GHz, $ N_{\mathrm{s}} = 94 $, $ 3 \times 3 $ array, $ \mathrm{SNR}_{\mathrm{1m}}^{\mathrm{in}} = 0.7\,$dB, $u_{\mathrm{de}}^{\mathrm{in}} = -16\,$dB; (ii) \emph{setup}-2: $ T_{\mathrm{p}} = 1$\,ns, $ B = 1\,$GHz, $ N_{\mathrm{s}} = 187 $, $ 5 \times 5 $ array, $ \mathrm{SNR}_{\mathrm{1m}}^{\mathrm{in}} = -6.7\,$dB, $u_{\mathrm{de}}^{\mathrm{in}} = -20\,$dB. Since no significant \ac{awgn} was observed from the filtered signal, artificial \ac{awgn} generated with $ \mathrm{SNR}_{\mathrm{1m}} $ was added. The simulation parameters are as follows: $ d_{\mathrm{max}} = 30 $\,m, $ \sigma_{\mathrm{d}} = 0.03\,\mathrm{m}/{\mathrm{s}^2} $, $ \sigma_{\mathrm{\varphi}} = 1^{\circ}/\mathrm{s}^2 $, $ {\underline{\sigma}_{\mathrm{u}}}_{k,n} = 0.1\, u_{k,n-1}^{\,\mathrm{MMSE}} $, $ \sigma_{\mathrm{fa}} = 1 $, $ \sigma_{\mathrm{v_{d}}} = 0.1\,\mathrm{m}/{\mathrm{s}} $, $ \sigma_{\mathrm{v_{\varphi}}} = 6^{\circ}/{\mathrm{s}} $.

\begin{figure*}[t]
	\centering
	\hspace*{10mm}\subfloat[\label{subfig:BP_realData_BF_dist_33array}]
	{\hspace*{-10.5mm}\includegraphics[width=0.37\textwidth,height=0.225\textwidth]{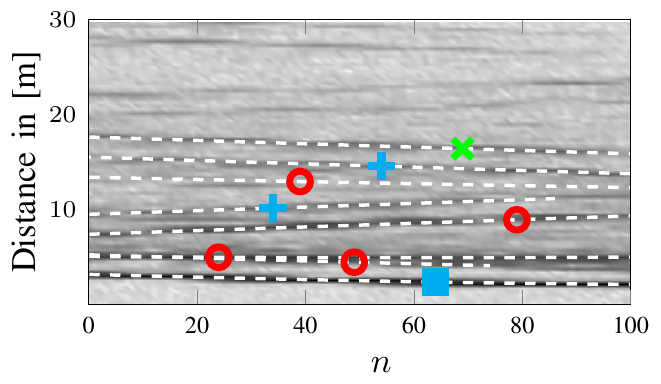}}
	\hspace*{10mm}\subfloat[\label{subfig:BP_realData_BF_AoA_33array}]
	{\hspace*{-8.6mm}\includegraphics[width=0.44\textwidth,height=0.225\textwidth]{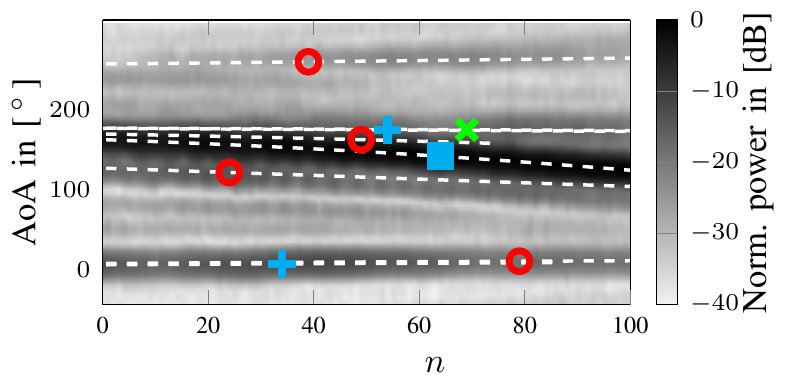}}\\[-6mm] 
	\hspace*{10mm}\subfloat[\label{subfig:BP_realData_3D_dist_SNR_33array}]
	{\hspace*{-10.5mm}\includegraphics[width=0.37\textwidth,height=0.225\textwidth]{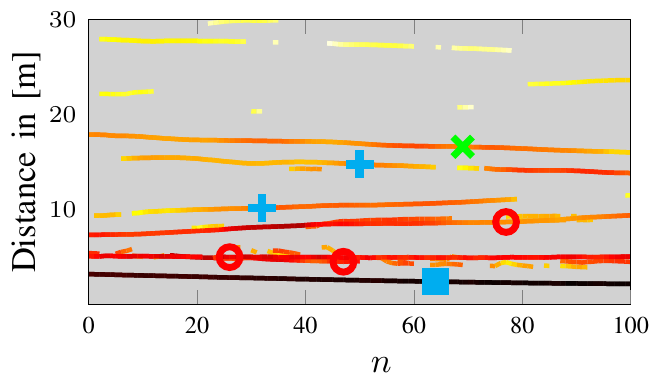}}
	\hspace*{10mm}\subfloat[\label{subfig:BP_realData_3D_AoA_SNR_33array}]
	{\hspace*{-8.3mm}\includegraphics[width=0.44\textwidth,height=0.237\textwidth]{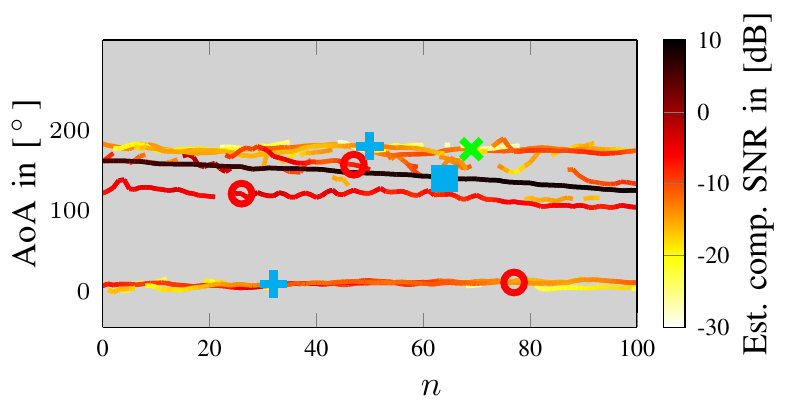}}\\[-6mm]
	\hspace*{10mm}\subfloat[\label{subfig:BP_realData_3D_dist_SNR_55array}]
	{\hspace*{-10.5mm}\includegraphics[width=0.37\textwidth,height=0.225\textwidth]{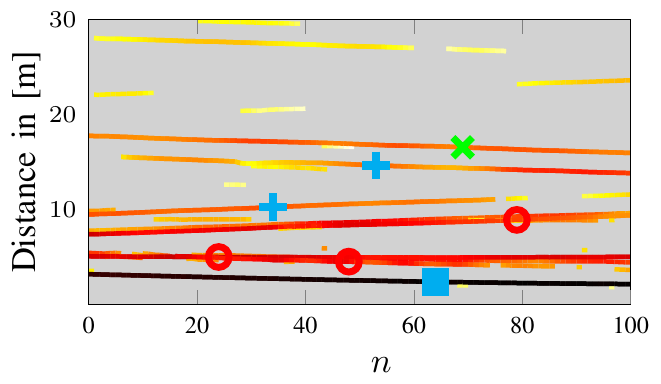}}
	\hspace*{10mm}\subfloat[\label{subfig:BP_realData_3D_AoA_SNR_55array}]
	{\hspace*{-8.5mm}\includegraphics[width=0.44\textwidth,height=0.237\textwidth]{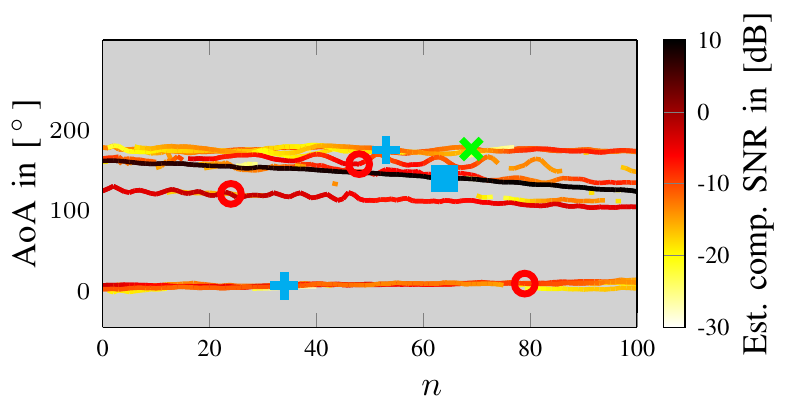}}\\[1mm]
	\caption{Results for real radio measurements. (a) and (b) depict the delay and angular power spectra versus time for \emph{setup}-2. Corresponding to the distinct peaks and their variations in the spectra, a few exemplary distance and \ac{aoa} paths (white dashed lines) are calculated with geometrically expected \acp{va}, of which the orders are highlighted with markers listed in Fig.~\ref{fig:Floorplan_realData}. The distance and \ac{aoa} estimates of the detected \acp{mpc} are shown for \emph{setup}-1 in (c) and (d), and for \emph{setup}-2 in (e) and (f), among which the estimates associated with the geometrically calculated paths in (a) and (c) are highlighted with markers.}	 
	\label{fig:BP_realData_setup1}
\end{figure*}

Fig.~\ref{fig:BP_realData_setup1} shows the estimated delays and \acp{aoa} of the detected \acp{mpc} for \emph{setup}-1 and \emph{setup}-2. The backgrounds of Fig.~\ref{subfig:BP_realData_BF_dist_33array} and Fig.~\ref{subfig:BP_realData_BF_AoA_33array} are the delay and angular power spectra versus time respectively for \emph{setup}-2, where the peaks and variations representing individual \ac{mpc} paths are readily visible in delay domain with good resolution capability of \ac{uwb} signals, but hardly resolved in angular domain with limited array aperture. As can be observed, some distinct peaks align with the \ac{mpc} paths predicted with up to $3\mathrm{rd}$-order geometrically expected \acp{va}, and related to the estimated \ac{mpc} paths highlighted with markers in Fig.~\ref{subfig:BP_realData_3D_dist_SNR_33array} and Fig.~\ref{subfig:BP_realData_3D_AoA_SNR_33array} for \emph{setup}-1 and in Fig.~\ref{subfig:BP_realData_3D_dist_SNR_55array} and Fig.~\ref{subfig:BP_realData_3D_AoA_SNR_55array} for \emph{setup}-2. The intersecting $1\mathrm{st}$-order \acp{mpc} and a few short-lived \acp{mpc} around $ 5\,$m are nearby in both delay and angular domains. They are better resolved for \emph{setup}-2 with larger signal bandwidth and array size, however oscillation on the estimates still exist especially in the angular domain as for \emph{setup}-1. Fig.~\ref{fig:Floorplan_realData} zooms into the performance of a single snapshot for \emph{setup}-2, where the MMSE estimates of the detected \ac{mpc} states at time $ n = 50$ after being transformed into two-dimensional coordinates are shown. Within the $ 10 $ detected \acp{mpc}, the LoS component and \acp{mpc} related to the three $1\mathrm{st}$-order \acp{va} w.r.t. surrounding walls are accurately estimated. Most of the other detected \acp{mpc} are also well explained by some higher order \acp{va} (up to order five). However, due to imperfect antenna-calibration leading to time-dispersive system response, ``ghost'' components are sometimes detected alongside significant \acp{mpc}. The estimated \acp{mpc} associated with the $2\mathrm{nd}$ and the upper $3\mathrm{rd}$-order \acp{va} are related to the complex room structure in the left-upper corner. Note that the lower wall has high attenuation coefficient and therefore no related \acp{mpc} are detected (as for example the \ac{mpc} related to the $1\mathrm{st}$-order \ac{va} of the lower wall which has parameters around $ 14\, $m and $ 270^{\circ} $). The estimated \acp{far} for both measurement setups converge rapidly and remain around one over time. The small and stable values can be explained by the high detection threshold $ u_{\mathrm{de}}^{\mathrm{in}} $ used in the \ac{sbl} channel estimator and the static measurement scenario. The proposed algorithm is capable of detecting \acp{mpc} and estimating their parameters that very well resemble the geometry, and capturing their dynamic behaviors related to the surrounding environment.

%% file: InputFiles/Conclusions.tex
We proposed a \ac{bp}-based algorithm for sequential detection and estimation of \ac{mpc} parameters based on radio signals, which adopts a two-stage structure combining a snapshot-based \ac{sr}-\ac{sbl} channel estimator with a BP-based sequential detection and estimation algorithm. Using amplitude information and the augmentation of \ac{pmpc} states with a binary existence variable enable the reliable detection of ``weak'' \acp{mpc} with very low component \acp{snr}. Simulation results using synthetic measurements show that the algorithm excellently copes with a high number of false alarm measurements and intersecting \acp{mpc} with parameters nearby in the dispersion space. It is capable of estimating the parameters of \acp{mpc} on \ac{pcrlb}-levels even for low \ac{snr} \acp{mpc}. We have shown that the performance of the proposed algorithm compares well to existing state-of-the-art algorithms for high \ac{snr} \acp{mpc}, but it significantly outperforms them for medium and low \ac{snr} \acp{mpc}. The results using real radio measurements show that the algorithm demonstrates excellent performance in challenging indoor-environments by detecting many geometry-related \acp{mpc} up to reflection-order five and estimating their dispersion parameters with high accuracy. Possible directions for future research include extending the proposed algorithm to a more general inhomogeneous false alarm intensity \cite{Alex_RadarConf2021} coping with false alarms resulting from model mismatches in the radio signal such as \ac{dmc} or incorporating correlations between measurements \cite{BerVoVoTSP2015}.

%% file: InputFiles/Acknowledgment.tex
The authors would like to thank Dr. Florian Meyer and Prof.	Bernard H. Fleury for carefully reading the manuscript and insightful comments.

%% file: InputFiles/AppendixJointPosteriorDist.tex
We provide derivations of the joint prior \ac{pdf} and the joint likelihood function in Section~\ref{app:JointPrior} and Section~\ref{app:JointLHF} respectively, which lead to the factorized expressions of the joint posterior \ac{pdf} \eqref{eq:TheJointPosteriorPDF_2} and the pseudo likelihood functions \eqref{eq:g} and \eqref{eq:h} in Section~\ref{app:JointPosterior}. In Section~\ref{app:normAmplitude}, we derive the squared scale parameter of the truncated Rician \ac{pdf} in \eqref{eq:pdf_normAmplitude}.

\subsection{Joint Prior \ac{pdf}}\label{app:JointPrior}

Before presenting derivations, we first define a few sets as follows: $ \mathcal{D}_{\V{a}_n,\underline{\V{r}}_n} \triangleq \{k \in \{1, \dots, K_{n-1}\}: \underline{r}_{k,n} = 1, a_{k,n} \neq 0 \} $ denotes the existing legacy \acp{pmpc} set, and $ \mathcal{N}_{\overline{\V{r}}_n} \triangleq \{m \in \{1,\dots,M_{n}\}: \overline{r}_{m,n} = 1, b_{m,n} = 0 \} $ denotes the existing new \acp{pmpc} set. Correspondingly, the sets of non-existing legacy \acp{pmpc} are given by $ \overline{\mathcal{D}}_{\V{a}_n, \underline{\V{r}}_n} \triangleq \{1, \dots, K_{n-1}\} \setminus \mathcal{D}_{\V{a}_n,\underline{\V{r}}_n}$, and the sets of non-existing new \acp{pmpc} are given as $ \overline{\mathcal{N}}_{\overline{\V{r}}_n} \triangleq \{1,\dots,M_{n}\} \setminus \mathcal{N}_{\overline{\V{r}}_n}$. Hence, the number of false alarm measurements can be represented with the sets as $ {k_\mathrm{fa}}_{n} = M_{n} - |\mathcal{D}_{\V{a}_n,\underline{\V{r}}_n}| - |\mathcal{N}_{\overline{\V{r}}_n}| $, and the number of \acp{pmpc} states is given as $ K_{n-1} + M_{n} = |\mathcal{D}_{\V{a}_n,\underline{\V{r}}_n} | + |\overline{\mathcal{D}}_{\V{a}_n,\underline{\V{r}}_n}| + |\mathcal{N}_{\overline{\V{r}}_n}| + |\overline{\mathcal{N}}_{\overline{\V{r}}_n}| $.

Assuming that the new \acp{pmpc} states $ \overline{\RV{y}}_{n} $ and the \ac{pmpc}-oriented association variables $ \RV{a}_{n} $ are conditionally independent given the legacy \acp{pmpc} state $ \underline{\RV{y}}_{n} $, the joint prior \ac{pdf} of $\underline{\RV{y}}_{1:n}$, $\overline{\RV{y}}_{1:n}$, $ \RV{a}_{1:n} $, $ \RV{b}_{1:n} $, ${\rv{\mu}_{\mathrm{fa}}}_{1:n} $, and the number of the measurements $\RV{m}_{1:n}$ factorizes as 
\begin{align}
&f(\V{y}_{1:n}, \V{a}_{1:n}, \V{b}_{1:n}, {\V{\mu}_{\mathrm{fa}}}_{1:n}, \V{m}_{1:n}) \nn \\ 
& \hspace{2mm} = f(\underline{\V{y}}_{1:n}, \overline{\V{y}}_{1:n}, \V{a}_{1:n}, \V{b}_{1:n}, {\V{\mu}_{\mathrm{fa}}}_{1:n}, \V{m}_{1:n}) \nn \\ 
& \hspace{2mm} = f({\mu_{\mathrm{fa}}}_{1}) f(\overline{\V{x}}_{1} | \overline{\V{r}}_{1}, M_{1}) p(\overline{\V{r}}_{1}, \V{a}_{1}, \V{b}_{1}, M_{1}|{\mu_{\mathrm{fa}}}_{1}) \nn \\
& \hspace{5mm} \times \prod_{ n' = 2 }^{n} f({\mu_{\mathrm{fa}}}_{n'}|{\mu_{\mathrm{fa}}}_{n'-1}) \left(\prod_{k = 1}^{K_{n'-1}} f(\underline{\V{y}}_{k,n'}|\V{y}_{k,n'-1})\right) \nn \\
& \hspace{5mm} \times f(\overline{\V{x}}_{n'} | \overline{\V{r}}_{n'}, M_{n'}) p(\overline{\V{r}}_{n'}, \V{a}_{n'}, \V{b}_{n'}, M_{n'}|{\mu_{\mathrm{fa}}}_{n'},\underline{\V{y}}_{n'})
\label{eq:jointPriorPDF_global} 
\end{align}
where $ p(\overline{\V{r}}_{1}, \V{a}_{1}, \V{b}_{1}, M_{1}|{\mu_{\mathrm{fa}}}_{1}, \underline{\V{y}}_{1}) = p(\overline{\V{r}}_{1}, \V{a}_{1}, \V{b}_{1}, M_{1}|{\mu_{\mathrm{fa}}}_{1}) $ since no legacy \acp{pmpc} exist at time $ n=1 $. We determine the prior \ac{pdf} of new \acp{pmpc} $ f(\overline{\V{x}}_{n} | \overline{\V{r}}_{n}, M_{n}) $ and the joint conditional prior \ac{pmf} $ p(\overline{\V{r}}_{n}, \V{a}_{n}, \V{b}_{n}, M_{n}|{\mu_{\mathrm{fa}}}_{n}, \underline{\V{y}}_{n}) $ as follows.

Before the current measurements are observed, the number of measurements $\rv{M}_{n}$ is random. The Poisson \ac{pmf} of the number of existing new \acp{pmpc} evaluated at $ |\mathcal{N}_{\overline{\V{r}}_n}| $ is given by $ p(|\mathcal{N}_{\overline{\V{r}}_n}|) = \mu_{\mathrm{n}}^{|\mathcal{N}_{\overline{\V{r}}_n}|}/|\mathcal{N}_{\overline{\V{r}}_n}|!\mathrm{e}^{\mu_{\mathrm{n}}} $. The prior \ac{pdf} of the new \ac{pmpc} state $ \overline{\RV{x}}_{n} $ conditioned on $\overline{\RV{r}}_{n}$ and $\rv{M}_{n}$ is expressed as 
\begin{align}
& f(\overline{\V{x}}_{n} | \overline{\V{r}}_{n}, M_{n}) = \prod_{m \in \mathcal{N}_{\overline{\V{r}}_n}} f_{\mathrm{n}}(\overline{\V{x}}_{m,n}) \prod_{m' \in \overline{\mathcal{N}}_{\overline{\V{r}}_n} } f_{\mathrm{D}}(\overline{\V{x}}_{m',n}).
\label{eq:priorPDF_newPSMC} 
\end{align}
The \ac{pmf} for the number of false alarm measurements is given by $ p({k_\mathrm{fa}}_{n}) = {\mu_{\mathrm{fa}}}_{n}^{{k_\mathrm{fa}}_{n}}/ {k_\mathrm{fa}}_{n}!\mathrm{e}^{{\mu_{\mathrm{fa}}}_{n}} $. The joint conditional prior \ac{pmf} of the binary existence variables of new \acp{pmpc} $\overline{\RV{r}}_{n} \triangleq [\overline{\rv{r}}_{1,n}\ist\cdots$ $\overline{\rv{r}}_{\rv{M}_n,n}]$, the \ac{da} vectors $ \RV{a}_{n} $ and $ \RV{b}_{n} $ and the number of the measurements $ \rv{M}_{n} $ conditioned on $ {\rv{\mu}_{\mathrm{fa}}}_{n} $ and $ \underline{\RV{y}}_{n} $ is obtained as \cite{BarShalom_AlgorithmHandbook, Florian_Proceeding2018, Erik_SLAM_TWC2019} 
\begin{align} 
& p(\overline{\V{r}}_{n}, \V{a}_{n}, \V{b}_{n}, M_{n}|{\mu_{\mathrm{fa}}}_{n},\underline{\V{y}}_{n}) \nn \\ 
& = \chi_{\overline{\V{r}}_{n}, \V{a}_{n}, M_{n}} \left(\prod_{ m\in \mathcal{N}_{\overline{\V{r}}_n} } \Gamma_{\V{a}_{n}}(\overline{r}_{m,n}) \right) \hspace*{-1mm} \left(\prod_{ k \in \mathcal{D}_{\V{a}_n,\underline{\V{r}}_n} } p_{\mathrm{d}}(\underline{u}_{k,n}) \right) \nn \\
& \hspace*{2mm} \times \Psi(\V{a}_n,\V{b}_n) \left(\prod_{k' \in \overline{\mathcal{D}}_{\V{a}_n,\underline{\V{r}}_n} } \hspace*{-3mm} \big( 1(a_{k',n}) - \underline{r}_{k',n} p_{\mathrm{d}}(\underline{u}_{k',n}) \big)\right)\ist .
\label{eq:priorPDF_DA}
\end{align}
The normalization constant $ \chi_{\overline{\V{r}}_{n}, \V{a}_{n}, M_{n}} $ combining the two Poisson \acp{pmf} above is given by 
\begin{align} 
\chi_{\overline{\V{r}}_{n}, \V{a}_{n}, M_{n}} & = \left(\mathrm{e}^{-\mu_{\mathrm{n}}} /M_n! \right) \left( (\mu_{\mathrm{n}}/{\mu_{\mathrm{fa}}}_{n})^{|\mathcal{N}_{\overline{\V{r}}_n}|}  {\mu_{\mathrm{fa}}}_{n}^{-|\mathcal{D}_{\V{a}_n,\underline{\V{r}}_n}|}  \right)  \nn \\
& \hspace*{30mm} \times \left( \mathrm{e}^{-{\mu_{\mathrm{fa}}}_{n}}  {\mu_{\mathrm{fa}}}_{n}^{M_{n}}\right)
\label{eq:normConst_FAR}
\end{align}
where the first part (the terms in the first brackets) is fixed after observing the current measurements given the assumption that the mean number of newly detected \acp{pmpc} $ \rv{\mu}_{\mathrm{n}} $ is a known constant. The second part can be merged with factors in the sets $ \mathcal{N}_{\overline{\V{r}}_n} $ and $ \mathcal{D}_{\V{a}_n,\underline{\V{r}}_n} $, respectively. The third part equals to $ n({\mu_{\mathrm{fa}}}_{n})^{(K_{n-1} + M_{n})} $ where $ n({\mu_{\mathrm{fa}}}_{n}) \triangleq ( \mathrm{e}^{-{\mu_{\mathrm{fa}}}_{n}} {\mu_{\mathrm{fa}}}_{n}^{M_{n}} )^{1/(K_{n-1} + M_{n})} $ is the \ac{far}-related normalization constant. The two exclusion functions $ \Psi(\V{a}_n,\V{b}_n) $ and $ \Gamma_{\V{a}_{n}}(\overline{r}_{m,n}) = 0 $ ensure that $ p(\overline{\V{r}}_{n}, \V{a}_{n}, \V{b}_{n}, M_{n}|{\mu_{\mathrm{fa}}}_{n},\underline{\V{y}}_{n}) \neq 0 $ if and only if a measurement is generated by only one \ac{pmpc} (either a legacy or a new one), and a \ac{pmpc} generates no more than one measurement. 

The product of the prior \ac{pdf} of new \acp{pmpc} \eqref{eq:priorPDF_newPSMC} and the joint conditional prior \ac{pmf} \eqref{eq:priorPDF_DA} after merging factors can be written up to the normalization constant as
\begin{align}
& p(\overline{\V{r}}_{n}, \V{a}_{n}, \V{b}_{n}, M_{n}|{\mu_{\mathrm{fa}}}_{n},\underline{\V{y}}_{n}) f(\overline{\V{x}}_{n} | \overline{\V{r}}_{n}, M_{n})\nn \\
& \propto \left( \psi(\bm{a}_n, \bm{b}_n) \prod_{ k \in \mathcal{D}_{\V{a}_n,\underline{\V{r}}_n} } \dfrac{n({\mu_{\mathrm{fa}}}_{n}) p_{\mathrm{d}}(\underline{u}_{k,n})} {{\mu_{\mathrm{fa}}}_{n}} 
\right. \nn \\
& \quad \left. \times 
\prod_{k' \in \overline{\mathcal{D}}_{\V{a}_n,\underline{\V{r}}_n} } n({\mu_{\mathrm{fa}}}_{n}) \big( \bar{1}(a_{k',n}) - \underline{r}_{k',n} p_{\mathrm{d}}(\underline{u}_{k',n}) \big) \right) \nn \\ 
& \quad \times \left( \prod_{ m\in \mathcal{N}_{\overline{\V{r}}_n} } \dfrac{n({\mu_{\mathrm{fa}}}_{n})\mu_{\mathrm{n}} f_{\mathrm{n}}(\overline{\V{x}}_{m,n})}{{\mu_{\mathrm{fa}}}_{n}} \Gamma_{\V{a}_{n}}(\overline{r}_{m,n}) \right. \nn \\
& \quad \left. \times 
\prod_{m' \in \overline{\mathcal{N}}_{\overline{\V{r}}_n} } n({\mu_{\mathrm{fa}}}_{n}) f_{\mathrm{D}}(\overline{\V{x}}_{m',n}) \right).
\label{eq:priorPDF_DA_newPSMC_combine} 
\end{align} 
With some simple manipulations using the definitions of exclusion functions $ \Psi(\V{a}_n,\V{b}_n) $ and $ \Gamma_{\V{a}_{n}}(\overline{r}_{m,n}) $ (see Section~\ref{sec:DA}), Eq.~\eqref{eq:priorPDF_DA_newPSMC_combine} can be rewritten as the product of factors related to the legacy \acp{pmpc} and to the new \acp{pmpc}, respectively, i.e.,
\begin{align}
& p(\overline{\V{r}}_{n}, \V{a}_{n}, \V{b}_{n}, M_{n}|{\mu_{\mathrm{fa}}}_{n},\underline{\V{y}}_{n}) f(\overline{\V{x}}_{n} | \overline{\V{r}}_{n}, M_{n}) \nn \\
& \hspace*{2mm} \propto \left(\prod_{k = 1}^{K_{n-1}} g_{1}(\underline{\V{y}}_{k,n}, a_{k,n}, {\mu_{\mathrm{fa}}}_{n}; M_{n}) \prod_{m = 1}^{M_{n}} \psi(a_{k,n},b_{m,n}) \right) \nn \\
& \hspace*{4mm}\times\left(\prod_{m' = 1}^{M_{n}} h_{1}(\overline{\V{y}}_{m',n}, b_{m',n}, {\mu_{\mathrm{fa}}}_{n}; M_{n})\right)
\label{eq:priorPDF_DA_newPSMC}
\end{align}
with $ g_{1}(\underline{\V{y}}_{k,n}, a_{k,n}, {\mu_{\mathrm{fa}}}_{n}; M_{n})\rmv\rmv=\rmv\rmv g_{1}(\underline{\V{x}}_{k,n}, \underline{r}_{k,n}, a_{k,n}, {\mu_{\mathrm{fa}}}_{n}; M_{n}) $ given by 
\begin{align}
& g_{1}(\underline{\V{x}}_{k,n}, \underline{r}_{k,n} = 1, a_{k,n}, {\mu_{\mathrm{fa}}}_{n}; M_{n}) \nn \\
& \hspace*{15mm} \triangleq
\begin{cases}
\dfrac{n({\mu_{\mathrm{fa}}}_{n}) p_{\mathrm{d}}(\underline{u}_{k,n})} {{\mu_{\mathrm{fa}}}_{n}}, 		& a_{k,n} = m \\
n({\mu_{\mathrm{fa}}}_{n}) \big(1 - p_{\mathrm{d}}(\underline{u}_{k,n})\big),								& a_{k,n} = 0
\end{cases}
\label{eq:g1}
\end{align}
and $ g_{1}(\underline{\V{x}}_{k,n}, \underline{r}_{k,n} = 0, a_{k,n}, {\mu_{\mathrm{fa}}}_{n}; M_{n}) \triangleq \bar{1}(a_{k,n}) n({\mu_{\mathrm{fa}}}_{n}) $, and $h_{1}(\overline{\V{y}}_{m,n}, b_{m,n}, {\mu_{\mathrm{fa}}}_{n}; M_{n}) = h_{1}(\overline{\V{x}}_{m,n}, \overline{r}_{m,n},b_{m,n}, {\mu_{\mathrm{fa}}}_{n}; M_{n}) $ is given by 
\begin{align}
& h_{1}(\overline{\V{x}}_{m,n}, \overline{r}_{m,n} = 1 , b_{m,n}, {\mu_{\mathrm{fa}}}_{n}; M_{n}) \nn \\
& \hspace*{15mm} \triangleq
\begin{cases}
0, 																							& b_{m,n} = k \\
\dfrac{n({\mu_{\mathrm{fa}}}_{n})\mu_{\mathrm{n}} f_{\mathrm{n}}(\overline{\V{x}}_{m,n})}{{\mu_{\mathrm{fa}}}_{n}},	& b_{m,n} = 0
\end{cases}
\label{eq:h1} 
\end{align}
and $ h_{1}(\overline{\V{x}}_{m,n}, \overline{r}_{m,n} = 0 , b_{m,n}, {\mu_{\mathrm{fa}}}_{n}; M_{n}) \triangleq n({\mu_{\mathrm{fa}}}_{n}) $. 

Finally, by inserting (\ref{eq:priorPDF_DA_newPSMC}) into (\ref{eq:jointPriorPDF_global}), the joint prior \ac{pdf} can be rewritten as 
\begin{align}
&f(\underline{\V{y}}_{1:n}, \overline{\V{y}}_{1:n}, \V{a}_{1:n}, \V{b}_{1:n}, {\V{\mu}_{\mathrm{fa}}}_{1:n}, \V{m}_{1:n}) \nn \\
& \propto f({\mu_{\mathrm{fa}}}_{1}) \prod_{l = 1}^{M_{1}} h_{1}(\overline{\V{y}}_{l,1}, b_{l,1} ,{\mu_{\mathrm{fa}}}_{1}; M_{1}) \nn \\
& \hspace*{2mm} \times \prod_{n' = 2}^{n} f({\mu_{\mathrm{fa}}}_{n'} | {\mu_{\mathrm{fa}}}_{n'-1}) \left(\prod_{k' = 1}^{K_{n'-1}} f(\underline{\V{y}}_{k',n'}|\V{y}_{k',n'-1})\right)  \nn \\
& \hspace*{2mm} \times \left(\prod_{k = 1}^{K_{n'-1}}g_{1}(\underline{\V{y}}_{k,n'}, a_{k,n'}, {\mu_{\mathrm{fa}}}_{n'}; M_{n'}) \prod_{m = 1}^{M_{n'}} \psi(a_{k,n'},b_{m,n'})\right) \nn \\
& \hspace*{2mm} \times \left(\prod_{m' = 1}^{M_{n'}} h_{1}(\overline{\V{y}}_{m',n'}, b_{m',n'}, {\mu_{\mathrm{fa}}}_{n'}; M_{n'})\right)\ist.
\label{eq:jointPriorPDF_global_factorized}
\end{align}

\subsection{Joint Likelihood Function}
\label{app:JointLHF}

Assume that the measurements $\RV{z}_{n}$ are independent across $n$, the conditional \ac{pdf} of $\RV{z}_{1:n}$ given $\underline{\RV{y}}_{1:n}$, $\overline{\RV{y}}_{1:n}$, $ \RV{a}_{1:n} $, $ \RV{b}_{1:n} $, and the number of measurements $\RV{m}_{1:n}$ is given by 
\begin{align}
&f(\V{z}_{1:n}|\underline{\V{y}}_{1:n}, \overline{\V{y}}_{1:n}, \V{a}_{1:n}, \V{b}_{1:n}, \V{m}_{1:n}) \nn \\
& \quad = \prod_{n'=1}^{n} f(\V{z}_{n'}|\underline{\V{y}}_{n'}, \overline{\V{y}}_{n'}, \V{a}_{n'}, \V{b}_{n'}, M_{n'})\ist.
\label{eq:LHF1} 
\end{align}
Note that $ f(\V{z}_{1}|\underline{\V{y}}_{1}, \overline{\V{y}}_{1}, \V{a}_{1}, \V{b}_{1}, M_{1}) = f(\V{z}_{1}|\overline{\V{y}}_{1}, \V{a}_{1}, \V{b}_{1}, M_{1}) $ since no legacy \acp{pmpc} exist at time $ n = 1 $. The conditional \ac{pdf} $ f(\V{z}_{m,n}|\V{x}_{k,n}) $ characterizing the statistical relation between the measurements $\RV{z}_{m,n}$ and the \ac{pmpc} states $\RV{x}_{k,n}$ is a central element in the conditional \ac{pdf} of the measurement vector $\RV{z}_{n}$ given $\underline{\RV{y}}_{n}$, $\overline{\RV{y}}_{n}$, $ \RV{a}_{n} $, $ \RV{b}_{n} $, and the number of the measurements $\rv{M}_{n}$. Assuming that the measurements $\RV{z}_{m,n}$ are conditionally independent across $m$ given $\underline{\RV{y}}_{k,n}$, $\overline{\RV{y}}_{m,n}$, $\rv{a}_{k,n}$, $\rv{b}_{m,n}$, and $\rv{M}_n$ \cite{BarShalom_AlgorithmHandbook}, Eq.~\eqref{eq:LHF1} factorizes as 
\begin{align}
& f(\V{z}_{1:n}|\underline{\V{y}}_{1:n}, \overline{\V{y}}_{1:n}, \V{a}_{1:n}, \V{b}_{1:n}, \V{m}_{1:n}) \nn \\
& \hspace*{2mm} = C(\V{z}_{1}) \left(\prod_{m \in \mathcal{N}_{\overline{\V{r}}_1}} \dfrac{f(\V{z}_{m,1}|\overline{\V{x}}_{m,1})}{f_{\mathrm{fa}}(\V{z}_{m,1})}\right) \nn \\ 
& \hspace*{6mm} \times \prod_{n' = 2}^{n} C(\V{z}_{n'}) \left(\prod_{k \in \mathcal{D}_{\V{a}_{n'},\underline{\V{r}}_{n'}}}  \dfrac{f(\V{z}_{a_{k,n'},n'}|\underline{\V{x}}_{k,n'})} {f_{\mathrm{fa}}(\V{z}_{a_{k,n'},n'})} \right) \nn \\
& \hspace*{6mm} \times \left(\prod_{m \in \mathcal{N}_{\overline{\V{r}}_{n'}}} \dfrac{f(\V{z}_{m,n'}|\overline{\V{x}}_{m,n'})} {f_{\mathrm{fa}}(\V{z}_{m,n'})}\right), 
\label{eq:LHF_global}
\end{align}
and the conditional \ac{pdf} at each time $ n \geq 2 $ factorizes as \cite{BarShalom_AlgorithmHandbook} 
\begin{align}
& f(\V{z}_{n}|\underline{\V{y}}_{n}, \overline{\V{y}}_{n}, \V{a}_{n}, \V{b}_{n}, M_{n}) \nn \\
& \hspace*{15mm} = C(\V{z}_{n}) \left(\prod_{k \in \mathcal{D}_{\V{a}_n,\underline{\V{r}}_n}}  \dfrac{f(\V{z}_{a_{k,n},n}|\underline{\V{x}}_{k,n})}{f_{\mathrm{fa}}(\V{z}_{a_{k,n},n})}\right)  \nn \\
& \hspace*{19mm} \times \left(\prod_{m \in \mathcal{N}_{\overline{\V{r}}_n}} \dfrac{f(\V{z}_{m,n}|\overline{\V{x}}_{m,n})}{f_{\mathrm{fa}}(\V{z}_{m,n})}\right)\ist.
\label{eq:LHF2} 
\end{align} 
Since the normalization factor $ C(\V{z}_{n}) = \prod_{m=1}^{M_n}f_{\mathrm{fa}}(\V{z}_{m,n}) $ depending on $ \V{z}_{n} $ and $ M_n $ is fixed after the current measurement $ \V{z}_{n} $ is observed, the likelihood function in \eqref{eq:LHF2} can be rewritten up to the normalization constant as
\begin{align}
& f(\V{z}_{n}|\underline{\V{y}}_{n}, \overline{\V{y}}_{n}, \V{a}_{n}, \V{b}_{n}, M_{n}) \nn \\
& \propto \left(\prod_{k = 1}^{K_{n-1}} g_{2}(\underline{\V{y}}_{k,n}, a_{k,n}; \V{z}_{n})\right) \hspace*{-1mm}
\vphantom{\prod_{k = 1}^{K_{n-1}}} \left( \prod_{m = 1}^{M_{n}} h_{2}(\overline{\V{y}}_{m,n}, b_{m,n}; \V{z}_{n}) \vphantom{\prod_{k = 1}^{K_{n-1}}}  \right) \rule[-1.8em]{0pt}{0pt}
\label{eq:LHF3} 
\end{align} 
where the factor related to legacy \ac{pmpc} states $ g_{2}(\underline{\V{y}}_{k,n}, a_{k,n};\\ \V{z}_{n}) \hspace*{-0.4mm} = \hspace*{-0.4mm} g_{2}(\underline{\V{x}}_{k,n}, \underline{r}_{k,n}, a_{k,n}; \V{z}_{n}) $ is given by 
\begin{align}
& g_{2}(\underline{\V{x}}_{k,n}, \underline{r}_{k,n} = 1, a_{k,n}; \V{z}_{n}) \nn \\
& \hspace*{20mm} \triangleq
\begin{cases}
\dfrac{f(\V{z}_{m,n}|\underline{\V{x}}_{k,n})}{f_{\mathrm{fa}}(\V{z}_{m,n})}, 	& a_{k,n} = m \\
1,																				& a_{k,n} = 0
\end{cases}
\label{eq:g2} 
\end{align}
and $ g_{2}(\underline{\V{x}}_{k,n}, \underline{r}_{k,n} = 0, a_{k,n}; \V{z}_{n}) \triangleq 1 $. The factor related to new \ac{pmpc} states $h_{2}(\overline{\V{y}}_{m,n}, b_{m,n}; \V{z}_{n})= h_{2}(\overline{\V{x}}_{m,n}, \overline{r}_{m,n}, b_{m,n};\\ \V{z}_{n}) $ is given by 
\begin{align}
& h_{2}(\overline{\V{x}}_{m,n}, \overline{r}_{m,n} = 1, b_{m,n}; \V{z}_{n}) \nn \\
& \hspace*{22mm} \triangleq 
\begin{cases}
1, 																				& b_{m,n} = k \\
\dfrac{f(\V{z}_{m,n}|\overline{\V{x}}_{m,n})}{f_{\mathrm{fa}}(\V{z}_{m,n})}, 	& b_{m,n} = 0
\end{cases}
\label{eq:h2}
\end{align}
and $ h_{2}(\overline{\V{x}}_{m,n}, \overline{r}_{m,n} = 0, b_{m,n}; \V{z}_{n}) \triangleq 1 $. Inserting \eqref{eq:LHF3} \eqref{eq:g2} and \eqref{eq:h2} into \eqref{eq:LHF1}, the conditional \ac{pdf} can be rewritten as the joint likelihood function 
\begin{align}
& f(\V{z}_{1:n}|\underline{\V{y}}_{1:n}, \overline{\V{y}}_{1:n}, \V{a}_{1:n}, \V{b}_{1:n}, \V{m}_{1:n})
\nn \\
&\hspace*{7mm}\propto \left(\prod_{m = 1}^{M_{1}} h_{2}(\overline{\V{x}}_{m,1}, \overline{r}_{m,1}, b_{m,1}; \V{z}_{1})\right) \nn \\
& \hspace*{11mm} \times \prod_{n' = 2}^{n} \left(\prod_{k = 1}^{K_{n'-1}} g_{2}(\underline{\V{x}}_{k,n'}, \underline{r}_{k,n'}, a_{k,n'}; \V{z}_{n'})\right) 
\nn \\ 
& \hspace*{11mm} \times \left(\prod_{m' = 1}^{M_{n'}} h_{2}(\overline{\V{x}}_{m',n'}, \overline{r}_{m',n'}, b_{m',n'}; \V{z}_{n'})\right)\ist.
\label{eq:LHF4} 
\end{align}

\subsection{Joint Posterior \ac{pdf}}
\label{app:JointPosterior}

Finally, by substituting the joint prior \ac{pdf} with \eqref{eq:jointPriorPDF_global_factorized} and the joint likelihood function with \eqref{eq:LHF4}, the joint posterior \ac{pdf} \eqref{eq:TheJointPosteriorPDF_1} can be rewritten as 
\begin{align}
& f(\V{y}_{1:n}, \V{a}_{1:n}, \V{b}_{1:n}, {\V{\mu}_{\mathrm{fa}}}_{1:n}, \V{m}_{1:n} | \V{z}_{1:n})\nn \\ 
& \hspace*{-1mm} \propto f({\mu_{\mathrm{fa}}}_{1}) \left(\prod_{l = 1}^{M_{1}} h_{1}(\overline{\V{y}}_{l,1}, b_{l,1}, {\mu_{\mathrm{fa}}}_{1}; M_{1}) h_{2}(\overline{\V{y}}_{l,1}, b_{l,1}; \V{z}_{1}) \right) \nn \\
& \times \prod_{n' = 2}^{n} f({\mu_{\mathrm{fa}}}_{n'}|{\mu_{\mathrm{fa}}}_{n'-1}) \left( \prod_{k' = 1}^{K_{n'-1}} f(\underline{\V{y}}_{k',n'}|\V{y}_{k',n'-1})\right) \nn \\
& \times \left( \prod_{k = 1}^{K_{n'-1}} g_{1}(\underline{\V{y}}_{k,n'}, a_{k,n'}, {\mu_{\mathrm{fa}}}_{n'}; M_{n'})g_{2}(\underline{\V{y}}_{k,n'}, a_{k,n'}; \V{z}_{n'}) \right. \nn \\
& \times \left. \prod_{m = 1}^{M_{n'}}\psi(a_{k,n'},b_{m,n'}) \hspace*{-0.5mm} \right) \hspace*{-1.5mm} \left( \prod_{m' = 1}^{M_{n'}} h_{1}(\overline{\V{y}}_{m',n'}, b_{m',n'}, {\mu_{\mathrm{fa}}}_{n'}; M_{n'}) \right. \nn \\
& \hspace*{1mm} \times \left.  h_{2}(\overline{\V{y}}_{m',n'}, b_{m',n'}; \V{z}_{n'}) \vphantom{\prod_{m' = 1}^{M_{n'}}} \right)\ist.
\label{eq:TheJointPosteriorPDF_complete} 
\end{align}
The factors related to the legacy \acp{pmpc} and to the new \acp{pmpc} can be simplified as $ g(\underline{\V{y}}_{k,n}, a_{k,n}, {\mu_{\mathrm{fa}}}_{n}; \V{z}_{n})\triangleq g_{1}(\underline{\V{y}}_{k,n}, a_{k,n}, {\mu_{\mathrm{fa}}}_{n}; M_{n}) g_{2}(\underline{\V{y}}_{k,n}, a_{k,n}; \V{z}_{n})$ and $ h(\overline{\V{y}}_{m,n}, b_{m,n},\\ {\mu_{\mathrm{fa}}}_{n}; \V{z}_{n}) \triangleq h_{1}(\overline{\V{y}}_{m,n}, b_{m,n}, {\mu_{\mathrm{fa}}}_{n}; M_{n}) h_{2}(\overline{\V{y}}_{m,n}, b_{m,n}; \V{z}_{n}) $, respectively (see Fig.~\ref{fig:factorGraph}).

\subsection{Squared Scale Parameter of the Truncated Rician \ac{pdf}}
\label{app:normAmplitude}

\begin{figure*}[htp]
	\centering
	\hspace*{8mm}\subfloat[\label{subfig:BP_distance_singleMC_fastFA}]
	{\hspace*{-11mm}\includegraphics[width=0.225\textwidth,height=0.218\textwidth]{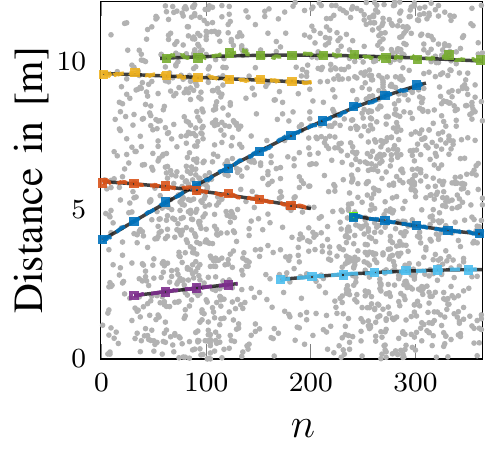}}
	\hspace*{11mm}\subfloat[\label{subfig:BP_azimuth_angle_singleMC_fastFA}]
	{\hspace*{-10mm}\includegraphics[width=0.246\textwidth,height=0.23\textwidth]{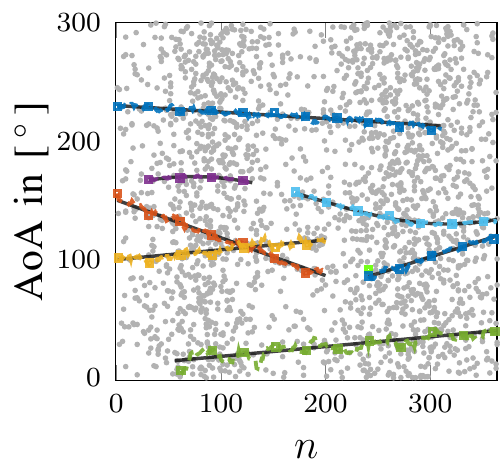}} 
	\hspace*{13mm}\subfloat[\label{subfig:BP_SNR_singleMC_fastFA}]
	{\hspace*{-11.5mm}\includegraphics[width=0.246\textwidth,height=0.225\textwidth]{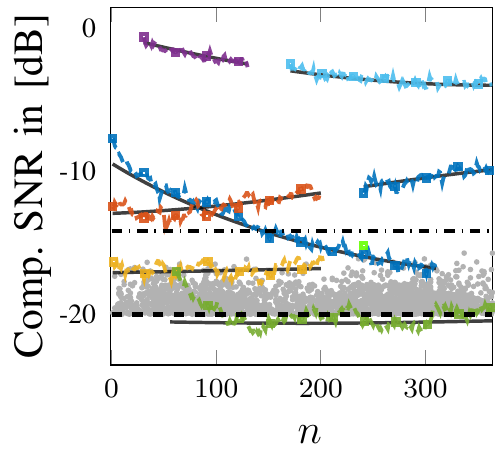}}
	\hspace*{10mm}\subfloat[\label{subfig:BP_meanFAR_singleMC_fastFA}]
	{\hspace*{-8.5mm}\includegraphics[width=0.235\textwidth,height=0.22\textwidth]{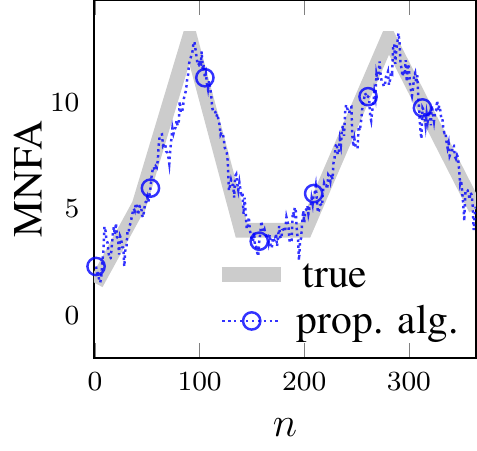}}\\[1mm]
	\caption{Results of the proposed algorithm for fully synthetic measurements given $ \mathrm{SNR}_{\mathrm{1m}}^{\mathrm{in}} = 5.4$\,dB. The gray dots denote the false alarm measurements. The black solid lines denote the true \ac{mpc} parameters. The estimates of different \acp{mpc} are denoted with densely-dashed lines with square markers in different colors. The horizontal dashed and dash-dotted lines in (c) indicate the thresholds of $ -20 $\,dB and $ -14.4 $\,dB respectively.}		 
	\label{fig:BP_fullSynMea_singleMC_fastFA}
\end{figure*}

Here, we derive the squared scale parameter of the truncated Rician \ac{pdf} in \eqref{eq:pdf_normAmplitude} in Section~\ref{subsec:LikelihoodFunctions}. We follow \cite[Ch. 3.8]{Kay_EstimationTheory} and approximate the squared scale parameter using the Fisher information which is sufficiently accurate for large amplitudes and large $N_\mathrm{s}H$.\footnote{Note that for unknown noise variance the distribution of the normalized amplitude measurement ${z_\mathrm{u}}_{m,n}$ is not described by a Rician distribution anymore. More specifically, the statistic of two times the squared \ac{pmpc}-oriented normalized amplitude measurements $2 {z^2_\mathrm{u}}_{m,n}$ is described by a non-central Fisher distribution \cite{WorsleyAIAP1994}, \cite[Ch.~15.10.3]{AdlerTaylor2007RandomFieldsGeometry}. For large $N_{\mathrm{s}}H$, the statistic of $2 {z ^2_\mathrm{u}}_{m,n} $ can be well approximated by a non-central $\chi^2$ distribution \cite[Ch.\,2.2]{Kay1998} and therefore the statistic of the normalized amplitude measurement ${z_\mathrm{u}}_{m,n}$ by a Rician distribution. However, the proof and the details are out-of-scope of this paper.}
For simplicity, we assume that the \acp{mpc} are well separated in dispersion space, so that their mutual correlations are negligible. We first determine the \ac{fim} for $ \V{\xi}_{l,n}\rmv\rmv=\rmv\rmv [ \Re\{\tilde{\alpha}_{l,n}\} \iist \Im\{\tilde{\alpha}_{l,n}\} \iist \sigma^2 ]^{\mathrm{T}}$, where $\Re\{\tilde{\alpha}_{l,n}\} $ and $ \Im\{\tilde{\alpha}_{l,n}\} $ denote the real and imaginary parts of the complex amplitudes $ \tilde{\alpha}_{l,n} $, and then apply the chain rule \cite{Kay_EstimationTheory} to get the \ac{fim} for $\tilde{u}_{l,n}$. According to \cite{Kay_EstimationTheory}, the elements of the \ac{fim} for $ \tilde{\V{\xi}}_{l,n} $ are given by
\begin{align}
\hspace*{-1mm}
[\V{J}_{l,n}(\V{\xi}_{l,n})]_{i,j} & = 2\Re \left\{ 
\frac{\partial \, \tilde{\alpha}_{l,n} \V{s}^{\mathrm{H}} (\tilde{\V{\theta}}_{l,n})} {\partial [\xi_{l,n}]_i}
\V{C}^{-1} 
\frac{\partial \V{s}(\tilde{\V{\theta}}_{l,n}) \tilde{\alpha}_{l,n}}{\partial [\xi_{l,n}]_j} \right\} \nn \\
& \quad +
\mathrm{tr} \left\{ 
\V{C}^{-1} 
\frac{\partial \V{C}^{-1} }{\partial [\xi_{l,n}]_i} 
\V{C}^{-1}
\frac{\partial \V{C}^{-1}}{\partial [\xi_{l,n}]_j} \right\}
\label{eq:FIM_complexAmp_NoiseVar1}
\end{align}
where $i,j \in \{1,2,3\}$. After some straightforward calculations \eqref{eq:FIM_complexAmp_NoiseVar1} can be rewritten as
\begin{align}
\V{J}_{l,n}(\V{\xi}_{l,n})
= 
\mathrm{diag} \left\{ \frac{2\norm{\V{s}(\tilde{\V{\theta}}_{l,n})}^2}{\sigma^2},  \frac{2\norm{\V{s}(\tilde{\V{\theta}}_{l,n})}^2}{\sigma^2}, \frac{N_{\mathrm{s}}H }{\sigma^4} \right\}\ist. \rule[-1.5em]{0pt}{0pt}
\label{eq:FIM_complexAmp_NoiseVar2}
\end{align}
The \ac{crlb} for $ \tilde{u}_{l,n} $ is obtained by applying the chain rule, i.e.,
\begin{align}
{\sigma^2_\mathrm{u}}_{l,n} \triangleq {J_{\tilde{\mathrm{u}}}}_{l,n}^{-1}(\tilde{u}_{l,n}) = \V{t}_{l,n}^{\mathrm{H}} 
\V{J}^{-1}_{l,n}(\V{\xi}_{l,n})
\V{t}_{l,n} = \frac{1}{2} + \frac{\tilde{u}_{l,n}^2}{4N_{\mathrm{s}}H}
\label{eq:crlb_ampVar}
\end{align}
where the Jacobian $ \V{t}_{l,n} $ containing the partial derivatives is 
\begin{align}
\V{t}_{l,n} & \triangleq \left[ \frac{\partial \tilde{u}_{l,n}}{\partial \Re\{\tilde{\alpha}_{l,n}\}} \quad \frac{\partial \tilde{u}_{l,n}}{\partial \Im\{\tilde{\alpha}_{l,n}\}} \quad  \frac{\partial \tilde{u}_{l,n}}{\partial \sigma^2} \right]^{\mathrm{T}} \nn \\
& = \left[ 
\frac{\Re\{\tilde{\alpha}_{l,n}\} \norm{\V{s}(\tilde{\V{\theta}}_{l,n})}}{|\tilde{\alpha}_{l,n}|\, \sigma} \quad
\frac{\Im\{\tilde{\alpha}_{l,n}\} \norm{\V{s}(\tilde{\V{\theta}}_{l,n})}}{|\tilde{\alpha}_{l,n}|\, \sigma} \right. \quad \nn \\
& \hspace*{40mm}
\left. \frac{-|\tilde{\alpha}_{l,n}|\, \norm{\V{s}(\tilde{\V{\theta}}_{l,n})}} {2\,\sigma^3}\right]^{\mathrm{T}}\ist. 
\end{align} 
Note that the second term $ \frac{\tilde{u}_{l,n}^2}{4N_{\mathrm{s}}H} $ in \eqref{eq:crlb_ampVar} characterizes the effect of the noise variance estimation, which becomes significant for high component \acp{snr}, and converges to zero for low component \acp{snr} or a large $N_\mathrm{s}H$. Thus, for \ac{pmpc}-oriented measurements the squared scale parameter of the truncated Rician \ac{pdf} in \eqref{eq:pdf_normAmplitude} is given by \eqref{eq:crlb_ampVar} and for false alarm measurements the squared scale parameter of the Rayleigh \ac{pdf} is given by $ 1/2 $.

\subsection{Additional Simulation Results}
\label{app:AddSimResults}

\vspace*{2mm}
\subsubsection{Fast Varying \ac{far}}
\label{app:AddSimResults_MNFA}

As shown in Section~\ref{subsec:fullSynMea}, the proposed algorithm accurately estimate the slow varying \ac{far}. With an increased standard deviation of the driving noise to the \ac{far} state-transition \ac{pdf} $\sigma_{\mathrm{fa}} =  0.5$, the results for fully synthetic measurements shown in Fig.~\ref{fig:BP_fullSynMea_singleMC_fastFA} further demonstrate the excellent performance of the proposed algorithm when handling fast varying \ac{far}.

\vspace*{2mm}
\subsubsection{Synthetic Radio Measurements}
\label{app:AddSimResults_synMea}

\begin{figure*}[t]
	\centering
	\hspace*{9.5mm}\subfloat[\label{subfig:BP_distance_singleMC}]
	{\hspace*{-11mm}\includegraphics[width=0.235\textwidth,height=0.218\textwidth]{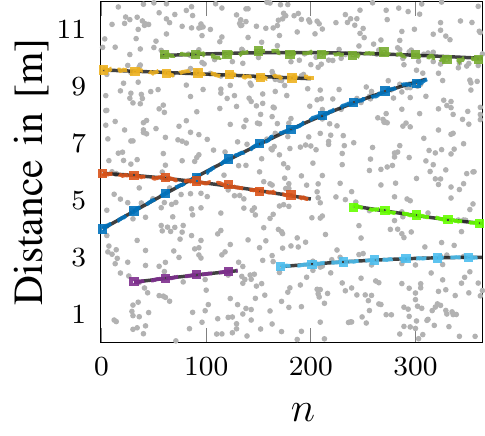}}
	\hspace*{11mm}\subfloat[\label{subfig:BP_azimuth_angle_singleMC}]
	{\hspace*{-10mm}\includegraphics[width=0.246\textwidth,height=0.23\textwidth]{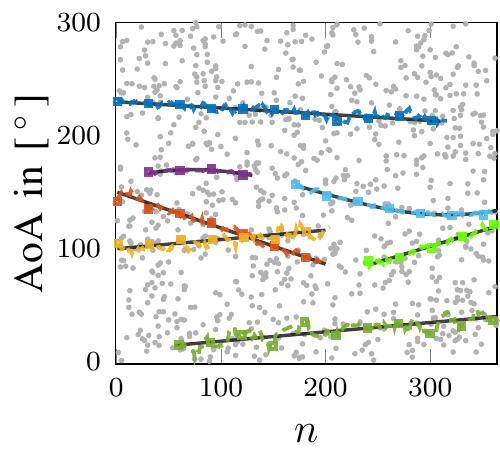}} 
	\hspace*{13mm}\subfloat[\label{subfig:BP_SNR_singleMC}]
	{\hspace*{-11.5mm}\includegraphics[width=0.246\textwidth,height=0.225\textwidth]{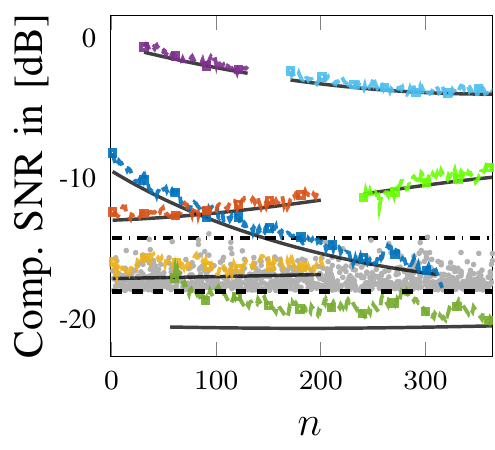}}
	\hspace*{12mm}\subfloat[\label{subfig:BP_detectedNumMPC_singleMC}]
	{\hspace*{-11mm}\includegraphics[width=0.235\textwidth,height=0.22\textwidth]{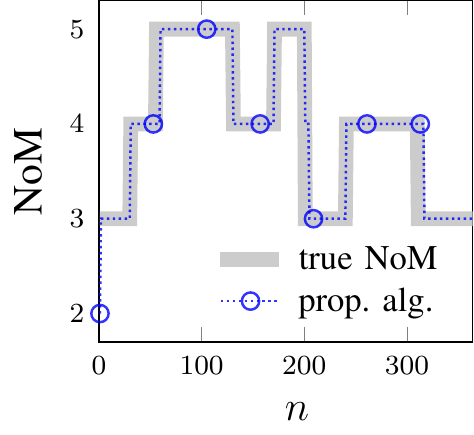}}\\[-2mm]
	\hspace*{9.5mm}\subfloat[\label{subfig:KEST_distance_singleMC}]
	{\hspace*{-11mm}\includegraphics[width=0.235\textwidth,height=0.2218\textwidth]{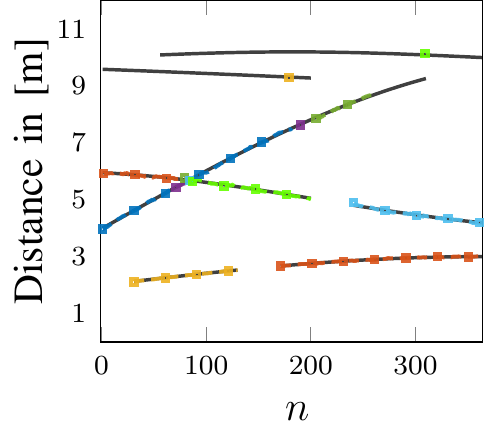}}
	\hspace*{11mm}\subfloat[\label{subfig:KEST_azimuth_angle_singleMC}]
	{\hspace*{-10mm}\includegraphics[width=0.246\textwidth,height=0.23\textwidth]{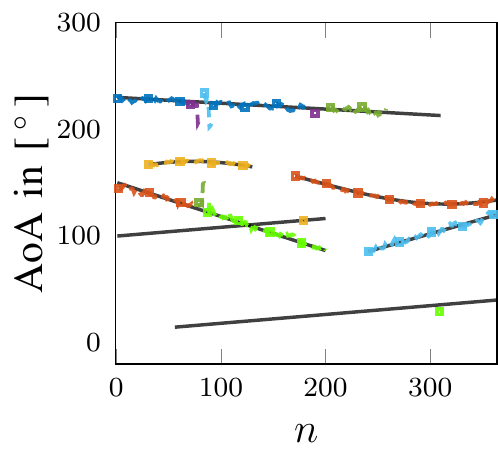}} 
	\hspace*{13mm}\subfloat[\label{subfig:KEST_SNR_singleMC}]
	{\hspace*{-11mm}\includegraphics[width=0.246\textwidth,height=0.225\textwidth]{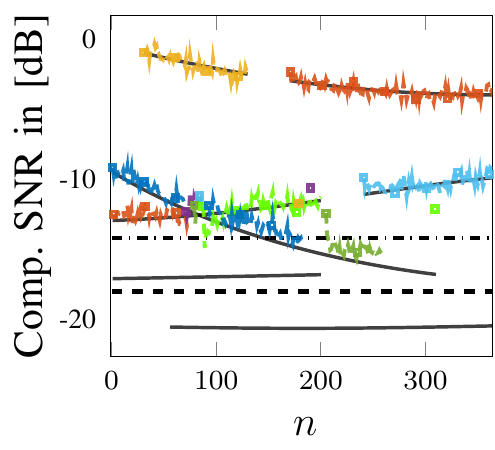}}
	\hspace*{12mm}\subfloat[\label{subfig:KEST_detectedNumMPC_singleMC}]
	{\hspace*{-11mm}\includegraphics[width=0.235\textwidth,height=0.22\textwidth]{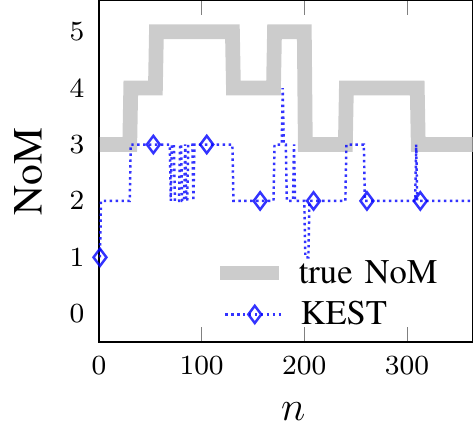}}\\[1mm]
	\caption{Results for synthetic radio measurements given $ \mathrm{SNR}_{\mathrm{1m}}^{\mathrm{in}} = 5.4$\,dB: (a) to (d), results of the proposed algorithm, the gray dots denote the false alarm measurements. (e) to (h), results of the \ac{kest} algorithm. The black solid lines denote the true \ac{mpc} parameters. The estimates of different \acp{mpc} are denoted with densely-dashed lines with square markers in different colors. The horizontal dashed and dash-dotted lines indicate the thresholds of $ -18 $\,dB and $ -14.4 $\,dB respectively.}	 
	\label{fig:BP_synChannelMea_singleMC}
\end{figure*}

Fig.~\ref{fig:BP_synChannelMea_singleMC} presents the results of the proposed algorithm and the \ac{kest} algorithm of an exemplary simulation run given $ \mathrm{SNR}_{\mathrm{1m}}^{\mathrm{in}} = 5.4$\,dB. When handling well-separated \acp{mpc} with high component \acp{snr}, the algorithms perform similarly good. The slightly over estimated component \acp{snr} shown in Fig.~\ref{subfig:BP_SNR_singleMC} are due to the Gaussian approximation on the likelihood function (see for Section~\ref{subsubsec:SimulationParameters}). Regarding the \ac{mpc} below the threshold of $-18$\,dB, it is immediately detected and accurately estimated in the proposed algorithm but totally missed in the \ac{kest} algorithm. Additionally, with probabilistic \ac{da} the proposed algorithm properly copes with intersecting \acp{mpc} around $ n = 83 $, while the association uncertainty which further leads to discontinuity of sequential estimations is clearly shown with the \ac{kest} algorithm.

\vspace*{2mm}
\subsubsection{Runtime Analysis}
\label{app:runtimeAnalysis}

\begin{figure*}[t]
	\centering
	\hspace*{8mm}\subfloat[\label{subfig:meanRunTimeAnalysis1}]
	{\hspace*{-10mm}\includegraphics[width=0.37\textwidth,height=0.22\textwidth]{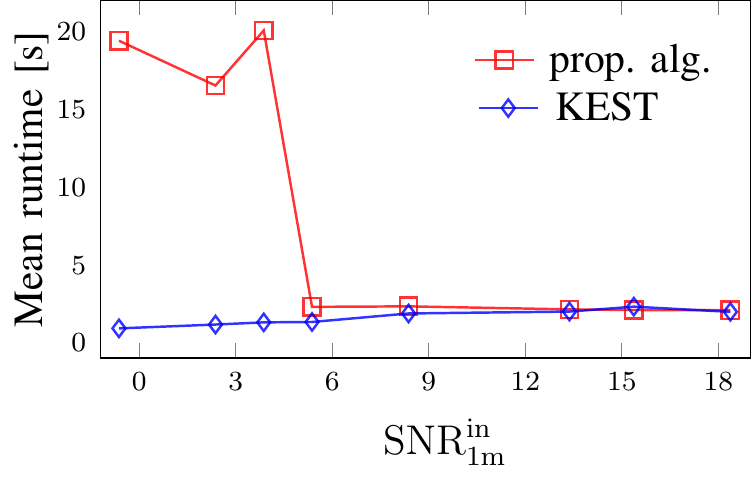}}
	\hspace*{19mm}\subfloat[\label{subfig:meanRunTimeAnalysis2}]
	{\hspace*{-10mm}\includegraphics[width=0.37\textwidth,height=0.22\textwidth]{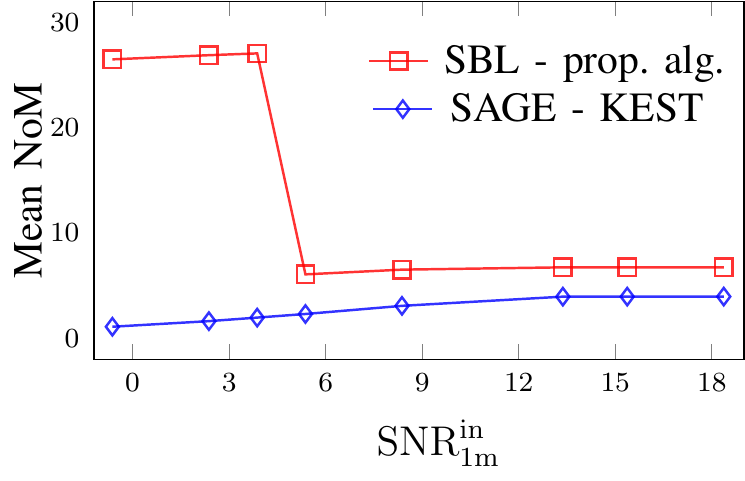}}\\[1mm] 
	\caption{Mean runtimes per time step of the proposed algorithm and the \ac{kest} algorithm using MATLAB implementations, and the mean \ac{nom} per time step estimated using the snapshot-based channel estimators \ac{sbl} and SAGE for synthetic radio measurements. The mean values are obtained by averaging over all simulation runs and time steps for each $ \mathrm{SNR}_{\mathrm{1m}}^{\mathrm{in}} \in \{-0.6, 2.4, 3.9, 5.4, 8.4, 13.4, 15.4, 18.4\} $\,dB.}	 
	\label{fig:meanRunTime}
\end{figure*}

The overall computational complexity of a sequential channel estimation method depends on many factors like the \ac{nom} and the number of measurements, and mostly importantly the complexity of snapshot-based channel estimator. Fig.~\ref{subfig:meanRunTimeAnalysis1} depicts the mean runtimes per time step of the proposed algorithm and the \ac{kest} algorithm for synthetic radio measurements.{\footnote{The simulation results were obtained by running MATLAB implementations on a cluster of computers with different hardware configurations. }} The mean runtimes are obtained by averaging over all simulation runs and time steps for each $ \mathrm{SNR}_{\mathrm{1m}}^{\mathrm{in}} \in \{-0.6, 2.4, 3.9, 5.4, 8.4, 13.4, 15.4, 18.4\} $\,dB. When the $ \mathrm{SNR}_{\mathrm{1m}}^{\mathrm{in}} $ is larger than $5\,$dB, the mean \acp{nom} estimated in the \ac{sbl} and the SAGE parametric channel estimators are comparable as shown in Fig.~\ref{subfig:meanRunTimeAnalysis2},{\footnote{At each time step, the \ac{kest} algorithm needs to run the SAGE channel estimator several times for testing different \ac{nom} hypotheses, the mean \ac{nom} of the SAGE shown in Fig.~\ref{subfig:meanRunTimeAnalysis2} depicts the mean \ac{nom} chosen by the MDL principle used in the \ac{kest}.}} and the mean runtimes of the proposed algorithm and the \ac{kest} algorithm are both approximately $ 2\,$s\,/\,step as depicted in Fig.~\ref{subfig:meanRunTimeAnalysis2}. For $ \mathrm{SNR}_{\mathrm{1m}}^{\mathrm{in}} $ smaller than $5\,$dB the detection threshold $u_\mathrm{de}^{\mathrm{in}} $ in the \ac{sbl} channel estimator is further relaxed from $-18\,$dB to $-20\,$dB to support the detection of low \ac{snr} \acp{mpc} (see for Section~\ref{sec:resultSynRadioMeas}), therefore the mean \ac{nom} per time step estimated in the \ac{sbl} increased sharply from 5 to 30, correspondingly the mean runtime increased from $ 2\,$s\,/\,step to $ 20\,$s\,/\,step. The results above illustrate that the mean runtime is approximately linear-proportional to the mean \ac{nom} estimated in the snapshot-based channel estimator.

%% file: InputFiles/bios.tex
\vspace*{-7mm}

\begin{IEEEbiography}[{\includegraphics[width=24mm,height=32.15mm,clip,keepaspectratio]{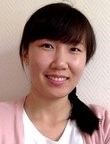}}]{Xuhong~Li} received her B.Sc. degree in telecommunication engineering from Jilin University, China, in 2010, and her M.Sc. degree in wireless communication from Lund University, Lund, Sweden, in 2014. Currently, she is pursuing her Ph.D. degree at the Department of Electrical and Information Technology, Lund University. Her research interests include radio-based localization and estimation of radio channels. 
\end{IEEEbiography}
\vspace*{-7mm}

\begin{IEEEbiography}[{\includegraphics[width=25mm,height=32.15mm,clip,keepaspectratio]{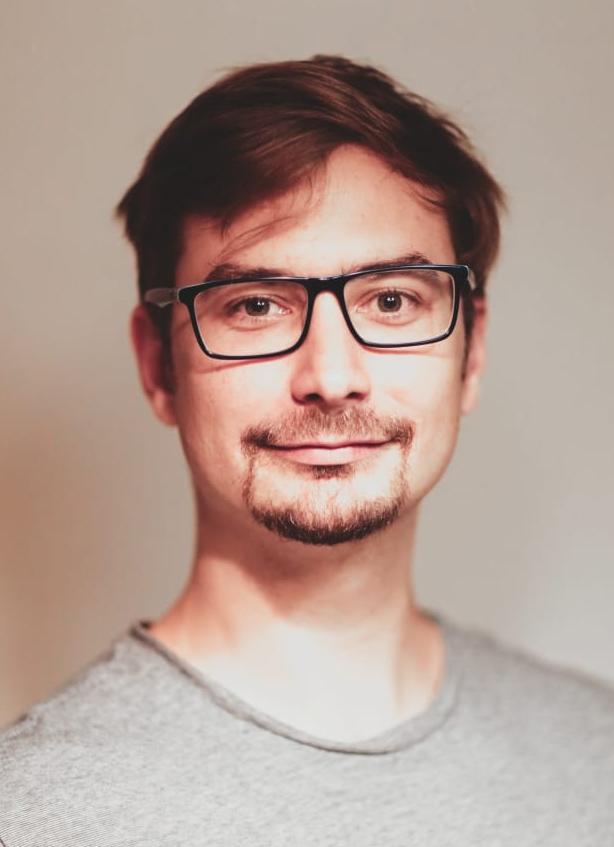}}]{Erik~Leitinger} (Member, IEEE) received the Dipl.\-Ing.\ (M.Sc.) and Ph.D.\ degrees (with highest Hons.) in electrical engineering from Graz University of Technology, Austria in 2012 and 2016. He is currently Senior Research Scientist with the Graz University of Technology. From 2016 to 2018, he was Postdoctoral Researcher at the department of Electrical and Information Technology at Lund University. He is an Erwin Schr\"{o}dinger Fellow.

Dr.\ Leitinger received an Award of Excellence from the Federal Ministry of Science, Research and Economy (BMWFW) for his Ph.D.\ Thesis. He served as co-chair of the special session ``Synergistic Radar Signal Processing and Tracking'' at the IEEE Radar Conference in 2021 and co-organizer of the special issue ``Graph-Based Localization and Tracking'' in the Journal of Advances in Information Fusion (JAIF). His research interests include statistical signal processing, nonlinear estimation and detection, inference on graphs, localization and navigation, SLAM, estimation of radio channels, and estimation/detection theory. 
	
\end{IEEEbiography}
\vspace*{-7mm}

\begin{IEEEbiography}[{\includegraphics[width=25mm,height=32.15mm,clip,keepaspectratio]{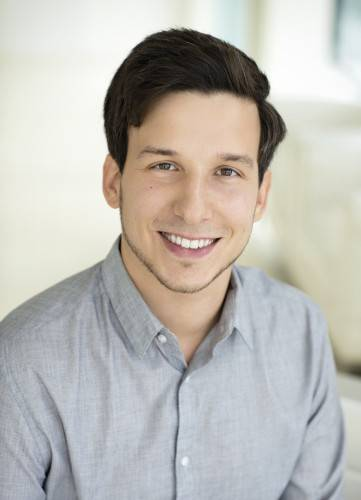}}]{Alexander~Venus} (S'20) received his B.Sc.\ and Dipl.-Ing.\ (M.Sc.) degrees (with highest Hons.) in biomedical engineering and information and communication engineering from Graz University of Technology, Austria in 2012 and 2015, respectively. He was a research and development engineer at Anton Paar GmbH, Graz from 2014 to 2019. He is currently a project assistant at Graz University of Technology, where he is pursuing his Ph.D.\ degree.
 	
His research interests include radio-based localization and navigation, estimation of radio channels, inference on graphs, stochastic modeling, and estimation/detection theory. 
\end{IEEEbiography}

\vspace*{-7mm}

\begin{IEEEbiography}[{\includegraphics[width=25mm,height=32.15mm,clip,keepaspectratio]{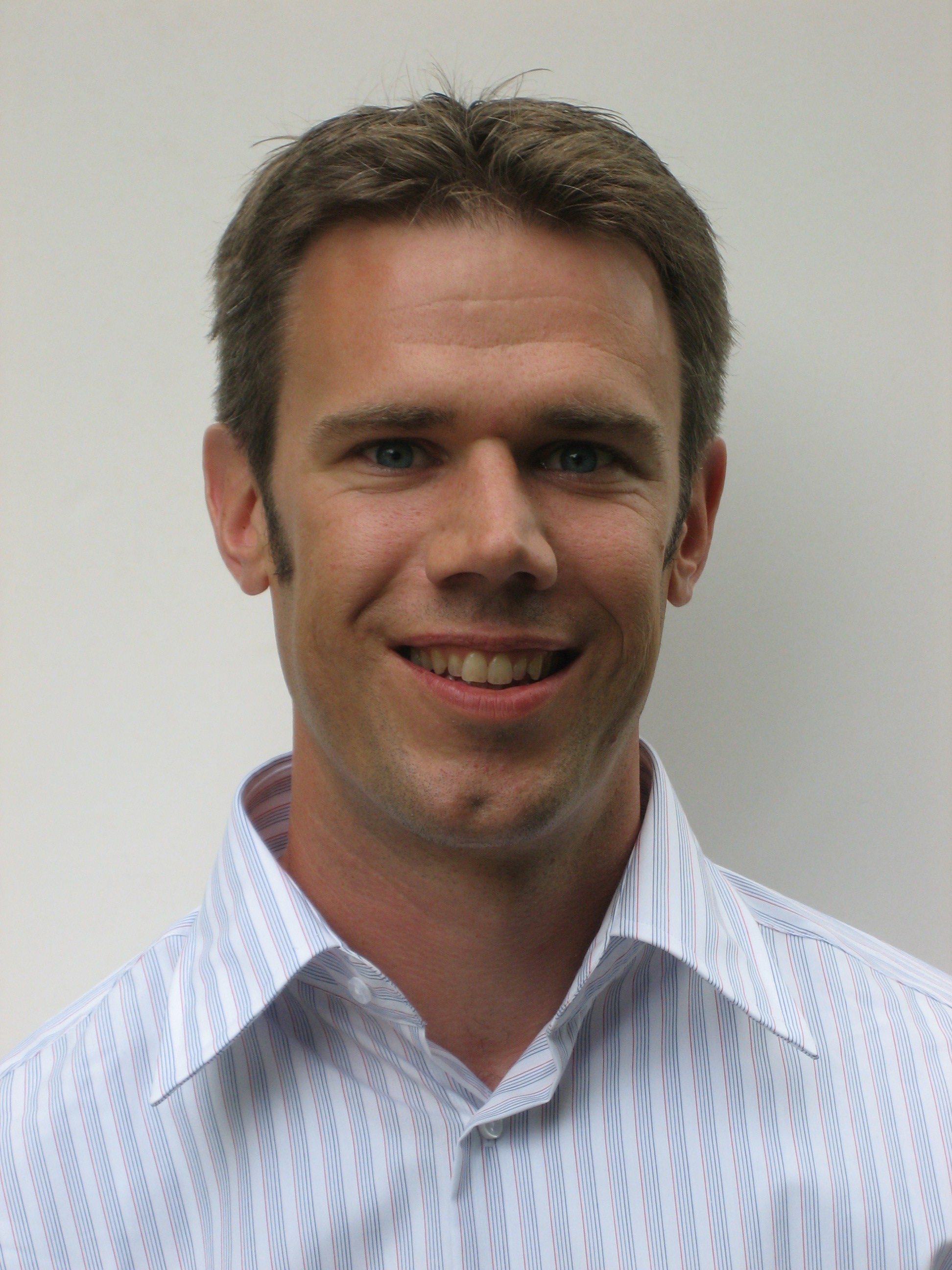}}]{Fredrik~Tufvesson} received his Ph.D. in 2000 from Lund University in Sweden. After two years at a startup company, he joined the department of Electrical and Information Technology at Lund University, where he is now professor of radio systems. His main research interest is the interplay between the radio channel and the rest of the communication system with various applications in 5G/B5G systems such as massive MIMO, mm wave communication, vehicular communication and radio based positioning.
	
Fredrik has authored around 100 journal papers and 150 conference papers, he is fellow of the IEEE and his research has been awarded with the Neal Shepherd Memorial Award (2015) for the best propagation paper in {\sc IEEE Transactions on Vehicular Technology} and the {\sc IEEE Communications Society} best tutorial paper award (2018, 2021).
	
\end{IEEEbiography}

%
%
%